\documentclass[11pt,a4paper]{article}

\usepackage[top=2.5cm, bottom=2.5cm, left=3cm, right=3cm]{geometry}

\usepackage{amsmath}
\usepackage{amssymb}

\usepackage{booktabs}
\usepackage{tabularx}

\usepackage{enumitem}

\usepackage{caption}
\usepackage{subcaption}

\usepackage{float}

\usepackage{pgfplots}
\pgfplotsset{compat=1.18}

\usepackage{tikz}
\usetikzlibrary{arrows.meta, positioning, decorations.pathreplacing, calc, backgrounds, fit}

\usepackage[T1]{fontenc}
\usepackage[utf8]{inputenc}
\usepackage{lmodern}

\usepackage{hyperref}
\hypersetup{
    colorlinks=true,
    linkcolor=blue!60!black,
    urlcolor=blue!60!black,
    citecolor=blue!60!black,
    pdfborder={0 0 0}
}

\usepackage{microtype}

\usepackage{array}

\usepackage{authblk}

\usepackage{fancyhdr}
\DeclareMathOperator{\clamp}{clamp}
\DeclareMathOperator{\median}{median}

\title{Predictive Autoscaling for Node.js on Kubernetes:\\ Lower Latency, Right-Sized Capacity}
\author{Ivan Tymoshenko}
\author{Luca Maraschi}
\author{Matteo Collina, Ph.D}
\affil{Platformatic}
\date{March 2026}

\begin{document}

\maketitle

\begin{abstract}
Kubernetes offers two default paths for scaling Node.js workloads, and both
have structural limitations. The Horizontal Pod Autoscaler scales on CPU
utilization, which does not directly measure event loop saturation: a Node.js
pod can queue requests and miss latency SLOs while CPU reports moderate usage.
KEDA extends HPA with richer triggers, including event-loop metrics, but
inherits the same reactive control loop, detecting overload only after it has
begun. By the time new pods start and absorb traffic, the system may already
be degraded. Lowering thresholds shifts the operating point but does not
change the dynamic: the scaler still reacts to a value it has already crossed,
at the cost of permanent over-provisioning.

We propose a predictive scaling algorithm that forecasts where load will be by
the time new capacity is ready and scales proactively based on that forecast.
Per-instance metrics are corrupted by the scaler's own actions: adding an
instance redistributes load and changes every metric, even if external traffic
is unchanged. We observe that operating on a cluster-wide aggregate that is
approximately invariant under scaling eliminates this feedback loop, producing
a stable signal suitable for short-term extrapolation.

We define a \emph{metric model} (a set of three functions that encode how a
specific metric relates to scaling) and a five-stage pipeline that transforms
raw, irregularly-timed, partial metric data into a clean prediction signal. In
benchmarks against HPA and KEDA under steady ramp and sudden spike, the
algorithm keeps per-instance load near the target threshold throughout.
Under the steady ramp, median latency is 26\,ms, compared to 154\,ms
for KEDA and 522\,ms for HPA.
\end{abstract}

\tableofcontents
\newpage

\section{Introduction \& Motivation}
\label{sec:introduction}

\subsection{The Problem}

Scalers typically react to the current value of a metric: when it crosses a threshold, they add instances.
Each evaluation is discrete: the scaler sees a single value, compares it to a target, and makes a decision.
There is no memory of prior state, no understanding of whether the metric is rising, falling, or stable.
The scaler cannot distinguish a sustained rise from a momentary spike (both look the same at the moment
of evaluation). It cannot anticipate overload, it detects it only after the threshold has been crossed.
And it cannot right-size the response, because without knowing how fast the metric is changing, it has no basis
to estimate how many instances will actually be needed. Existing solutions and their limitations are
compared in Section~\ref{sec:existing-solutions}.

These problems are amplified by the startup gap: the delay between a scaling decision and the moment new
capacity is ready. A new instance must start, initialize, and begin absorbing traffic, a process that
takes seconds to minutes. During this window, the existing instances carry the full load. A scaler that
only reacts when the threshold is crossed is already behind: by the time the new instances are ready, the system
may already be degraded.

\subsection{Predict and Act Ahead}

The \emph{predictive scaler} uses time-series forecasting to estimate where the load will be in the near future ---
by the time a new instance would start and absorb its share of the traffic.

If the predicted load at that future point exceeds the capacity of the current instance count, the scaler adds
instances \emph{now}, so they are ready exactly when the extra capacity is needed.

This shifts the scaling decision from ``we are overloaded, add capacity'' to ``we will be overloaded in $T$
seconds, add capacity now so it's ready in time.''

\subsection{The Core Idea}

The main idea is to take per-instance metrics, combine them into a single value that represents the total load
on the cluster, predict where that value is heading, and convert the prediction back into an instance count.

This approach requires two properties of the aggregation to work.

First, the aggregate must \textbf{encode the number of instances}. A per-instance
statistic (whether an average, median, or percentile) describes load intensity but says nothing about
scale: an average of 0.5 across 2 instances is a fundamentally different cluster state than 0.5 across 10.
To understand cluster health at any moment, the metric value alone is not enough, you also need to know
how many instances produced it. A cluster-wide aggregate that sums across all instances captures both
in a single number: the load intensity and the number of instances carrying it are folded together.

Second, the aggregate must be \textbf{approximately invariant under scaling}. Scaling changes
how load is distributed across instances, not how much load there is. A per-instance metric depends on both ---
it moves when load changes and when instances are added or removed. A prediction based on the per-instance
metric would be corrupted by the algorithm's own actions.

An aggregate with the invariance property doesn't have this problem. When instances are added and load
redistributes, individual metrics shift but the aggregate stays close to where it was. This lets the algorithm
base its predictions on the load itself, not on the side effects of its own scaling decisions.

The simplest aggregation with both properties is a sum: $\mathcal{A}(\mathbf{v}) = \sum v_i$. The examples
and figures below use it for concreteness, but the algorithm works with any aggregation that satisfies
these two properties (the general form is defined in Section~\ref{sec:architecture}).

\begin{figure}[H]
\centering
\begin{tikzpicture}
\begin{axis}[
    width=0.92\textwidth,
    height=5.5cm,
    xlabel={Time},
    ylabel={Metric value},
    xmin=0, xmax=62,
    ymin=0, ymax=0.95,
    xtick={0, 10, 20, 30, 40, 50, 60},
    xticklabels={$t_1$, $t_2$, $t_3$, $t_4$, $t_5$, $t_6$, $t_7$},
    ytick={0, 0.2, 0.4, 0.6, 0.8},
    yticklabel style={text width=2em, align=right},
    grid=major,
    major grid style={dashed, black!15},
    legend style={at={(0.03,0.97)}, anchor=north west, font=\small},
    every axis x label/.style={at={(ticklabel* cs:1.0)}, anchor=west},
    clip=false,
]

\draw[dashed, black!50] (axis cs:20,0) -- (axis cs:20,0.95);
\node[anchor=south, font=\small] at (axis cs:20,0.95) {scale-up};
\draw[dashed, black!50] (axis cs:40,0) -- (axis cs:40,0.95);
\node[anchor=south, font=\small] at (axis cs:40,0.95) {scale-up};

\addplot[thick, blue!70!black, mark=none] coordinates {
    (0,0.30) (4,0.40) (8,0.50) (12,0.60) (16,0.70)
    (20,0.75) (24,0.65) (28,0.55) (32,0.62) (36,0.68)
    (40,0.65) (44,0.53) (48,0.58) (52,0.65) (56,0.65) (60,0.65)
};
\addlegendentry{Instance A}

\addplot[thick, orange!80!black, mark=none] coordinates {
    (20,0.05) (24,0.25) (28,0.45) (32,0.48) (36,0.52)
    (40,0.60) (44,0.72) (48,0.62) (52,0.55) (56,0.57) (60,0.60)
};
\addlegendentry{Instance B}

\addplot[thick, green!50!black, mark=none] coordinates {
    (40,0.05) (44,0.15) (48,0.30) (52,0.40) (56,0.48) (60,0.55)
};
\addlegendentry{Instance C}

\end{axis}
\end{tikzpicture}
\caption{Per-instance metrics as instances are added over time. Each curve shows the load on
one instance, but from these values alone it is not possible to determine whether the total
load on the cluster is growing, shrinking, or stable.}
\label{fig:mechanism-per-instance}
\end{figure}
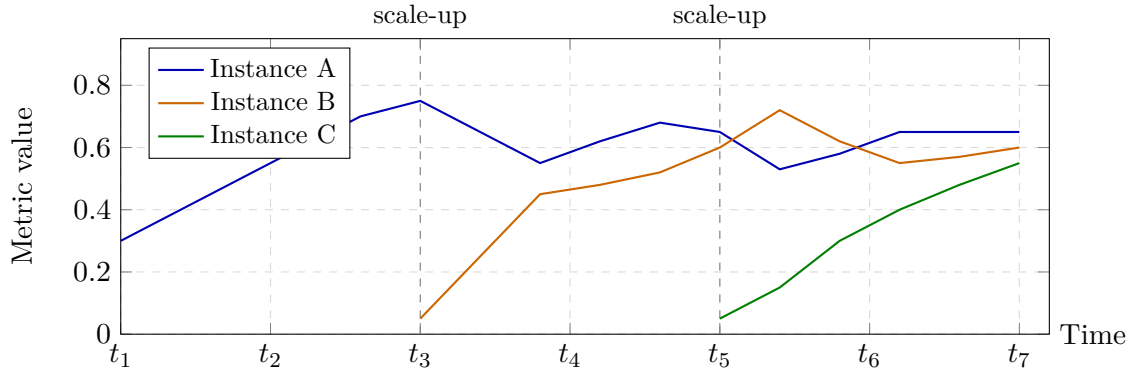

\begin{figure}[H]
\centering
\begin{tikzpicture}
\begin{axis}[
    width=0.92\textwidth,
    height=5.5cm,
    xlabel={Time},
    ylabel={Metric value},
    xmin=0, xmax=62,
    ymin=0, ymax=2.0,
    xtick={0, 10, 20, 30, 40, 50, 60},
    xticklabels={$t_1$, $t_2$, $t_3$, $t_4$, $t_5$, $t_6$, $t_7$},
    ytick={0, 0.4, 0.8, 1.2, 1.6, 2.0},
    yticklabel style={text width=2em, align=right},
    grid=major,
    major grid style={dashed, black!15},
    legend style={at={(0.03,0.97)}, anchor=north west, font=\small},
    every axis x label/.style={at={(ticklabel* cs:1.0)}, anchor=west},
]

\addplot[thick, black, mark=none] coordinates {
    (0,0.30) (4,0.40) (8,0.50) (12,0.60) (16,0.70)
    (20,0.80) (24,0.90) (28,1.00) (32,1.10) (36,1.20)
    (40,1.30) (44,1.40) (48,1.50) (52,1.60) (56,1.70) (60,1.80)
};
\addlegendentry{Aggregate (sum)}

\end{axis}
\end{tikzpicture}
\caption{The aggregate (sum) over the same period. Despite the chaotic
per-instance behavior, the total load grows as a steady linear trend.}
\label{fig:mechanism-aggregate}
\end{figure}
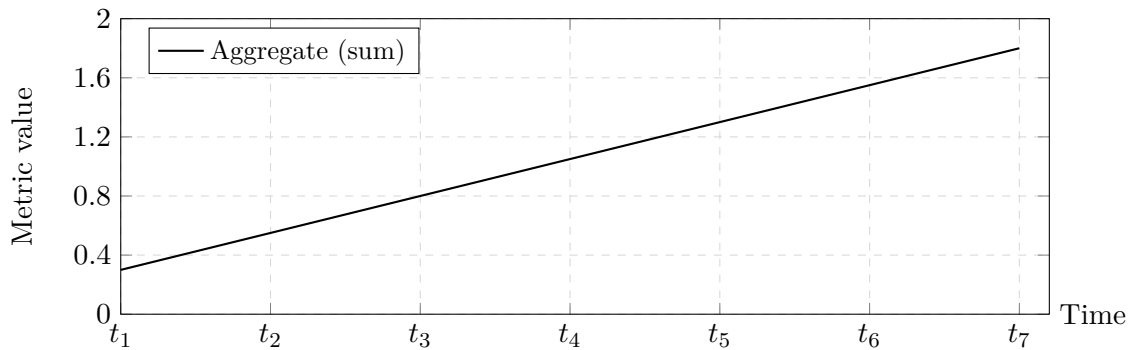

\paragraph{The metric model.}
The relationship between per-instance values and the cluster-wide aggregate is encapsulated
in a \emph{metric model}: a set of three functions specific to each metric.

\begin{itemize}[leftmargin=*, itemsep=3pt]
    \item $\mathcal{A}$ (\textbf{aggregation}), combines per-instance values into a single number representing
        the aggregate load on the cluster. Accepts optional per-instance weights for partial inclusion
        (used during redistribution, Section~\ref{sec:redistribution}).
    \item $\mathcal{P}$ (\textbf{projection}), the reverse: given an aggregate load and an instance count,
        produces the expected per-instance metric value.
    \item $\mathcal{N}$ (\textbf{required count}), given an aggregate load and a per-instance threshold,
        returns the number of instances needed to keep the per-instance metric at or below the threshold.
\end{itemize}

The metric model must satisfy two properties. First, \textbf{scaling invariance}: the aggregate stays
approximately constant when load redistributes across instances after scaling, so the prediction signal
reflects external load rather than the algorithm's own actions. Second, \textbf{per-instance separability}:
each instance's contribution to the aggregate can be independently scaled by a weight, so the redistribution
stage can gradually include new instances without creating artifacts.

The simplest metric model is sum and average:
$\mathcal{A}(\mathbf{v}) = \sum v_i$, \;
$\mathcal{P}(S, N) = S / N$, \;
$\mathcal{N}(S, \tau) = S / \tau$.
This works when the metric redistributes equally across instances and the entire measured value represents
load-related work. When the metric has a more complex relationship with scaling (for example, a fixed
per-instance baseline that does not redistribute, or a non-linear sensitivity to load), the model can be
adjusted accordingly. The formal interface, its requirements, and examples of non-trivial models are given
in Section~\ref{sec:architecture} and Section~\ref{sec:metric-model-examples}.

\subsection{The Metric}

The algorithm works with any numeric, per-instance metric that satisfies these properties:

\begin{itemize}[leftmargin=*, itemsep=4pt]
    \item \textbf{Monotonically related to load.} Higher values mean more load on the instance.
    \item \textbf{Distributed across instances.} The metric reflects each instance's share of the total
        workload. When a new instance is added and absorbs traffic, the metric on existing instances decreases.
    \item \textbf{Has a meaningful threshold.} There is a per-instance value above which the instance is
        considered overloaded.
\end{itemize}

For example:

\begin{itemize}[leftmargin=*, itemsep=4pt]
    \item \textbf{ELU (Event Loop Utilization)}~\cite{nodejs-elu}, how busy the Node.js event loop is, on a 0--1 scale.
        A primary indicator of CPU-bound load in Node.js applications.
    \item \textbf{Heap usage}, memory pressure per instance. Useful for detecting memory-bound workloads.
\end{itemize}

Each metric is processed independently. When multiple metrics are used, the highest
target instance count wins.

\subsection{Challenges}

Predicting load is straightforward in theory (fit a trend line and extrapolate). In practice, several
complications arise:

\begin{itemize}[leftmargin=*, itemsep=4pt]
    \item \textbf{Working with the latest data.}
        To react fast and build accurate predictions, the algorithm needs to work with the newest data
        possible. But in most systems, the scaler doesn't have real-time access to instance metrics ---
        continuous monitoring of every instance is impractical. Instead, metrics are delivered in batches:
        instances collect measurements and send them periodically. Since batches from different instances
        arrive at different times, the algorithm must be able to work with partial data.

    \item \textbf{The redistribution problem.}
        After scaling up, existing instances don't shed load immediately: queues need to drain, in-flight
        requests must complete, garbage collection and buffers take time to settle. Meanwhile, the new instance
        may start receiving traffic right away. During this transition, per-instance metrics don't reflect the
        actual distribution of load, which can mislead any algorithm that relies on them for prediction.

    \item \textbf{Noise vs.\ trend.}
        With historical data, noise is easy to spot (it's a momentary spike that came back down). At the edge
        of the data, a new movement could be noise or the start of a real trend, and the difference isn't yet
        visible. The algorithm must determine when it's confident enough to treat a movement as a real trend and
        act on it.
\end{itemize}

\subsection{Applicability}

The algorithm is deliberately abstract. It operates on \emph{instances} and \emph{metric samples},
without assuming anything about the infrastructure that manages them. An instance can be a Kubernetes
pod, an OS process, a worker thread, or a virtual machine; the algorithm treats them identically.

How instances are created, destroyed, or routed traffic is outside the algorithm's scope. So is how
metrics are delivered. The algorithm includes stages that handle the constraints of batch-based
delivery: alignment (Section~\ref{sec:alignment}) places irregularly-timed samples onto a uniform
grid, and imputation (Section~\ref{sec:imputation}) estimates values for instances that haven't
reported yet. In environments where the scaler can read metrics from all instances simultaneously ---
for example, a runtime that scales its own processes or threads, these stages may be unnecessary
and can be skipped as an optimization. Whether this applies depends on the specific implementation
and should be verified case by case.

Section~\ref{sec:icc-implementation} describes an implementation for scaling Kubernetes application
pods with the Platformatic Intelligent Command Center.

\section{Existing Scaling Solutions}
\label{sec:existing-solutions}

Several scaling approaches are widely deployed. They differ in trigger mechanisms and
details, but most share a fundamental pattern: evaluate the metric at a point in time,
compare it to a target, and adjust the instance count. (Vertical scaling, e.g.\ the
Vertical Pod Autoscaler, adjusts per-instance resources rather than instance count
and is outside the scope of this comparison.)

\paragraph{Reactive approaches.}
The Kubernetes Horizontal Pod Autoscaler~\cite{k8s-hpa} (HPA) computes the desired
replica count from the ratio of the current metric value to the target value.
KEDA~\cite{keda} extends HPA with a wide range of event-driven triggers (queue length,
HTTP request rate, custom metrics), but the scaling logic is the same. The Knative
Autoscaler~\cite{knative-autoscaler} targets per-instance request concurrency and adds a
shorter panic window for rapid scale-up, but the panic mode reacts faster, not earlier.

\paragraph{Pattern-based prediction.}
AWS Predictive Scaling~\cite{aws-predictive} analyzes historical load patterns using
machine learning and pre-provisions capacity hours ahead. It is effective for recurring,
predictable workloads (daily or weekly traffic cycles), but does not handle sudden
unexpected spikes. A reactive scaler is still needed as a fallback.

\paragraph{Limitations of existing approaches.}
All reactive solutions treat the metric as a discrete value sampled at evaluation time.
Each decision is independent, with no understanding of direction, velocity, or
dynamics. Pattern-based prediction adds foresight, but on a different timescale: it learns
recurring cycles over hours and days, not emerging trends over seconds and minutes.

This algorithm takes a fundamentally different approach. It treats the metric as a
continuous signal, where every sample contributes to a running estimate of the level and
its rate of change. This continuous understanding provides dynamic awareness: the
algorithm knows not just where the metric is, but where it is heading. Short-term prediction
follows naturally: the trend is already known, so extrapolation is straightforward.

Existing solutions apply the same formula to every metric regardless of its nature. This
algorithm operates with a metric model that captures how the specific metric relates to
scaling. This allows it to treat each metric according to its actual behavior, producing
more precise scaling decisions rather than applying a single generic rule to every problem.

The architecture is described in Section~\ref{sec:architecture} and the details follow in
Sections~\ref{sec:alignment}--\ref{sec:scaling-decision}.

\begin{table}[H]
\centering
\begin{tabularx}{\textwidth}{@{} l X X @{}}
\toprule
& \textbf{Where it excels} & \textbf{Where it falls short} \\
\midrule
\textbf{HPA}
& Simple to configure: one target value per metric, no parameter tuning.
& Uses per-instance average as its signal. When a pod is added, the average drops even
  if external load hasn't changed; the signal is corrupted by the scaler's own actions,
  causing oscillation. Relies on cooldowns as a blunt workaround. \\
\midrule
\textbf{KEDA}
& Wide ecosystem of event-driven triggers out of the box (message queues,
  HTTP rates, databases).
& The scaling logic is the same as HPA\@: snapshot-based, no trend awareness, same
  oscillation and cooldown limitations. Richer inputs, but the same decision model. \\
\midrule
\textbf{Knative}
& Can scale to zero and wake on first request. Operates at request-level concurrency
  granularity.
& The panic mode shortens the evaluation window to react faster, but still detects
  overload after it begins. Each evaluation is independent, with no memory of prior
  state. \\
\midrule
\textbf{AWS Predictive}
& Learns recurring daily/weekly patterns from historical data and pre-provisions capacity
  hours ahead.
& Cannot handle sudden unexpected spikes, needs a reactive scaler running alongside
  as fallback. Operates on a timescale of hours, not seconds. \\
\bottomrule
\end{tabularx}
\caption{Comparison with existing scaling solutions, relative to the algorithm presented
in this document.}
\label{tab:comparison}
\end{table}

A direct performance comparison against HPA and KEDA under identical conditions
is presented in Section~\ref{sec:performance-comparison}.

\newpage
\section{High-Level Architecture}
\label{sec:architecture}

\subsection{Core Algorithm}

The algorithm is a pipeline of five stages. It processes metric samples and computes the number of instances
needed to handle the load:

\bigskip
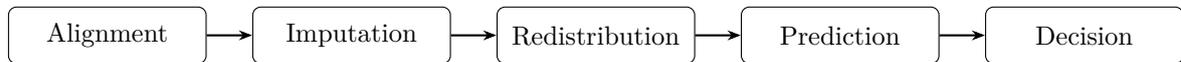
\begin{center}
\begin{tikzpicture}[
    node distance=0.6cm,
    box/.style={draw, rounded corners=3pt, minimum width=2.6cm, minimum height=0.75cm,
                text centered, font=\small},
    arr/.style={-{Stealth[length=5pt]}, thick}
]
    \node[box] (A) {Alignment};
    \node[box, right=of A] (B) {Imputation};
    \node[box, right=of B] (C) {Redistribution};
    \node[box, right=of C] (D) {Prediction};
    \node[box, right=of D] (E) {Decision};
    \draw[arr] (A) -- (B);
    \draw[arr] (B) -- (C);
    \draw[arr] (C) -- (D);
    \draw[arr] (D) -- (E);
\end{tikzpicture}
\end{center}
\bigskip

\begin{table}[H]
\centering
\begin{tabularx}{\textwidth}{@{} l X @{}}
\toprule
\textbf{Stage} & \textbf{Role} \\
\midrule
\textbf{Alignment}       & Snaps irregularly-timed metric samples onto a uniform time grid \\
\textbf{Imputation}  & Estimates values for instances that didn't report in a given tick \\
\textbf{Redistribution}  & Compensates for newly-added instances that haven't absorbed their share of load \\
\textbf{Prediction} & Detects the metric trend and projects the value forward in time \\
\textbf{Decision}        & Converts the forecast into a target instance count \\
\bottomrule
\end{tabularx}
\caption{Pipeline stages and their roles.}
\end{table}

Each stage enriches the data and passes it to the next. The pipeline runs independently for each metric.

\subsection{Data Format and Delivery}

Instances send every individual measurement to the scaler with no client-side aggregation, so the scaler sees
the full picture with no data lost.

However, the scaler cannot monitor measurements in real time. Sending a request for every individual measurement
would be too expensive. Instead, instances collect samples locally and send them to the scaler in batches.

With fixed batch intervals, there is a tradeoff: long intervals reduce network overhead but delay spike
detection; short intervals provide fresh data but waste resources when the instance is idle.
\emph{Dynamic batch timing} resolves this. Each instance uses two timeouts based on the metric values it has
collected:

\begin{itemize}[leftmargin=*, itemsep=4pt]
    \item \textbf{Short batch timeout} (e.g.\ 5\,s), used when a batch contains high metric values. The
        instance is under load and the scaler needs fresh data.
    \item \textbf{Long batch timeout} (e.g.\ 40\,s), used when metric values are normal. There is no
        urgency to report.
\end{itemize}

This delivers fresh data when it matters (under load) and reduces overhead when the instance is healthy.

\subsection{Processing Cadence}

The pipeline is triggered by batch arrivals, but running it on every arrival would be wasteful: batches from
different instances often arrive in close succession, and each run would repeat most of the work with only
marginally more data. A scaler uses a \emph{processing cooldown} (e.g.\ 10\,s) after each run: batches that
arrive during the cooldown are stored, and the pipeline runs once over all of them when the cooldown expires.

\subsection{Inputs}

The algorithm requires the following inputs:

\begin{itemize}[leftmargin=*, itemsep=4pt]
    \item \textbf{Instances metric samples}, per-instance metric values with timestamps: tuples $(t, v)$
        where $t$ is the timestamp and $v$ is the metric value. These arrive in batches (see Section~\ref{sec:introduction}).

    \item \textbf{Instance state}, for each instance $i$, the algorithm needs:
        \begin{itemize}
            \item $t_i^0$, when the instance started
            \item $t_i^{\text{end}}$, when the instance terminated
        \end{itemize}

    \item \textbf{Init timeout} $T_I$, how long a new instance takes to become ready. This determines the
        prediction horizon: how far into the future the algorithm looks. This value can be a fixed constant,
        or it can be adaptively estimated from observed startup times
        (see Section~\ref{sec:adaptive-init-timeout}).

    \item \textbf{Threshold} $\tau$, the per-instance metric value above which an instance is considered
        overloaded.

    \item \textbf{Instance count bounds}, min and max allowed instance counts $N_{\min}$, $N_{\max}$.
\end{itemize}

\subsection{Notation}

The following symbols are used throughout the document. Stage-specific symbols are introduced
in their respective sections.

\vspace{-0.5em}
\begin{center}
\begin{tabular}{@{} l p{9cm} @{}}
\toprule
\textbf{Symbol} & \textbf{Meaning} \\
\midrule
$t$                         & Discrete tick index on the uniform time grid \\
$\Delta t$                  & Sample interval, grid spacing (e.g.\ 1000\,ms) \\
$i$                         & Instance identifier \\
$\tau$                      & Per-instance overload threshold \\
$N_{\min},\; N_{\max}$      & Instance count bounds \\
$T_I$                       & Init timeout (time for a new instance to become ready) \\
$T_R$                       & Redistribution timeout (expected time for full stabilization) \\
$\eta$                      & Prediction horizon multiplier \\
$H_{\min},\; H_{\max}$     & Prediction horizon bounds ($H = \clamp(\eta \cdot T_I,\; H_{\min},\; H_{\max})$) \\
$N_{\text{step}}$           & Maximum scale-up step per decision \\
$t_i^0$                     & Start time of instance $i$ \\
\midrule
\textbf{Function} & \textbf{Meaning} \\
\midrule
$\mathcal{A}(\cdot)$\; {\small(default $\sum w_j v_j$)}
    & Aggregation: per-instance values $\to$ cluster aggregate \\
$\mathcal{P}(S, N)$\; {\small(default $S / N$)}
    & Projection: aggregate $+$ count $\to$ per-instance value \\
$\mathcal{N}(S, \tau)$\; {\small(default $S / \tau$)}
    & Required count: aggregate $+$ threshold $\to$ instance count \\
\bottomrule
\end{tabular}
\captionof{table}{Notation and metric model functions (Section~\ref{sec:introduction}).}
\end{center}

\section{Alignment}
\label{sec:alignment}

\subsection{The Problem}

Metric samples are timestamped locally on each instance at the moment of measurement. Each instance samples on
its own schedule, so timestamps across instances never coincide. The rest of the pipeline needs to compare and
aggregate values across instances at the same points in time, which requires placing them onto a uniform time
grid.

\subsection{The Solution: Grid Snapping with Interpolation}

Alignment processes each instance's samples independently, placing them onto a uniform grid defined by
$\Delta t$ (e.g.\ 1000\,ms).

\textbf{Grid ticks} are computed by flooring raw timestamps to the interval.
Let $t^r_j$ and $v^r_j$ denote the raw (unaligned) timestamp and value of sample $j$:

\[
t = \left\lfloor \frac{t^r_j}{\Delta t} \right\rfloor \cdot \Delta t
\]

For each grid tick $t$ that falls between two consecutive raw samples $(t^r_j, v^r_j)$ and
$(t^r_{j+1}, v^r_{j+1})$, the value is \textbf{linearly interpolated}:

\begin{align*}
\lambda &= \frac{t - t^r_j}{t^r_{j+1} - t^r_j} \\
v_t &= v^r_j + (v^r_{j+1} - v^r_j) \cdot \lambda
\end{align*}

For example, given raw samples $(1001, 0.4)$ and $(2003, 0.6)$ with $\Delta t = 1000$\,ms, the grid tick at
$t = 2000$:

\begin{align*}
\lambda &= \frac{2000 - 1001}{2003 - 1001} = 0.998 \\
v_{2000} &= 0.4 + (0.6 - 0.4) \cdot 0.998 \approx 0.5996
\end{align*}

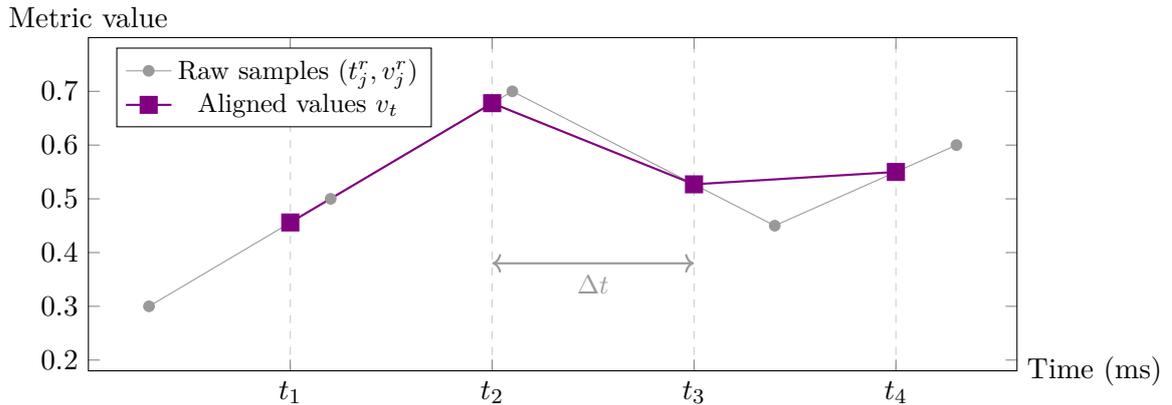
\begin{figure}[H]
\centering
\begin{tikzpicture}
\begin{axis}[
    width=0.92\textwidth,
    height=6cm,
    xlabel={Time (ms)},
    ylabel={Metric value},
    xmin=0, xmax=4600,
    ymin=0.18, ymax=0.80,
    xtick={1000, 2000, 3000, 4000},
    xticklabels={$t_1$, $t_2$, $t_3$, $t_4$},
    ytick={0.2, 0.3, 0.4, 0.5, 0.6, 0.7},
    xmajorgrids=true,
    major grid style={dashed, black!20},
    legend style={at={(0.03,0.97)}, anchor=north west, font=\small},
    every axis x label/.style={at={(ticklabel* cs:1.0)}, anchor=west},
    every axis y label/.style={at={(ticklabel* cs:1.0)}, anchor=south},
]

\addplot[thin, black!40, mark=*, mark options={fill=black!40}, mark size=2pt] coordinates {
    (300,0.30) (1200,0.50) (2100,0.70) (3400,0.45) (4300,0.60)
};
\addlegendentry{Raw samples $(t^r_j, v^r_j)$}

\addplot[thick, violet, mark=square*, mark size=3pt] coordinates {
    (1000,0.456) (2000,0.678) (3000,0.527) (4000,0.550)
};
\addlegendentry{Aligned values $v_t$}

\draw[<->, thick, black!40] (axis cs:2000,0.38) -- node[below, font=\small, black!40] {$\Delta t$} (axis cs:3000,0.38);

\end{axis}
\end{tikzpicture}
\caption{Grid alignment by interpolation. Raw samples (grey circles) arrive at irregular
timestamps and are connected by line segments. Aligned values (purple squares) are computed by
linear interpolation at uniform grid ticks (dashed vertical lines, spaced $\Delta t = 1000$\,ms
apart).}
\label{fig:alignment}
\end{figure}

\subsection{Continuity Across Batches}

When samples arrive in batches, the last sample from the previous batch is preserved. On the next batch,
interpolation uses this previous sample as the starting point, producing multiple grid ticks across the gap.
For example, if a batch ends at 5200\,ms and the next starts at 8100\,ms with $\Delta t = 1000$\,ms,
alignment produces grid ticks at $t = 6000, 7000, 8000$, all interpolated between the last sample of the
previous batch and the first of the new one. This ensures a continuous, gap-free aligned series regardless of
batch timing.

\subsection{Output}

For each instance, alignment produces a sequence of grid values $v_t$ on the uniform time grid. These become
the input to the next stage, where the instance index $i$ is introduced to compare across instances:
$v_t^i$.

\section{Imputation}
\label{sec:imputation}

\subsection{The Problem}

After alignment, we have a uniform time series per instance, but different instances send their metric
batches at different times with varying offsets. At any given processing moment, some instances have reported
data up to timestamp $T$, others only up to $T - 3\,\text{s}$, and others up to $T - 7\,\text{s}$. There is
no moment when you have complete data from all instances up to ``now.''

The prediction pipeline downstream needs a continuous estimate of the metric at every tick.
Imputation's job is to produce that estimate from incomplete data, and to do so in a way that is
accurate enough for good scaling decisions, self-correcting as more data arrives, and resilient to instance
failures.

\subsection{Overview}

The imputation maintains a running estimate of the total metric sum across all instances. At each
tick, it updates this estimate using whatever data is available:

\begin{itemize}[leftmargin=*, itemsep=3pt]
    \item \textbf{Known instances} contribute their real, measured values.
    \item \textbf{Unknown instances} (those that haven't reported yet) are estimated by carrying
        forward their previous contribution to the total. The algorithm takes the previous total,
        removes the previous values of instances that are now known, and what remains is the estimated
        contribution of the instances that are still missing.
\end{itemize}

The key property of this approach is that \textbf{changes on known instances are immediately reflected
in the total, while unknown instances are held at their previous estimated level}. Combined with the
core assumption that load changes happen similarly across all instances, this produces accurate
estimates across all load scenarios:

\begin{itemize}[leftmargin=*, itemsep=4pt]
    \item \textbf{Steady load.} Unknown instances haven't changed, so estimating them from the
        previous total is accurate. The imputed sum matches reality.

    \item \textbf{During a spike.} All instances are rising, but we only see some of them. The known
        instances' increase is fully captured in the total. The unknown instances are estimated at
        their pre-spike level, so the total rises but less than reality. The underestimate is
        proportional to the fraction of missing instances: if 3 out of 4 report, we capture
        roughly 3/4 of the spike immediately. This is usually enough for the smoothing stage to
        detect the upward trend and begin predicting further increase.

    \item \textbf{During a drop.} The mirror case. Known instances show the decrease, but unknown
        instances are estimated at their previous (higher) level. The imputed total drops
        slower than reality. This is conservative in the safe direction: it prevents the algorithm
        from thinking load has dropped faster than it actually has, avoiding premature scale-down.
\end{itemize}

\subsection{The Imputation Step}

At each tick $t$, the algorithm has:

\begin{itemize}[leftmargin=*, itemsep=3pt]
    \item $I^k_t$, the set of instances that reported at tick $t$ (the ``known'' instances)
    \item $v_t^i$, the aligned metric value of instance $i$ at tick $t$, defined for $i \in I^k_t$
    \item $N_t$, the total number of instances that were active at time $t$
    \item $s_{t-1}$, the imputed metric sum across all instances from the previous tick
\end{itemize}

From these, it computes:

\textbf{1.\ Count unknowns:}

\[
\tilde{N}_t = N_t - |I^k_t|
\]

\textbf{2.\ Compute the known sum:}

\[
s^k_t = \sum_{i \in I^k_t} v_t^i
\]

\textbf{3.\ Previous contribution of known instances.} Every instance known at $t$ was also known
at $t{-}1$, so we can look up what they contributed to the previous total:

\[
s^{*}_{t-1} = \sum_{i \,\in\, I^k_t} v_{t-1}^i
\]

\textbf{4.\ Estimate the unknown contribution.} Since aligned data is continuous within each
instance's range, the set of known instances can only shrink over time: $I^k_t \subseteq I^k_{t-1}$.
Subtract the accounted portion from the previous total. What remains is the estimated contribution
of the instances that are no longer reporting:

\[
s^u_t = s_{t-1} - s^{*}_{t-1}
\]

\textbf{5.\ Impute per-instance values.} Known instances retain their real measurements.
Unknown instances are each assigned an equal share of the estimated unknown contribution:

\[
\hat{v}_t^i = \begin{cases}
    v_t^i & \text{if } i \in I^k_t \\[4pt]
    s^u_t \,/\, \tilde{N}_t & \text{if } i \notin I^k_t
\end{cases}
\]

\textbf{6.\ Compute the tick's total} (carried forward internally as $s_{t-1}$ for the next tick):

\[
s_t = s^k_t + s^u_t
\]

\bigskip
\textbf{Example.} Three instances report batches at different times. Instance~C has the latest
data (up to $t_6$), instance~A up to $t_4$, and instance~B only up to $t_2$:

\begin{table}[H]
\centering
\begin{tabular}{@{} l *{6}{>{\centering\arraybackslash}p{1.1cm}} @{}}
\toprule
& $t_1$ & $t_2$ & $t_3$ & $t_4$ & $t_5$ & $t_6$ \\
\midrule
Instance A & 0.3 & 0.4 & 0.5 & 0.6 & ? & ? \\
Instance B & 0.2 & 0.3 & ? & ? & ? & ? \\
Instance C & 0.4 & 0.5 & 0.6 & 0.7 & 0.6 & 0.5 \\
\midrule
$s^k_t$ & 0.9 & 1.2 & 1.1 & 1.3 & 0.6 & 0.5 \\
$s^u_t$ & 0 & 0 & 0.3 & 0.3 & 0.9 & 0.9 \\
\midrule
$s_t$ & 0.9 & 1.2 & 1.4 & 1.6 & 1.5 & 1.4 \\
\midrule
\multicolumn{7}{@{}l}{\textit{Imputed values $\hat{v}_t^i$ (\textcolor{gray}{estimated}):}} \\[2pt]
Instance A & 0.3 & 0.4 & 0.5 & 0.6 & \textcolor{gray}{0.45} & \textcolor{gray}{0.45} \\
Instance B & 0.2 & 0.3 & \textcolor{gray}{0.30} & \textcolor{gray}{0.30} & \textcolor{gray}{0.45} & \textcolor{gray}{0.45} \\
Instance C & 0.4 & 0.5 & 0.6 & 0.7 & 0.6 & 0.5 \\
\bottomrule
\end{tabular}
\caption{Imputation example. At $t_1$--$t_2$ all instances are known and $s^u_t = 0$.
At $t_3$, instance~B has not yet reported; its contribution is estimated from
$s_{t_2}$ minus the known values at $t_2$: $s^u_{t_3} = 1.2 - 0.4 - 0.5 = 0.3$.
As more instances drop off toward the right, the estimated portion $s^u_t$ grows while
the total $s_t$ gradually degrades; these estimates are temporary and self-correct
as late batches arrive (Section~\ref{sec:self-correction}).}
\label{tab:imputation-example}
\end{table}

\subsection{Self-Correction}
\label{sec:self-correction}

The estimates are temporary. Each processing cycle reruns imputation as a forward pass
over the entire window using all currently available data. When a late batch arrives from
a previously-missing instance, its values enter $I^k_t$ for the ticks it covers ---
what were previously imputed values are replaced by real measurements, and $s_t$ is
recomputed from scratch. \textbf{Spikes are never missed}: an imputed tick may
undercount a spike initially, but when the real data arrives, the full spike is captured.

Each new batch makes the imputation more accurate. As batches arrive from more instances,
the fraction of imputed vs.\ real data shrinks and $s_t$ converges toward the true value.

\subsection{Resilience}

Imputation is inherently resilient to instance failures. If an instance stops reporting entirely, the
algorithm continues working with the remaining instances. The failed instance's contribution remains
embedded in $s^u_t$ temporarily. When the instance is confirmed terminated, $N_t$ decreases; on the
next forward pass, the imputation recomputes without the terminated instance, and the estimates adjust
accordingly.

No single instance can block or break the pipeline. The algorithm produces imputed values from whatever
data is available, whether that's all instances, half of them, or just one.

\subsection{Cold Start}

When no previous data exists (first tick ever), there is no previous total to estimate from: $s_{t-1}$ is
undefined, so $s^u_t = 0$. Unknown instances are assigned a value of 0. This means the first tick may
underestimate the total, but as more ticks arrive and more batches come in, the estimates converge quickly.

\section{Redistribution}
\label{sec:redistribution}

\subsection{The Problem}

The algorithm must be sensitive to changes in the external load on the cluster (that's how it detects
traffic spikes and triggers scaling before overload occurs). But scaling itself creates changes in the metrics
that have nothing to do with external load; they are internal artifacts of the redistribution process.

When the algorithm scales up, new instances don't absorb traffic instantly, and the existing instances don't
shed load instantly either. Queues need to drain, garbage collection cycles complete, buffers flush. For a
period after scale-up, the old instances remain at their elevated metric levels while the new instance starts
accumulating additional load on top.

The timing and pace of the actual redistribution is unpredictable. Different load balancers
behave differently, network conditions vary, application warm-up times differ. We don't know exactly when the major
part of the redistribution will happen, or even whether it will complete cleanly at all. But we do know that
the metric changes it causes are internal artifacts, not signals the algorithm should react to.

\begin{figure}[H]
\centering
\begin{tikzpicture}
\begin{axis}[
    ybar stacked,
    width=0.92\textwidth,
    height=7.5cm,
    bar width=18pt,
    ymin=0, ymax=3.9,
    ylabel={Aggregated metric value},
    xtick=data,
    xticklabels={$t_1$, $t_2$, $t_3$, $t_4$, $t_5$, $t_6$},
    xlabel={Time},
    ytick={0, 0.5, 1.0, 1.5, 2.0, 2.5, 3.0, 3.5},
    grid=major,
    major grid style={dashed, black!15},
    legend style={at={(0.5,-0.15)}, anchor=north, font=\small, legend columns=4,
        column sep=8pt},
    every axis x label/.style={at={(ticklabel* cs:1.0)}, anchor=west},
    nodes near coords={\empty},
    clip=false,
]

\addplot[fill=blue!60!black, draw=black,
    nodes near coords, point meta=explicit symbolic,
    every node near coord/.style={font=\scriptsize, color=white}] coordinates {
    (1,0.90) [.90] (2,0.90) [.90] (3,0.82) [.82] (4,0.75) [.75] (5,0.70) [.70] (6,0.68) [.68]
};
\addlegendentry{Instance A}

\addplot[fill=blue!40!white, draw=black,
    nodes near coords, point meta=explicit symbolic,
    every node near coord/.style={font=\scriptsize, color=black}] coordinates {
    (1,0.90) [.90] (2,0.90) [.90] (3,0.82) [.82] (4,0.75) [.75] (5,0.70) [.70] (6,0.68) [.68]
};
\addlegendentry{Instance B}

\addplot[fill=orange!60!white, draw=black,
    nodes near coords, point meta=explicit symbolic,
    every node near coord/.style={font=\scriptsize, color=black}] coordinates {
    (1,0.90) [.90] (2,0.90) [.90] (3,0.82) [.82] (4,0.75) [.75] (5,0.70) [.70] (6,0.68) [.68]
};
\addlegendentry{Instance C}

\addplot[fill=green!40!white, draw=black,
    nodes near coords, point meta=explicit symbolic,
    every node near coord/.style={font=\scriptsize, color=black}] coordinates {
    (1,0) [] (2,0.70) [.70] (3,0.70) [.70] (4,0.68) [.68] (5,0.68) [.68] (6,0.68) [.68]
};
\addlegendentry{Instance D (new)}

\draw[dashed, black!50] (axis cs:1.5,0) -- (axis cs:1.5,3.9);
\node[anchor=south, font=\small] at (axis cs:1.5,3.9) {scale-up};

\node[anchor=south, font=\scriptsize] at (axis cs:1,2.70) {2.70};
\node[anchor=south, font=\scriptsize] at (axis cs:2,3.40) {3.40};
\node[anchor=south, font=\scriptsize] at (axis cs:3,3.16) {3.16};
\node[anchor=south, font=\scriptsize] at (axis cs:4,2.93) {2.93};
\node[anchor=south, font=\scriptsize] at (axis cs:5,2.78) {2.78};
\node[anchor=south, font=\scriptsize] at (axis cs:6,2.72) {2.72};

\end{axis}
\end{tikzpicture}
\caption{Redistribution after scaling from 3 to 4 instances. Before scale-up ($t_1$),
three instances carry 0.9 each. The new instance immediately receives traffic ($t_2$),
but the existing instances don't shed load instantly: queues must drain, in-flight
requests must complete, garbage collection must settle. The sum spikes from 2.70 to 3.40
before gradually settling as redistribution completes.}
\label{fig:redistribution-example}
\end{figure}
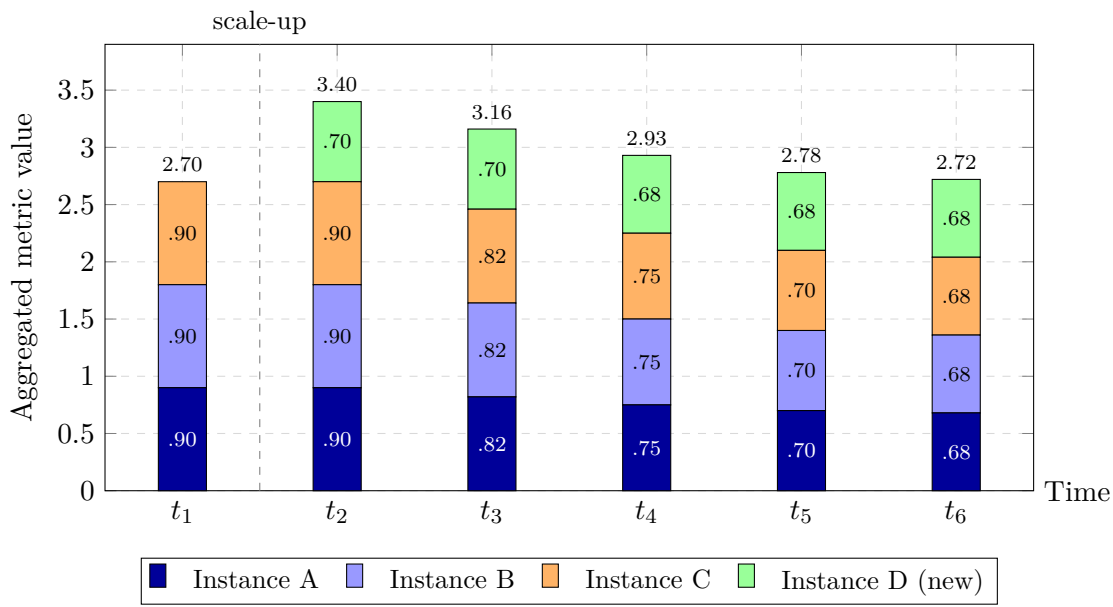

The raw aggregation makes the system look \emph{more} stressed after scaling up, not less. If the algorithm
acts on this, it could trigger yet another scale-up, which would compound the problem. We can't include the new instance's metric into the aggregated value as is.

\paragraph{Why not use a cooldown?}
Redistribution might take some time. The algorithm just scaled because load was rising, the
external pressure that triggered the scale-up is still there and may continue to grow. If the
added capacity turns out to be insufficient, the algorithm needs to detect this and scale again.
A cooldown would prevent it from making any decisions until redistribution completes, by which
point the system may already be overloaded.

\paragraph{Why not ignore new instances?}

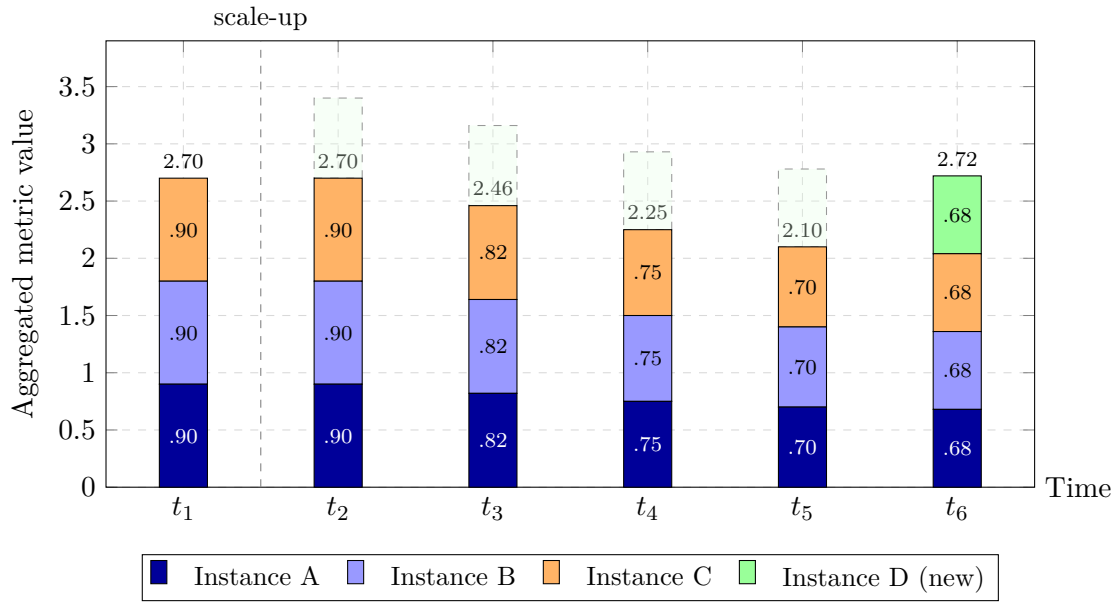
\begin{figure}[H]
\centering
\begin{tikzpicture}
\begin{axis}[
    ybar stacked,
    width=0.92\textwidth,
    height=7.5cm,
    bar width=18pt,
    ymin=0, ymax=3.9,
    ylabel={Aggregated metric value},
    xtick=data,
    xticklabels={$t_1$, $t_2$, $t_3$, $t_4$, $t_5$, $t_6$},
    xlabel={Time},
    ytick={0, 0.5, 1.0, 1.5, 2.0, 2.5, 3.0, 3.5},
    grid=major,
    major grid style={dashed, black!15},
    legend style={at={(0.5,-0.15)}, anchor=north, font=\small, legend columns=4,
        column sep=8pt},
    every axis x label/.style={at={(ticklabel* cs:1.0)}, anchor=west},
    nodes near coords={\empty},
    clip=false,
]

\addplot[fill=blue!60!black, draw=black,
    nodes near coords, point meta=explicit symbolic,
    every node near coord/.style={font=\scriptsize, color=white}] coordinates {
    (1,0.90) [.90] (2,0.90) [.90] (3,0.82) [.82] (4,0.75) [.75] (5,0.70) [.70] (6,0.68) [.68]
};
\addlegendentry{Instance A}

\addplot[fill=blue!40!white, draw=black,
    nodes near coords, point meta=explicit symbolic,
    every node near coord/.style={font=\scriptsize, color=black}] coordinates {
    (1,0.90) [.90] (2,0.90) [.90] (3,0.82) [.82] (4,0.75) [.75] (5,0.70) [.70] (6,0.68) [.68]
};
\addlegendentry{Instance B}

\addplot[fill=orange!60!white, draw=black,
    nodes near coords, point meta=explicit symbolic,
    every node near coord/.style={font=\scriptsize, color=black}] coordinates {
    (1,0.90) [.90] (2,0.90) [.90] (3,0.82) [.82] (4,0.75) [.75] (5,0.70) [.70] (6,0.68) [.68]
};
\addlegendentry{Instance C}

\addplot[fill=green!10!white, fill opacity=0.3, draw=black!40, dashed, forget plot,
    nodes near coords, point meta=explicit symbolic,
    every node near coord/.style={font=\scriptsize, color=black!40}] coordinates {
    (1,0) [] (2,0.70) [] (3,0.70) [] (4,0.68) [] (5,0.68) [] (6,0) []
};

\addplot[fill=green!40!white, draw=black,
    nodes near coords, point meta=explicit symbolic,
    every node near coord/.style={font=\scriptsize, color=black}] coordinates {
    (1,0) [] (2,0) [] (3,0) [] (4,0) [] (5,0) [] (6,0.68) [.68]
};
\addlegendentry{Instance D (new)}

\draw[dashed, black!50] (axis cs:1.5,0) -- (axis cs:1.5,3.9);
\node[anchor=south, font=\small] at (axis cs:1.5,3.9) {scale-up};

\node[anchor=south, font=\scriptsize] at (axis cs:1,2.70) {2.70};
\node[anchor=south, font=\scriptsize] at (axis cs:2,2.70) {2.70};
\node[anchor=south, font=\scriptsize] at (axis cs:3,2.46) {2.46};
\node[anchor=south, font=\scriptsize] at (axis cs:4,2.25) {2.25};
\node[anchor=south, font=\scriptsize] at (axis cs:5,2.10) {2.10};
\node[anchor=south, font=\scriptsize] at (axis cs:6,2.72) {2.72};

\end{axis}
\end{tikzpicture}
\caption{The problem with ignoring new instances. Instance~D exists from $t_2$ onward
(shown dashed) but is excluded from the aggregated value until $t_6$.}
\label{fig:ignore-new-instances}
\end{figure}

The simplest fix would be to completely ignore new instances for a stabilization period. But this creates its
own problem (see Figure~\ref{fig:ignore-new-instances}):

\begin{itemize}[leftmargin=*, itemsep=3pt]
    \item While a new Instance D is ignored ($t_2$--$t_5$), the stable instances gradually offload traffic to it.
        The aggregated value that the algorithm sees drops: $2.70 \to 2.46 \to 2.25 \to 2.10$.
    \item When the stabilization period expires ($t_6$) and Instance D is included, the
        value jumps: $2.10 + 0.68 = 2.72$.
    \item The algorithm interprets this jump as a real load increase and may trigger an
        unnecessary scale-up.
\end{itemize}

The goal of the redistribution stage is to \textbf{smooth the metric values redistribution
during the scale up, while preserving full sensitivity to changes in the external load}.
Redistribution artifacts are smoothed away; real signals pass through.

\subsection{The Solution: Gradual Weighted Inclusion}

The idea is to include new instances from the moment they appear, but gradually increase their contribution to the
aggregated value over a redistribution period. When a new instance $i$ starts at time $t_i^0$,
it is assigned a weight $w_i$ that starts at 0 and increases linearly to 1 over the redistribution timeout $T_R$.

Redistribution only applies after a scale-up, when new instances exist. When all instances are stable, this
stage is a pass-through: the aggregated value and count equal the raw aggregation and full instance count.

Note that there is no single global redistribution period. Each instance increases its contribution
independently based on its own age. The drop absorption works across all of them: as long as any instance
has $w < 1$, drops are caught.

\begin{figure}[H]
\centering
\begin{tikzpicture}
\begin{axis}[
    ybar stacked,
    width=0.92\textwidth,
    height=7.5cm,
    bar width=18pt,
    ymin=0, ymax=3.9,
    ylabel={Aggregated metric value},
    xtick=data,
    xticklabels={$t_1$, $t_2$, $t_3$, $t_4$, $t_5$, $t_6$},
    xlabel={Time},
    ytick={0, 0.5, 1.0, 1.5, 2.0, 2.5, 3.0, 3.5},
    grid=major,
    major grid style={dashed, black!15},
    legend style={at={(0.5,-0.15)}, anchor=north, font=\small, legend columns=4,
        column sep=8pt},
    every axis x label/.style={at={(ticklabel* cs:1.0)}, anchor=west},
    nodes near coords={\empty},
    clip=false,
]

\addplot[fill=blue!60!black, draw=black,
    nodes near coords, point meta=explicit symbolic,
    every node near coord/.style={font=\scriptsize, color=white}] coordinates {
    (1,0.90) [.90] (2,0.90) [.90] (3,0.82) [.82] (4,0.75) [.75] (5,0.70) [.70] (6,0.68) [.68]
};
\addlegendentry{Instance A}

\addplot[fill=blue!40!white, draw=black,
    nodes near coords, point meta=explicit symbolic,
    every node near coord/.style={font=\scriptsize, color=black}] coordinates {
    (1,0.90) [.90] (2,0.90) [.90] (3,0.82) [.82] (4,0.75) [.75] (5,0.70) [.70] (6,0.68) [.68]
};
\addlegendentry{Instance B}

\addplot[fill=orange!60!white, draw=black,
    nodes near coords, point meta=explicit symbolic,
    every node near coord/.style={font=\scriptsize, color=black}] coordinates {
    (1,0.90) [.90] (2,0.90) [.90] (3,0.82) [.82] (4,0.75) [.75] (5,0.70) [.70] (6,0.68) [.68]
};
\addlegendentry{Instance C}

\addplot[fill=green!40!white, draw=black,
    nodes near coords, point meta=explicit symbolic,
    every node near coord/.style={font=\scriptsize, color=black}] coordinates {
    (1,0) [] (2,0.09) [] (3,0.33) [.33] (4,0.54) [.54] (5,0.68) [.68] (6,0.68) [.68]
};
\addlegendentry{Instance D (new)}

\addplot[fill=green!10!white, fill opacity=0.3, draw=black!40, dashed, forget plot,
    nodes near coords, point meta=explicit symbolic,
    every node near coord/.style={font=\scriptsize, color=black!40}] coordinates {
    (1,0) [] (2,0.61) [] (3,0.37) [] (4,0.14) [] (5,0) [] (6,0) []
};

\draw[dashed, black!50] (axis cs:1.5,0) -- (axis cs:1.5,3.9);
\node[anchor=south, font=\small] at (axis cs:1.5,3.9) {scale-up};

\node[anchor=south, font=\scriptsize] at (axis cs:1,2.70) {2.70};
\node[anchor=south, font=\scriptsize] at (axis cs:2,2.79) {2.79};
\node[anchor=south, font=\scriptsize] at (axis cs:3,2.79) {2.79};
\node[anchor=south, font=\scriptsize] at (axis cs:4,2.79) {2.79};
\node[anchor=south, font=\scriptsize] at (axis cs:5,2.78) {2.78};
\node[anchor=south, font=\scriptsize] at (axis cs:6,2.72) {2.72};

\end{axis}
\end{tikzpicture}
\caption{The same scenario with gradual weighted inclusion ($\kappa = 1$, $T_R = 5\,\Delta t$).
Instance~D's contribution is included proportionally to its stabilization weight~$w$.
The aggregated value stays nearly flat (2.70--2.79) compared to the raw value
(Figure~\ref{fig:redistribution-example}) which spiked to 3.40.}
\label{fig:redistribution-solution}
\end{figure}
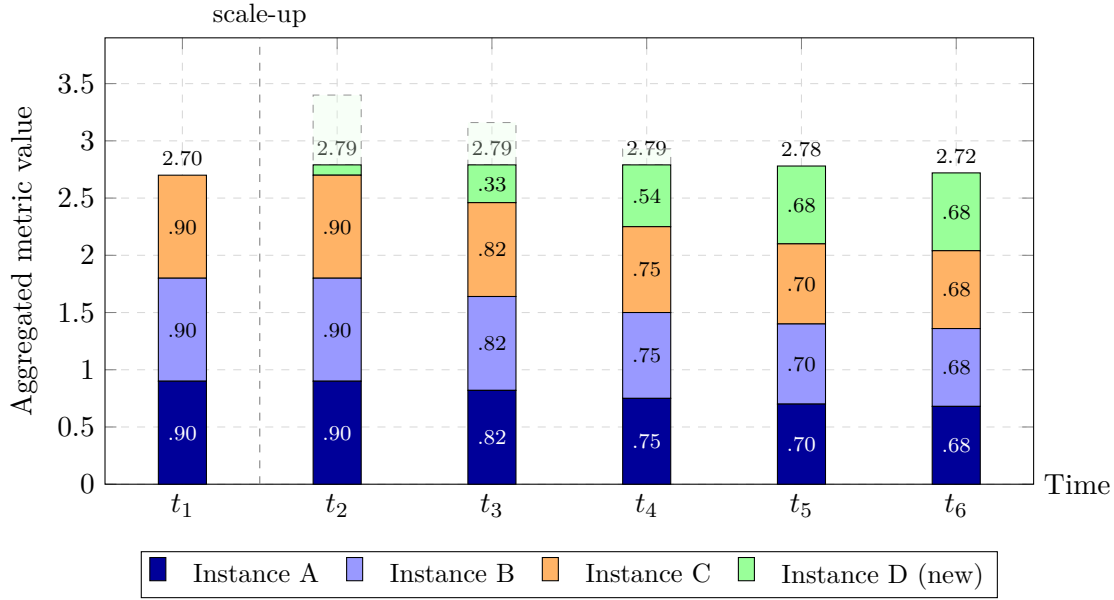

\subsection{Stabilization Weight}

The weight models how quickly a new instance absorbs its share of traffic. It uses an exponential curve:

\[
w(a) = \frac{e^{\kappa \cdot a \,/\, T_R} - 1}{e^{\kappa} - 1}
\]

Where:
\begin{itemize}[leftmargin=*, itemsep=3pt]
    \item $a$, how long the instance has been running (age)
    \item $T_R$, the expected time for full stabilization (redistribution timeout)
    \item $\kappa$, shape parameter (default 1), controls the curve's steepness
\end{itemize}

The curve goes from $w(0) = 0$ to $w(T_R) = 1$. The exponential shape allows starting
with small weights expecting the redistribution to happen earlier, but still reaching
full weight by $T_R$ even if the redistribution is slow. Changing $\kappa$ allows tuning
the weight to match different redistribution patterns.

Each instance has its own independent ramp-up timeline based on its own age. If the algorithm scales up
multiple times, there may be instances at different stages of ramp-up simultaneously, each increasing its
contribution at its own pace.

\begin{figure}[H]
\centering
\begin{tikzpicture}
\begin{axis}[
    width=0.82\textwidth,
    height=6cm,
    xlabel={Age $a$ (s)},
    ylabel={Weight $w(a)$},
    xmin=0, xmax=30,
    ymin=0, ymax=1.05,
    xtick={0, 5, 10, 15, 20, 25, 30},
    ytick={0, 0.2, 0.4, 0.6, 0.8, 1.0},
    grid=major,
    major grid style={dashed, black!20},
    every axis x label/.style={at={(ticklabel* cs:1.0)}, anchor=west},
    every axis y label/.style={at={(ticklabel* cs:1.0)}, anchor=south},
    domain=0:30,
    samples=100,
]

\addplot[thick, violet] {(exp(x/30) - 1) / (exp(1) - 1)};

\addplot[thin, dashed, black!40] {x/30};

\end{axis}
\end{tikzpicture}
\caption{Stabilization weight $w(a)$ with $\kappa = 1$ and $T_R = 30$\,s (solid) compared to
linear ramp-up (dashed). The curve starts slowly, reflecting the initial period where load
balancers gradually route traffic to the new instance, then accelerates as the instance
proves healthy.}
\label{fig:stabilization-weight}
\end{figure}
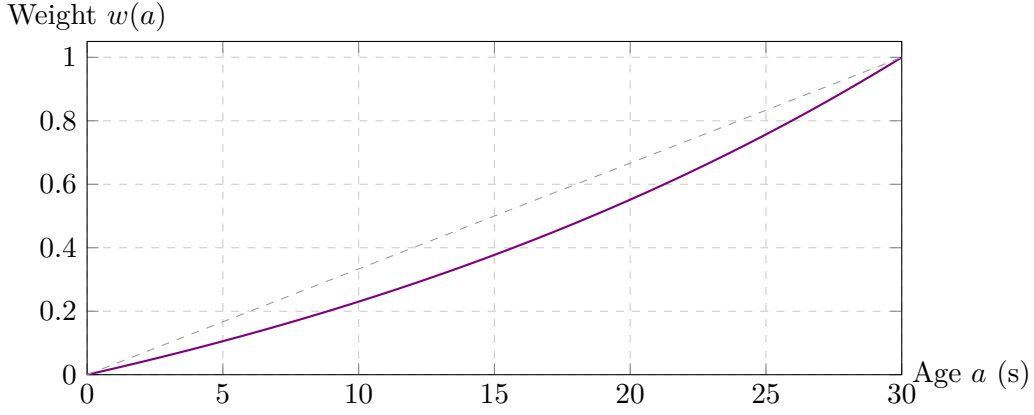

\subsection{The Calculation}

By this stage, every instance $i$ active at tick $t$ has a value $\hat{v}_t^i$ (measured or imputed) and a
start time $t_i^0$. The instance's \textbf{age} is $a_t^i = t - t_i^0$. Let $I_t$ denote the set of all active instances at tick $t$. They are classified
into two subsets:

\begin{itemize}[leftmargin=*, itemsep=3pt]
    \item $I_t^s$ (\textbf{stable}), instances with $a_t^i \geq T_R$, contributing at
        full weight
    \item $I_t^n$ (\textbf{new}), instances with $a_t^i < T_R$, contributing at partial
        weight $w(a_t^i)$
\end{itemize}

The redistributed aggregated value is computed as follows.

\textbf{1.\ Raw aggregated value.} Aggregate all instance values through the metric model's
aggregation function $\mathcal{A}$ (Section~\ref{sec:introduction}). This is the raw cluster-wide
metric, the value the algorithm would use if it did not account for redistribution at all:

\[
A_t^r = \mathcal{A}\!\left(\left\{
    \left(\hat{v}_t^i,\; 1\right) : i \in I_t\right\}\right)
\qquad \left(\,= \sum_{i \,\in\, I_t} \hat{v}_t^i \text{ when } \mathcal{A} \text{ is sum}\right)
\]

\textbf{2.\ Weighted aggregated value.} Apply the aggregation function
with per-instance weights: stable instances contribute at full weight, new instances at their
stabilization weight:

\[
\hat{A}_t = \mathcal{A}\!\left(\left\{
    \left(\hat{v}_t^i,\; w_t^i\right) : i \in I_t\right\}\right)
\qquad \left(\,= \sum_{i \,\in\, I_t} w_t^i \cdot \hat{v}_t^i \text{ when } \mathcal{A} \text{ is sum}\right)
\]

where

\[
w_t^i = \begin{cases} 1 & i \in I_t^s \\ w(a_t^i) & i \in I_t^n \end{cases}
\]

\textbf{3.\ Drop absorption} (see Section~\ref{sec:drop-absorption}):

\[
A_t = \begin{cases} \min\!\big(A_t^r,\;\; A_{t-1}\big) & \text{if } A_{t-1} > \hat{A}_t \\ \hat{A}_t & \text{otherwise} \end{cases}
\]

On the first tick, when no previous $A_{t-1}$ exists, drop absorption is skipped:
$A_t = \hat{A}_t$.

\subsection{Absorbing the Redistribution Drop}
\label{sec:drop-absorption}

The redistribution can happen \emph{rapidly} at any point after scale-up. If the algorithm
ignores it and uses the weighted value $\hat{A}_t$, it will create artificial
fluctuations that may trigger unnecessary scaling.

\textbf{The redistribution stage allows the aggregated value to rise to not miss the real load increase,
but prevents it from dropping until the new instances are fully included.}

If the weighted aggregation value $\hat{A}_t$ is lower than the previous value $A_{t-1}$, the
algorithm increases new instances' contributions until the new aggregated value $A_t$
matches the previous value $A_{t-1}$ or the new instances' values are fully included.

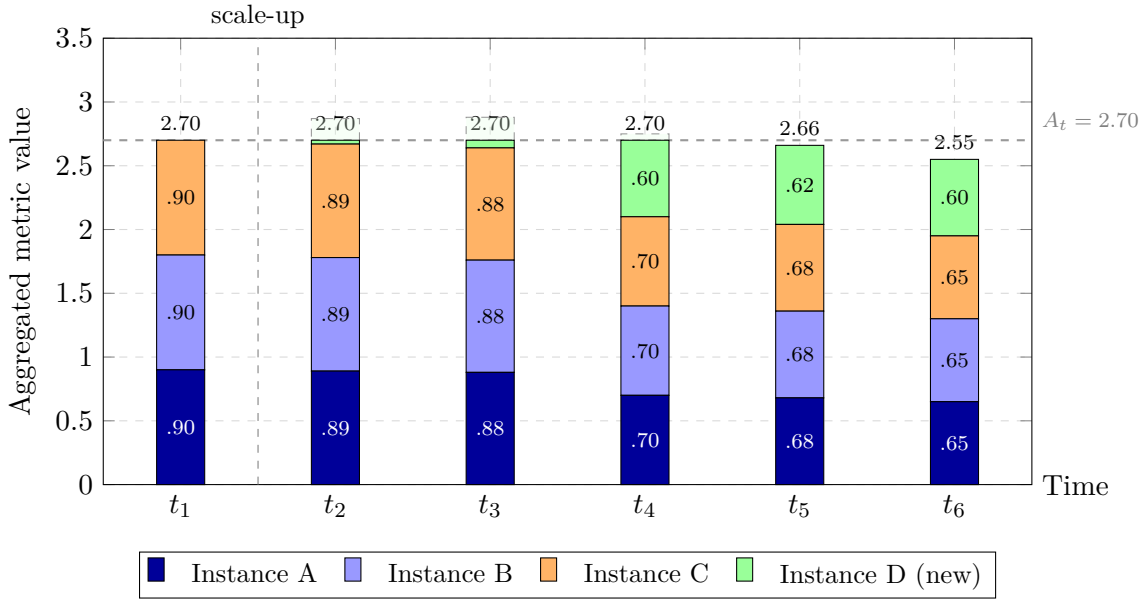
\begin{figure}[H]
\centering
\begin{tikzpicture}
\begin{axis}[
    ybar stacked,
    width=0.92\textwidth,
    height=7.5cm,
    bar width=18pt,
    ymin=0, ymax=3.5,
    ylabel={Aggregated metric value},
    xtick=data,
    xticklabels={$t_1$, $t_2$, $t_3$, $t_4$, $t_5$, $t_6$},
    xlabel={Time},
    ytick={0, 0.5, 1.0, 1.5, 2.0, 2.5, 3.0, 3.5},
    grid=major,
    major grid style={dashed, black!15},
    legend style={at={(0.5,-0.15)}, anchor=north, font=\small, legend columns=4,
        column sep=8pt},
    every axis x label/.style={at={(ticklabel* cs:1.0)}, anchor=west},
    nodes near coords={\empty},
    clip=false,
]

\addplot[fill=blue!60!black, draw=black,
    nodes near coords, point meta=explicit symbolic,
    every node near coord/.style={font=\scriptsize, color=white}] coordinates {
    (1,0.90) [.90] (2,0.89) [.89] (3,0.88) [.88] (4,0.70) [.70] (5,0.68) [.68] (6,0.65) [.65]
};
\addlegendentry{Instance A}

\addplot[fill=blue!40!white, draw=black,
    nodes near coords, point meta=explicit symbolic,
    every node near coord/.style={font=\scriptsize, color=black}] coordinates {
    (1,0.90) [.90] (2,0.89) [.89] (3,0.88) [.88] (4,0.70) [.70] (5,0.68) [.68] (6,0.65) [.65]
};
\addlegendentry{Instance B}

\addplot[fill=orange!60!white, draw=black,
    nodes near coords, point meta=explicit symbolic,
    every node near coord/.style={font=\scriptsize, color=black}] coordinates {
    (1,0.90) [.90] (2,0.89) [.89] (3,0.88) [.88] (4,0.70) [.70] (5,0.68) [.68] (6,0.65) [.65]
};
\addlegendentry{Instance C}

\addplot[fill=green!40!white, draw=black,
    nodes near coords, point meta=explicit symbolic,
    every node near coord/.style={font=\scriptsize, color=black}] coordinates {
    (1,0) [] (2,0.03) [] (3,0.06) [] (4,0.60) [.60] (5,0.62) [.62] (6,0.60) [.60]
};
\addlegendentry{Instance D (new)}

\addplot[fill=green!10!white, fill opacity=0.3, draw=black!40, dashed, forget plot,
    nodes near coords, point meta=explicit symbolic,
    every node near coord/.style={font=\scriptsize, color=black!40}] coordinates {
    (1,0) [] (2,0.17) [] (3,0.18) [] (4,0.05) [] (5,0) [] (6,0) []
};

\draw[dashed, black!50] (axis cs:1.5,0) -- (axis cs:1.5,3.5);
\node[anchor=south, font=\small] at (axis cs:1.5,3.5) {scale-up};

\draw[dashed, black!40, thick] (axis cs:0.5,2.70) -- (axis cs:6.5,2.70);
\node[anchor=south west, font=\scriptsize, black!50] at (axis cs:6.5,2.70) {$A_t = 2.70$};

\node[anchor=south, font=\scriptsize] at (axis cs:1,2.70) {2.70};
\node[anchor=south, font=\scriptsize] at (axis cs:2,2.70) {2.70};
\node[anchor=south, font=\scriptsize] at (axis cs:3,2.70) {2.70};
\node[anchor=south, font=\scriptsize] at (axis cs:4,2.70) {2.70};
\node[anchor=south, font=\scriptsize] at (axis cs:5,2.66) {2.66};
\node[anchor=south, font=\scriptsize] at (axis cs:6,2.55) {2.55};

\end{axis}
\end{tikzpicture}
\caption{Drop absorption with a load decrease at $t_5$. Instance~D is added at $t_2$
and begins absorbing traffic. At $t_4$, redistribution happens rapidly ---
stable instances drop from~0.88 to~0.70. The dashed line shows the clamped
level: $A_t$ is held at~2.70 during $t_2$--$t_4$ by increasing Instance~D's
effective contribution (solid green) beyond its weighted share. At $t_5$--$t_6$,
external load decreases and $A_t$ drops below the clamped level ---
real load changes pass through immediately.}
\label{fig:drop-absorption}
\end{figure}

\subsection{Catching the Spike}

An essential requirement is that the redistribution step should not hide a spike caused by the external load. When traffic keeps rising after a scale-up, two things happen in parallel:

\begin{itemize}[leftmargin=*, itemsep=3pt]
    \item \textbf{Stable instances keep climbing.} They are counted at full weight, so any continued increase
        is reflected immediately in $A_t$.
    \item \textbf{New instances ramp in.} Their contribution grows with $w(a)$, adding more of the rising
        signal over time rather than delaying it.
\end{itemize}

Drop absorption only blocks \emph{downward} movement during the redistribution window. It never clamps a rise.
So if the spike continues, $A_t$ keeps increasing and the trend estimate sees the sustained growth.

\subsection{Contribution-Weighted Count}

$N_t^w$ is the \emph{contribution-weighted count}: the number of instances weighted by how much
they contribute to the aggregated $A_t$. During redistribution, new instances are only partially
counted in $A_t$, and the count should reflect that same partial participation.

Redistribution smooths the \emph{aggregated} value by partially including new instances;
if the count stayed integer while the aggregate was weighted, the per-instance estimate
would jump downward whenever a new instance appears in the count. Those artificial drops
would leak into the decision logic even though the total signal was smoothed. Weighting
the count in the same way keeps the per-instance estimate $\mathcal{P}(A_t,\, N_t^w)$ smooth, so the per-instance view
reflects only real load changes, not redistribution artifacts.

Each stable instance contributes~1; each new instance contributes its stabilization
weight $w_t^i$. With $N_t^S = |I_t^s|$:

\begin{figure}[H]
\centering
\begin{minipage}[c]{0.45\textwidth}
\raggedleft
\begin{tikzpicture}[
    lbrace/.style={decorate, decoration={brace, amplitude=6pt, raise=3pt},
                   thick},
    rbrace/.style={decorate, decoration={brace, mirror, amplitude=6pt, raise=3pt},
                   thick},
]


\fill[blue!60!black, draw=black] (0,0) rectangle (2.2,1.64);
\node[font=\scriptsize, white] at (1.1,0.82) {A};

\fill[blue!40!white, draw=black] (0,1.64) rectangle (2.2,3.28);
\node[font=\scriptsize] at (1.1,2.46) {B};

\fill[orange!60!white, draw=black] (0,3.28) rectangle (2.2,4.92);
\node[font=\scriptsize] at (1.1,4.10) {C};

\fill[green!40!white, draw=black] (0,4.92) rectangle (2.2,5.58);
\node[font=\scriptsize] at (1.1,5.25) {D};

\fill[green!10!white, fill opacity=0.3, draw=black!40, dashed]
    (0,5.58) rectangle (2.2,6.32);

\draw[lbrace] (-0.15,0) -- (-0.15,5.58)
    node[midway, left=10pt, font=\small] {$A_t$};

\draw[rbrace] (2.35,0) -- (2.35,4.92)
    node[midway, right=10pt, font=\small] {$A_t^S$};

\draw[rbrace] (3.5,0) -- (3.5,6.32)
    node[midway, right=10pt, font=\small] {$A_t^r$};

\end{tikzpicture}
\end{minipage}%
\hspace{0.3cm}
\begin{minipage}[c]{0.45\textwidth}
\[
N_t^w = N_t^S + \sum_{i \,\in\, I_t^n} w_t^i
\]
\end{minipage}
\caption{Decomposition of the aggregated value at a single tick.}
\label{fig:sigma-count}
\end{figure}
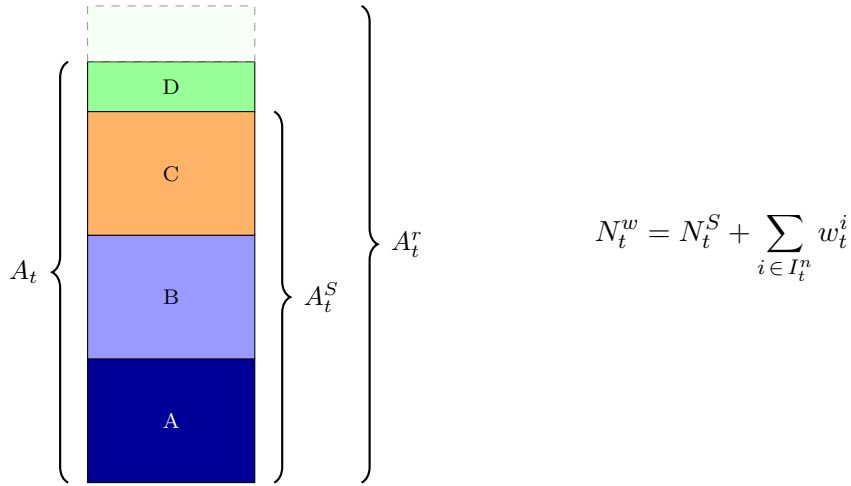

The count is derived from the weights alone, not from the metric values or from $A_t$.
This means drop absorption (Section~\ref{sec:drop-absorption}) does not affect $N_t^w$
--- even when $A_t$ is clamped, the count continues to grow smoothly with the stabilization
weights. This is important because $A_t$ is smoothed through the prediction stage and
changes slowly; if the count reacted to the clamp faster than the smoothed value, the
per-instance estimate $\mathcal{P}(A_t,\, N_t^w)$ would oscillate.

\subsection{Redistribution Delta}
\label{sec:redistribution-delta}

At every tick during redistribution, the aggregated value $A_t$ changes for two reasons:

\begin{itemize}[leftmargin=*, itemsep=3pt]
    \item \textbf{Load change}, the changes in the per-instance metric values driven by
        external load changes.
    \item \textbf{Redistribution}, the redistribution process itself moves $A_t$ by
        gradually including new instances' values; drop absorption may clamp the value.
        These changes are internal and happen even when external traffic is unchanged.
\end{itemize}

The value $A_t$ (with both effects combined) is the correct representation of the
application's metric state at each tick. The next stage in the pipeline, prediction
(Section~\ref{sec:prediction}), needs to know the correct current state to
know where to start the prediction. But it also tracks how fast the metric changes over time.

The redistribution component creates its own impact on the $A_t$ change rate that has nothing
to do with changes in the external load. To filter it out, the \emph{delta} $\Delta_t^R$ is computed
alongside $A_t$. It shows the change in the aggregated value driven only by new instances' weight changes.

\[
\Delta_t^R = \mathcal{A}\!\left(\left\{
    \left(\hat{v}_{t-1}^i,\; \boldsymbol{w_t^i}\right) : i \in I_{t-1}^n\right\}\right)
\;-\;
\mathcal{A}\!\left(\left\{
    \left(\hat{v}_{t-1}^i,\; \boldsymbol{w_{t-1}^i}\right) : i \in I_{t-1}^n\right\}\right)
\]

The delta is a difference between two aggregated values calculated for the previous tick
with different weights. The first aggregates the previous metric values with current weights;
the second with previous weights. The difference is purely the effect of weights changing.

For instances that become fully redistributed at time~$t$ ($w_t^i = 1$),
the difference captures the final step from partial to full inclusion.

On the first tick after new instances appear ($t{-}1$ had no new instances),
the previous aggregate has no new-instance component and $\Delta_t^R = 0$.

\paragraph{Interaction with drop absorption.}

When drop absorption holds the aggregate $A_t$ at previous level $A_{t-1}$ (Section~\ref{sec:drop-absorption})
--- the redistribution impact is compensated by the drop in stable instances. The
$A_t$ does not change so the delta is 0.

\[
\Delta_t^R = 0 \qquad \text{when drop absorption is active}
\]

This ensures the two mechanisms never conflict. Drop absorption holds the input flat;
the redistribution delta reports that no weight-driven change occurred in the input.

\subsection{Output}

Redistribution produces:

\begin{itemize}[leftmargin=*, itemsep=3pt]
    \item $A_t$ (every tick), the redistributed aggregated value, representing the effective total load.
        Passed to the prediction stage, which sees a clean signal: real traffic changes are preserved,
        redistribution artifacts are smoothed out.
    \item $\Delta_t^R$ (every tick), the redistribution delta, representing the expected
        change in $A_t$ from weight growth alone. Passed to the prediction stage so it can
        separate the weight-growth ramp from real load changes
        (Section~\ref{sec:redistribution-delta}).
    \item $N_t^w$ (last tick only), the contribution-weighted count, representing the
        effective number of contributing instances. Passed directly to the decision stage.
\end{itemize}

\section{Prediction}
\label{sec:prediction}

\subsection{Purpose}

This stage takes aggregated values $A_t$ from the redistribution step,
detects the current metric trend and uses it to extrapolate the value to the future.
To do this, it applies the double exponential smoothing method, also known as
\emph{Holt's method}~\cite{holt2004}.

\subsection{Holt's Method}

Holt's method is a time series forecasting technique that extends simple exponential
smoothing by incorporating a trend component. It is designed to handle data with a
linear trend, making it suitable for predicting future values based on past observations.

Although a metric's behavior can be complex and non-linear, Holt's method is still effective for short-term forecasting.
The key is that the trend component captures the general direction of change, which is
more important than modeling the exact shape of the curve when making predictions over a short horizon.

Holt's method maintains two state variables, \textbf{level} $l_t$ (the smoothed value) and
\textbf{trend} $t_t$ (the smoothed rate of change), and updates them at each tick
using two parameters, $\alpha$ and $\beta$:

\begin{itemize}[leftmargin=*, itemsep=3pt]
    \item $\alpha$ controls how much each individual data point moves the \textbf{level}.
        At $\alpha = 1$ the level equals the raw input; at $\alpha = 0$ it ignores new data
        entirely and follows the forecast. In practice, a moderate $\alpha$ lets the level
        track real movement while absorbing single-tick noise.
    \item $\beta$ controls how quickly the \textbf{trend} changes direction.
        When the signal reverses (say, load was rising and starts falling), a high $\beta$
        pivots the trend immediately, while a low $\beta$ keeps extrapolating in the old direction
        until several ticks of sustained change accumulate.
\end{itemize}

\textbf{Initialization} (first tick, no prior state):

\[
l_0 = A_0, \qquad t_0 = 0
\]

\textbf{For each subsequent tick} $t \geq 1$:

One-step-ahead forecast:

\[
F_t = l_{t-1} + t_{t-1} + \Delta_t^R
\]

The delta $\Delta_t^R$ from Section~\ref{sec:redistribution-delta}
accounts for redistribution impact. When no new instances exist,
$\Delta_t^R = 0$ and the forecast reduces to the standard Holt form. The motivation and
mechanics are detailed in Section~\ref{sec:delta-compensation}.

Parameter selection: the algorithm uses direction-dependent parameters:
$(\alpha_{\uparrow}, \beta_{\uparrow})$ when the input overshoots the forecast and
$(\alpha_{\downarrow}, \beta_{\downarrow})$ otherwise (see
Section~\ref{sec:asymmetric-reaction}).

\[
(\alpha_t,\; \beta_t) = \begin{cases} (\alpha_{\uparrow},\; \beta_{\uparrow}) & \text{if } A_t > F_t \\ (\alpha_{\downarrow},\; \beta_{\downarrow}) & \text{otherwise} \end{cases}
\]

Level update:

\[
l_t = \alpha_t \cdot A_t + (1 - \alpha_t) \cdot F_t
\]

Trend update:

\[
t_t = \beta_t \cdot (l_t - l_{t-1} - \Delta_t^R) + (1 - \beta_t) \cdot t_{t-1}
\]

The redistribution delta is subtracted from the level difference to prevent it from leaking
into the trend (Section~\ref{sec:delta-compensation}).

On the first tick the algorithm starts by tracking the input directly, with no prediction. Over subsequent
ticks, the trend converges to the actual rate of change.

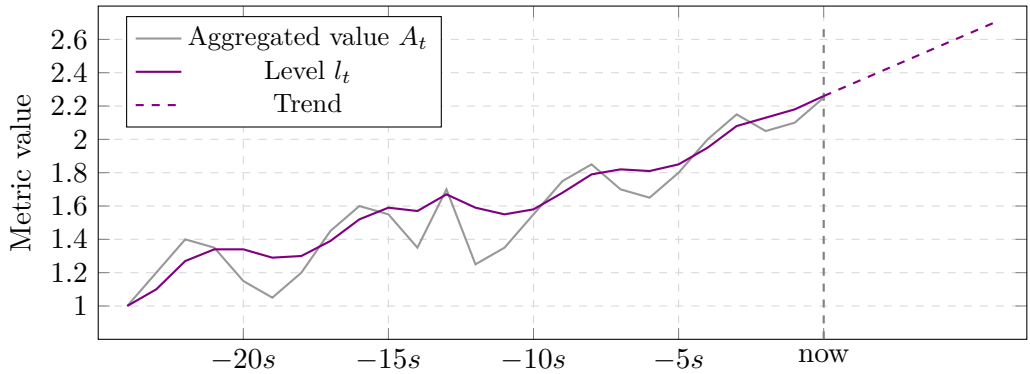
\begin{figure}[H]
\centering
\begin{tikzpicture}
\begin{axis}[
    width=0.92\textwidth,
    height=6cm,
    ylabel={Metric value},
    xmin=-25, xmax=7,
    ymin=0.8, ymax=2.8,
    xtick={-20, -15, -10, -5, 0},
    xticklabels={$-20s$, $-15s$, $-10s$, $-5s$, now},
    ytick={1.0, 1.2, 1.4, 1.6, 1.8, 2.0, 2.2, 2.4, 2.6},
    grid=major,
    major grid style={dashed, black!15},
    legend style={at={(0.03,0.97)}, anchor=north west, font=\small},
    clip=false,
]

\draw[dashed, black!50, thick] (axis cs:0,0.8) -- (axis cs:0,2.7);

\addplot[thick, black!40, mark=none] coordinates {
    (-24,1.00) (-23,1.20) (-22,1.40) (-21,1.35) (-20,1.15) (-19,1.05)
    (-18,1.20) (-17,1.45) (-16,1.60) (-15,1.55) (-14,1.35) (-13,1.70)
    (-12,1.25) (-11,1.35) (-10,1.55) (-9,1.75) (-8,1.85)
    (-7,1.70) (-6,1.65) (-5,1.80) (-4,2.00) (-3,2.15)
    (-2,2.05) (-1,2.10) (0,2.25)
};
\addlegendentry{Aggregated value $A_t$}

\addplot[thick, violet, mark=none] coordinates {
    (-24,1.00) (-23,1.10) (-22,1.27) (-21,1.34) (-20,1.34) (-19,1.29)
    (-18,1.30) (-17,1.39) (-16,1.52) (-15,1.59) (-14,1.57) (-13,1.67)
    (-12,1.59) (-11,1.55) (-10,1.58) (-9,1.68) (-8,1.79)
    (-7,1.82) (-6,1.81) (-5,1.85) (-4,1.95) (-3,2.08)
    (-2,2.13) (-1,2.18) (0,2.26)
};
\addlegendentry{Level $l_t$}

\addplot[thick, dashed, violet, mark=none] coordinates {
    (0,2.26) (6,2.71)
};
\addlegendentry{Trend}

\end{axis}
\end{tikzpicture}
\caption{Holt's method applied to the aggregated value. The grey line shows the raw input
$A_t$ with wave-like fluctuations and a noisy spike around $-13s$; the solid purple line shows
the smoothed level $l_t$, which tracks the general movement while filtering both the waves
and the spike. The dashed purple line shows the trend extrapolation beyond the current moment.}
\label{fig:holt-method}
\end{figure}

\subsection{Smoothing}

The $\alpha$ parameter controls how much the level reacts to each new input. A low $\alpha$ means the level
changes slowly: noise is filtered, but real changes are detected late. A high $\alpha$ means the level
tracks the input closely: real changes are detected immediately, but noise passes through.

Holt's method resolves this tension by combining the level with a trend. Even with a moderate $\alpha$, the
forecast $F_t = l_{t-1} + t_{t-1}$ anticipates where the signal is heading. If the input keeps pushing in
the same direction across multiple ticks, the trend accumulates and the level follows. A single noisy tick
pushes the level only slightly and the trend absorbs the rest. This way the smoothing can be aggressive enough
to filter noise while still reacting quickly to sustained changes.

\subsection{Trend Generation}

The trend is a natural byproduct of the Holt update. At each tick, $\beta$ controls how much the trend adjusts
to the latest level change. Over successive ticks, the trend converges to the actual rate of change of the
signal.

The prediction stage uses this trend to extrapolate the aggregated value forward to the prediction horizon
(Section~\ref{sec:prediction-horizon}), producing $A_H$ (Section~\ref{sec:extrapolation}).

\subsection{Asymmetric Reaction}
\label{sec:asymmetric-reaction}

The algorithm treats upward and downward movements differently. This is a crucial design choice that
reflects the different risks of missing a spike versus overreacting to a drop.

The whole point of the prediction stage is to catch spikes early and scale up before overload happens.
A drop, on the other hand, only means the cluster is temporarily over-provisioned, which costs resources but doesn't
affect users. There is no urgency to scale down immediately; in fact, scaling down too eagerly on a
momentary dip risks having to scale right back up.

This asymmetry motivates using different parameters for each direction. When the forecast
underestimates the input ($F_t < A_t$), the algorithm uses the aggressive pair
$(\alpha_{\uparrow}, \beta_{\uparrow})$ so the level and trend react quickly, catching the spike
faster. When the forecast meets or exceeds the input ($F_t \geq A_t$), it uses the conservative pair
$(\alpha_{\downarrow}, \beta_{\downarrow})$, letting the downward trend build slowly over many ticks
before the algorithm becomes confident that the drop is real.

\[
(\alpha_t,\; \beta_t) = \begin{cases} (\alpha_{\uparrow},\; \beta_{\uparrow}) & \text{if } A_t > F_t \\ (\alpha_{\downarrow},\; \beta_{\downarrow}) & \text{otherwise} \end{cases}
\]

Typically $\alpha_{\uparrow} > \alpha_{\downarrow}$ and $\beta_{\uparrow} > \beta_{\downarrow}$.

\subsection{Trend Dampening}
\label{sec:trend-dampening}

Trend dampening addresses a fundamental problem with any smoothing that tracks a trend:
\textbf{downward overshoot}.

When the real signal drops and then levels off, the smoothed level follows the drop and the trend becomes
increasingly negative. When the real signal flattens, the smoothed level can't stop immediately: the
accumulated negative trend carries it below the real value. To recover, the level must then rise back up
toward the real signal, which creates a positive trend. The algorithm could interpret this upward recovery as
a real load increase and trigger a false scale-up.

The dampening mechanism prevents this overshoot by reducing the trend whenever the level is above the input.
Whenever the smoothed level is above the actual input ($l_t > A_t$), the trend is progressively reduced:

\begin{align*}
g_t &= l_t - A_t \\
t_t &= t_t \cdot \frac{g_t}{g_t + |t_t| + \varepsilon}
\end{align*}

The dampening factor $g_t \,/\, (g_t + |t_t| + \varepsilon)$ is self-regulating: when the trend is large
relative to the gap, it's dampened heavily; when the gap is large relative to the trend, the dampening is
light. This decelerates the level as it approaches the real value from above: instead of crossing below and
bouncing, it merges smoothly into the real curve. The level is allowed to overshoot upward but not downward.

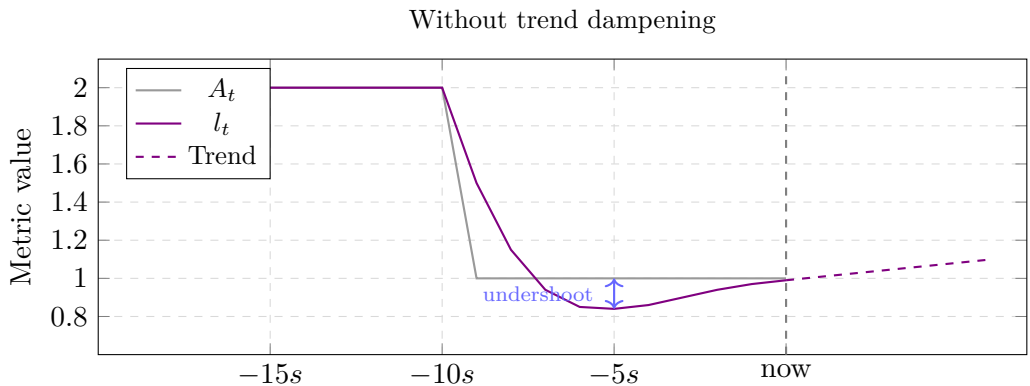
\begin{figure}[H]
\centering
\begin{tikzpicture}
\begin{axis}[
    width=0.92\textwidth,
    height=5.5cm,
    title={\small Without trend dampening},
    ylabel={Metric value},
    xmin=-20, xmax=7,
    ymin=0.6, ymax=2.15,
    xtick={-15, -10, -5, 0},
    xticklabels={$-15s$, $-10s$, $-5s$, now},
    ytick={0.8, 1.0, 1.2, 1.4, 1.6, 1.8, 2.0},
    grid=major,
    major grid style={dashed, black!15},
    legend style={at={(0.03,0.97)}, anchor=north west, font=\small},
    clip=false,
]

\draw[dashed, black!50, thick] (axis cs:0,0.6) -- (axis cs:0,2.1);

\addplot[thick, black!40, mark=none] coordinates {
    (-19,2.00) (-18,2.00) (-17,2.00) (-16,2.00) (-15,2.00) (-14,2.00)
    (-13,2.00) (-12,2.00) (-11,2.00) (-10,2.00) (-9,1.00) (-8,1.00)
    (-7,1.00) (-6,1.00) (-5,1.00) (-4,1.00) (-3,1.00) (-2,1.00)
    (-1,1.00) (0,1.00)
};
\addlegendentry{$A_t$}

\addplot[thick, violet, mark=none] coordinates {
    (-19,2.00) (-18,2.00) (-17,2.00) (-16,2.00) (-15,2.00) (-14,2.00)
    (-13,2.00) (-12,2.00) (-11,2.00) (-10,2.00) (-9,1.50) (-8,1.15)
    (-7,0.94) (-6,0.85) (-5,0.84) (-4,0.86) (-3,0.90) (-2,0.94)
    (-1,0.97) (0,0.99)
};
\addlegendentry{$l_t$}

\addplot[thick, dashed, violet, mark=none] coordinates {
    (0,0.99) (6,1.10)
};
\addlegendentry{Trend}

\draw[<->, thick, blue!60] (axis cs:-5,0.84) -- (axis cs:-5,1.00);
\node[anchor=east, font=\scriptsize, blue!60] at (axis cs:-5.3,0.92) {undershoot};

\end{axis}
\end{tikzpicture}
\caption{Without trend dampening: the level undershoots below the real value and then
recovers upward, creating a false positive trend (a potential false scale-up signal).}
\label{fig:no-dampening}
\end{figure}

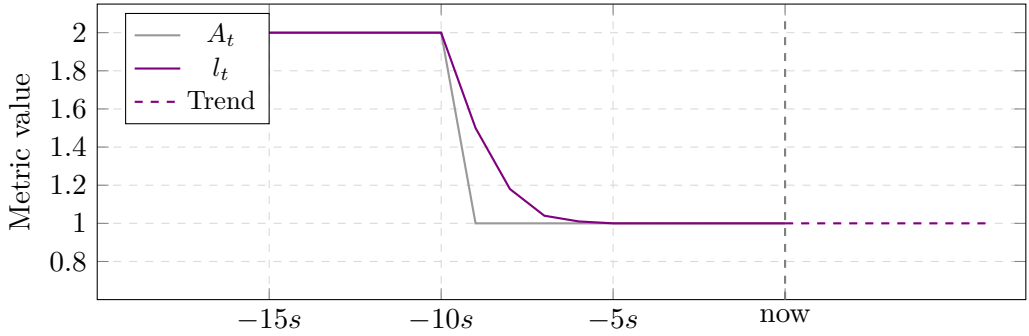
\begin{figure}[H]
\centering
\begin{tikzpicture}
\begin{axis}[
    width=0.92\textwidth,
    height=5.5cm,
    ylabel={Metric value},
    xmin=-20, xmax=7,
    ymin=0.6, ymax=2.15,
    xtick={-15, -10, -5, 0},
    xticklabels={$-15s$, $-10s$, $-5s$, now},
    ytick={0.8, 1.0, 1.2, 1.4, 1.6, 1.8, 2.0},
    grid=major,
    major grid style={dashed, black!15},
    legend style={at={(0.03,0.97)}, anchor=north west, font=\small},
    clip=false,
]

\draw[dashed, black!50, thick] (axis cs:0,0.6) -- (axis cs:0,2.1);

\addplot[thick, black!40, mark=none] coordinates {
    (-19,2.00) (-18,2.00) (-17,2.00) (-16,2.00) (-15,2.00) (-14,2.00)
    (-13,2.00) (-12,2.00) (-11,2.00) (-10,2.00) (-9,1.00) (-8,1.00)
    (-7,1.00) (-6,1.00) (-5,1.00) (-4,1.00) (-3,1.00) (-2,1.00)
    (-1,1.00) (0,1.00)
};
\addlegendentry{$A_t$}

\addplot[thick, violet, mark=none] coordinates {
    (-19,2.00) (-18,2.00) (-17,2.00) (-16,2.00) (-15,2.00) (-14,2.00)
    (-13,2.00) (-12,2.00) (-11,2.00) (-10,2.00) (-9,1.50) (-8,1.18)
    (-7,1.04) (-6,1.01) (-5,1.00) (-4,1.00) (-3,1.00) (-2,1.00)
    (-1,1.00) (0,1.00)
};
\addlegendentry{$l_t$}

\addplot[thick, dashed, violet, mark=none] coordinates {
    (0,1.00) (6,1.00)
};
\addlegendentry{Trend}

\end{axis}
\end{tikzpicture}
\caption{With trend dampening: the level converges to the real value from above and the
trend settles to zero. The extrapolation is flat, correctly reflecting the steady load.}
\label{fig:with-dampening}
\end{figure}

\subsection{Redistribution Delta Compensation}
\label{sec:delta-compensation}

During redistribution (Section~\ref{sec:redistribution}), the aggregated value $A_t$ changes
for two reasons: changes in the external load to the application and gradual inclusion of
new instances' values. The prediction stage needs the redistributed value to maintain an accurate level,
but the trend should reflect only real load changes; the redistribution component
must be accounted for separately.

The delta $\Delta_t^R$ (Section~\ref{sec:redistribution-delta}) captures the
expected redistribution component of the change in $A_t$. The smoother uses it in two places
to neutralize the impact:

\paragraph{In the forecast.}
Adding $\Delta_t^R$ to the one-step-ahead forecast makes the redistribution growth
``expected.'' When the actual input matches the forecast, the trend is unchanged.
Only changes in the per-instance metrics, whether from stable or new instances,
--- affect the trend:

\[
F_t = l_{t-1} + t_{t-1} + \Delta_t^R
\]

\paragraph{In the trend update.}
Because $\Delta_t^R$ appears in the forecast, it inflates $l_t$ via the level update
$l_t = \alpha_t A_t + (1 - \alpha_t) F_t$. To prevent $\Delta_t^R$ from leaking into the
trend through the level, it is subtracted from the level difference:

\[
t_t = \beta_t \cdot (l_t - l_{t-1} - \Delta_t^R) + (1 - \beta_t) \cdot t_{t-1}
\]

Without this correction, the level difference $l_t - l_{t-1}$ would carry the delta's
contribution, gradually building a false positive trend of approximately
$\beta_t \cdot \Delta_t^R$ per tick.

\paragraph{Effect on the trend.}

After the compensation, the trend captures only real load changes. Changes from stable
instances impact the trend at full weight. Changes from new instances impact it
proportionally to their current stabilization weight. The redistribution itself is
invisible to the trend.

When $\Delta_t^R = 0$ (either because no new instances exist, or because drop absorption
is active, Section~\ref{sec:redistribution-delta}), both equations reduce to the standard
Holt form.

\subsection{Metric Saturation}
\label{sec:saturation}

Some metrics have a natural upper bound (for example, Event Loop Utilization (ELU) is capped
at 1.0). When the metric approaches this bound, the signal is \emph{clipped}: the real load may
be growing, but the metric cannot rise further to reflect it. The trend, which is derived from
level changes, would decay toward zero even though load is still increasing. The algorithm
would underestimate future load and fail to scale up.

\paragraph{Detection.}
The algorithm detects saturation by comparing the raw (unweighted) aggregate $A_t^r$
against a per-instance maximum $v_{\max}$, with a configurable saturation zone $\sigma_z$
(e.g.\ 0.02):

\[
A_t^r \;\geq\; N_t \cdot v_{\max} \cdot (1 - \sigma_z)
\]

\paragraph{Behavior during saturation.}
When saturated, the level is clamped at the aggregate maximum
$N_t \cdot v_{\max}$ to prevent it from drifting above the clipped input.
The trend is constrained: it may increase but never decrease:

\begin{align*}
l_t &\leftarrow \min(l_t,\;\; N_t \cdot v_{\max}) \\
t_t &\leftarrow \max(t_t,\;\; t_{t-1})
\end{align*}

When saturation ends (the raw aggregate drops below the threshold), normal Holt updates
resume and the trend self-corrects within a few ticks.

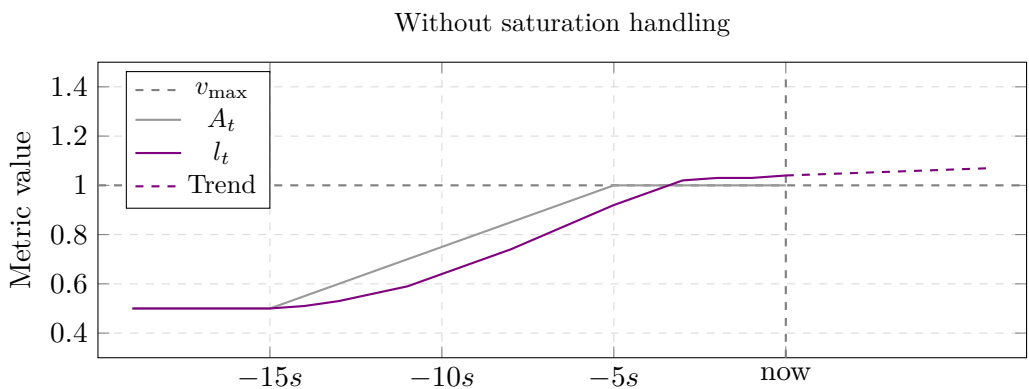
\begin{figure}[H]
\centering
\begin{tikzpicture}
\begin{axis}[
    width=0.92\textwidth,
    height=5.5cm,
    title={\small Without saturation handling},
    ylabel={Metric value},
    xmin=-20, xmax=7,
    ymin=0.3, ymax=1.5,
    xtick={-15, -10, -5, 0},
    xticklabels={$-15s$, $-10s$, $-5s$, now},
    ytick={0.4, 0.6, 0.8, 1.0, 1.2, 1.4},
    grid=major,
    major grid style={dashed, black!15},
    legend style={at={(0.03,0.97)}, anchor=north west, font=\small},
    clip=false,
]

\draw[dashed, black!50, thick] (axis cs:0,0.3) -- (axis cs:0,1.45);

\addplot[thick, dashed, black!50, mark=none] coordinates {(-20,1.0) (7,1.0)};
\addlegendentry{$v_{\max}$}

\addplot[thick, black!40, mark=none] coordinates {
    (-19,0.50) (-18,0.50) (-17,0.50) (-16,0.50) (-15,0.50)
    (-14,0.55) (-13,0.60) (-12,0.65) (-11,0.70) (-10,0.75)
    (-9,0.80) (-8,0.85) (-7,0.90) (-6,0.95) (-5,1.00)
    (-4,1.00) (-3,1.00) (-2,1.00) (-1,1.00) (0,1.00)
};
\addlegendentry{$A_t$}

\addplot[thick, violet, mark=none] coordinates {
    (-19,0.50) (-18,0.50) (-17,0.50) (-16,0.50) (-15,0.50)
    (-14,0.51) (-13,0.53) (-12,0.56) (-11,0.59) (-10,0.64)
    (-9,0.69) (-8,0.74) (-7,0.80) (-6,0.86) (-5,0.92)
    (-4,0.97) (-3,1.02) (-2,1.03) (-1,1.03) (0,1.04)
};
\addlegendentry{$l_t$}

\addplot[thick, dashed, violet, mark=none] coordinates {
    (0,1.04) (6,1.07)
};
\addlegendentry{Trend}

\end{axis}
\end{tikzpicture}
\caption{Without saturation handling: the metric is clamped at $v_{\max} = 1.0$ from
$-5s$ onward, but load is still growing behind the clipped signal. The trend decays
toward zero and the extrapolation is nearly flat. The algorithm fails to predict
continued growth.}
\label{fig:no-saturation}
\end{figure}

\begin{figure}[H]
\centering
\begin{tikzpicture}
\begin{axis}[
    width=0.92\textwidth,
    height=5.5cm,
    title={\small With saturation handling},
    ylabel={Metric value},
    xmin=-20, xmax=7,
    ymin=0.3, ymax=1.5,
    xtick={-15, -10, -5, 0},
    xticklabels={$-15s$, $-10s$, $-5s$, now},
    ytick={0.4, 0.6, 0.8, 1.0, 1.2, 1.4},
    grid=major,
    major grid style={dashed, black!15},
    legend style={at={(0.03,0.97)}, anchor=north west, font=\small},
    clip=false,
]

\draw[dashed, black!50, thick] (axis cs:0,0.3) -- (axis cs:0,1.45);

\addplot[thick, dashed, black!50, mark=none] coordinates {(-20,1.0) (7,1.0)};
\addlegendentry{$v_{\max}$}

\addplot[thick, black!40, mark=none] coordinates {
    (-19,0.50) (-18,0.50) (-17,0.50) (-16,0.50) (-15,0.50)
    (-14,0.55) (-13,0.60) (-12,0.65) (-11,0.70) (-10,0.75)
    (-9,0.80) (-8,0.85) (-7,0.90) (-6,0.95) (-5,1.00)
    (-4,1.00) (-3,1.00) (-2,1.00) (-1,1.00) (0,1.00)
};
\addlegendentry{$A_t$}

\addplot[thick, violet, mark=none] coordinates {
    (-19,0.50) (-18,0.50) (-17,0.50) (-16,0.50) (-15,0.50)
    (-14,0.51) (-13,0.53) (-12,0.56) (-11,0.59) (-10,0.64)
    (-9,0.69) (-8,0.74) (-7,0.80) (-6,0.86) (-5,0.92)
    (-4,0.97) (-3,1.00) (-2,1.00) (-1,1.00) (0,1.00)
};
\addlegendentry{$l_t$}

\addplot[thick, dashed, violet, mark=none] coordinates {
    (0,1.00) (6,1.28)
};
\addlegendentry{Trend}

\end{axis}
\end{tikzpicture}
\caption{With saturation handling: the trend is preserved during the clipped period and
the level is clamped at $v_{\max}$. The extrapolation continues upward, correctly predicting
that more capacity is needed.}
\label{fig:with-saturation}
\end{figure}
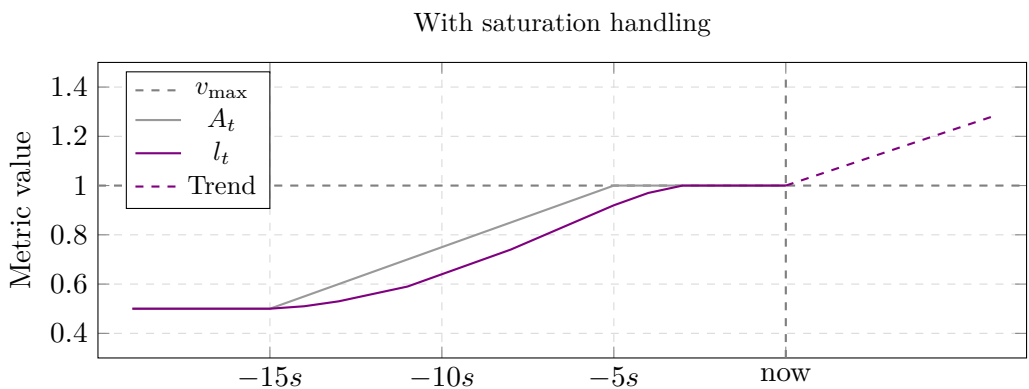

Saturation is enabled for metrics that have a natural upper bound (e.g.\ ELU, capped at 1.0).
When $v_{\max}$ is not configured, the saturation check is skipped entirely.

\subsection{The Prediction Horizon}
\label{sec:prediction-horizon}

The prediction horizon is derived from the init timeout, scaled by a configurable multiplier $\eta$,
and clamped to configurable bounds:

\[
H = \clamp(\eta \cdot T_I,\;\; H_{\min},\;\; H_{\max})
\]

The multiplier acts as a safety buffer on top of the measured init timeout. At $\eta = 1$ the algorithm
predicts exactly one startup time ahead. In practice, a value between 1 and 2 is recommended: this
accounts for cases where the next scale-up takes slightly longer than average and covers some of the
redistribution time after the new instance starts.

The floor $H_{\min}$ ensures a useful prediction horizon even when $T_I$ is very short. A fast-starting
application may produce a small $T_I$, which is an accurate reflection of startup time, but a very
short horizon makes the algorithm less effective. The floor guarantees the
algorithm always looks far enough ahead to make meaningful predictions, regardless of how quickly
instances start.

The ceiling $H_{\max}$ prevents the horizon from growing too long when $T_I$ is large.
With a long prediction horizon, the algorithm tries to predict too far into the future,
which is inherently uncertain. The trend extrapolation becomes less reliable and the
algorithm may end up scaling up unnecessarily. The ceiling caps this risk even if
$T_I$ or $\eta$ are configured high.

\subsection{Extrapolation}
\label{sec:extrapolation}

In its final step, the prediction stage takes the level $l_{\text{now}}$ and trend $t_{\text{now}}$
from the most recent tick and extrapolates forward to the prediction horizon, producing the predicted
total load at the time a new instance would be ready:

\[
A_H = l_{\text{now}} + t_{\text{now}} \cdot \frac{H}{\Delta t}
\]

\begin{figure}[H]
\centering
\begin{tikzpicture}
\begin{axis}[
    width=0.92\textwidth,
    height=6cm,
    ylabel={Metric value},
    xmin=-25, xmax=9,
    ymin=0.8, ymax=2.9,
    xtick={-20, -15, -10, -5, 0, 6},
    xticklabels={$-20s$, $-15s$, $-10s$, $-5s$, now, $H$},
    ytick={1.0, 1.2, 1.4, 1.6, 1.8, 2.0, 2.2, 2.4, 2.6, 2.8},
    grid=major,
    major grid style={dashed, black!15},
    legend style={at={(0.03,0.97)}, anchor=north west, font=\small},
    clip=false,
]

\draw[dashed, black!50, thick] (axis cs:0,0.8) -- (axis cs:0,2.8);

\draw[dashed, black!50] (axis cs:6,0.8) -- (axis cs:6,2.71);

\addplot[thick, black!40, mark=none] coordinates {
    (-24,1.00) (-23,1.20) (-22,1.40) (-21,1.35) (-20,1.15) (-19,1.05)
    (-18,1.20) (-17,1.45) (-16,1.60) (-15,1.55) (-14,1.35) (-13,1.70)
    (-12,1.25) (-11,1.35) (-10,1.55) (-9,1.75) (-8,1.85)
    (-7,1.70) (-6,1.65) (-5,1.80) (-4,2.00) (-3,2.15)
    (-2,2.05) (-1,2.10) (0,2.25)
};
\addlegendentry{Aggregated value $A_t$}

\addplot[thick, violet, mark=none] coordinates {
    (-24,1.00) (-23,1.10) (-22,1.27) (-21,1.34) (-20,1.34) (-19,1.29)
    (-18,1.30) (-17,1.39) (-16,1.52) (-15,1.59) (-14,1.57) (-13,1.67)
    (-12,1.59) (-11,1.55) (-10,1.58) (-9,1.68) (-8,1.79)
    (-7,1.82) (-6,1.81) (-5,1.85) (-4,1.95) (-3,2.08)
    (-2,2.13) (-1,2.18) (0,2.26)
};
\addlegendentry{Level $l_t$}

\addplot[thick, dashed, violet, mark=none] coordinates {
    (0,2.26) (6,2.71)
};
\addlegendentry{Trend extrapolation}

\node[fill=violet, circle, inner sep=1.5pt] at (axis cs:6,2.71) {};
\node[anchor=west, font=\small] at (axis cs:6.3,2.71) {$A_H$};

\end{axis}
\end{tikzpicture}
\caption{The complete prediction output. The level $l_t$ (solid purple) tracks the aggregated
input $A_t$ (grey), then the trend extrapolation (dashed purple) projects the level forward
from now to the prediction horizon~$H$, producing the predicted load~$A_H$.}
\label{fig:extrapolation}
\end{figure}
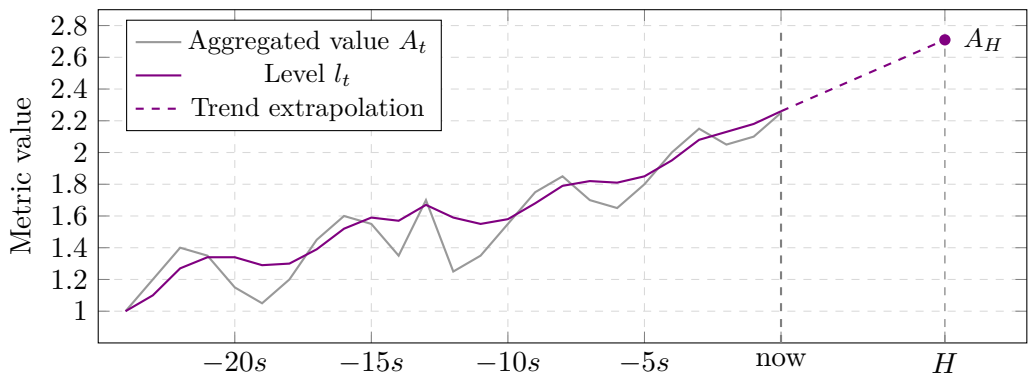

\subsection{Output}

The prediction stage passes the following to the decision stage:

\begin{itemize}[leftmargin=*, itemsep=3pt]
    \item $A_H$, the predicted aggregated value at the prediction horizon, representing the forecasted total load when a new instance would be ready.
    \item $\bar{A}_{\text{now}} = l_{\text{now}}$, the current smoothed aggregated value (the Holt level at time $t = \text{now}$).
    \item $t_{\text{now}}$, the current trend of the aggregated value.
\end{itemize}

\section{Scaling Decision}
\label{sec:scaling-decision}

\subsection{Purpose}

The decision stage takes $A_H$, $\bar{A}_{\text{now}}$, and $t_{\text{now}}$ from the prediction stage and
converts them into a target instance count: how many instances need to be scaled up or down. The main
purpose of the algorithm is to keep the per-instance load below a certain threshold $\tau$ with the
minimum number of instances.

To decide whether to scale, it needs to know the \emph{per-instance} metric value.
It applies a projection function~$\mathcal{P}$ (by default, division) to
convert the aggregated values into per-instance values. For a given tick,
the current per-instance load is:
\[
P_t = \mathcal{P}(A_t,\, N_t)
\]
where $N_t$ is the instance count at time $t$.
The current per-instance metric projection $P_{\text{now}}$ is computed using the
current level $\bar{A}_{\text{now}}$ and the contribution-weighted count $N_{\text{now}}^w$
from the redistribution stage:
\[
P_{\text{now}} = \mathcal{P}(\bar{A}_{\text{now}},\, N_{\text{now}}^w)
\]
The same transformation is applied to the predicted value, but at the horizon the
divisor is the \emph{target} instance count $N_{\text{target}}$, the number of
instances the scaler is currently aiming for. This is needed because the scaler can
be triggered with a new batch before the previously scaled instances are initialized:
\[
P_H = \mathcal{P}(A_H,\, N_{\text{target}})
\]
This distinction matters: $P_{\text{now}}$ uses the weighted count because it
reflects the load each instance is \emph{currently} handling, while $P_H$ uses
the target count because the question is whether the planned fleet will be
overloaded.

\begin{figure}[H]
\centering
\begin{tikzpicture}
\begin{axis}[
    width=0.92\textwidth,
    height=6cm,
    ylabel={Per-instance metric},
    xmin=-25, xmax=9,
    ymin=0.2, ymax=0.9,
    xtick={-20, -15, -10, -5, 0, 6},
    xticklabels={$-20s$, $-15s$, $-10s$, $-5s$, now, $H$},
    ytick={0.3, 0.4, 0.5, 0.6, 0.7, 0.8},
    grid=major,
    major grid style={dashed, black!15},
    legend style={at={(0.5,-0.15)}, anchor=north, font=\small, legend columns=2},
    clip=false,
]

\draw[dashed, black!50, thick] (axis cs:0,0.2) -- (axis cs:0,0.85);

\draw[dashed, black!50] (axis cs:6,0.2) -- (axis cs:6,0.78);

\addplot[thick, violet, mark=none] coordinates {
    (-24,0.33) (-23,0.37) (-22,0.42) (-21,0.45) (-20,0.45) (-19,0.43)
    (-18,0.43) (-17,0.46) (-16,0.50) (-15,0.52) (-14,0.50) (-13,0.54)
    (-12,0.51) (-11,0.49) (-10,0.51) (-9,0.55) (-8,0.59)
    (-7,0.60) (-6,0.59) (-5,0.60) (-4,0.63) (-3,0.68)
    (-2,0.70) (-1,0.70) (0,0.73)
};
\addlegendentry{Per-instance metric value $P_t$}

\addplot[thick, dashed, red, mark=none] coordinates {
    (-25,0.75) (9,0.75)
};
\addlegendentry{Overload threshold $\tau$}

\addplot[thick, dashed, violet, mark=none] coordinates {
    (0,0.73) (6,0.78)
};
\addlegendentry{Per-instance metric prediction}

\node[fill=violet, circle, inner sep=1.5pt] at (axis cs:0,0.73) {};
\node[anchor=south west, font=\small] at (axis cs:-2.9,0.75) {$P_{\text{now}}$};

\node[fill=violet, circle, inner sep=1.5pt] at (axis cs:6,0.78) {};
\node[anchor=west, font=\small] at (axis cs:6.3,0.78) {$P_H$};

\end{axis}
\end{tikzpicture}
\caption{The per-instance metric projection and prediction.}
\label{fig:decision-input}
\end{figure}
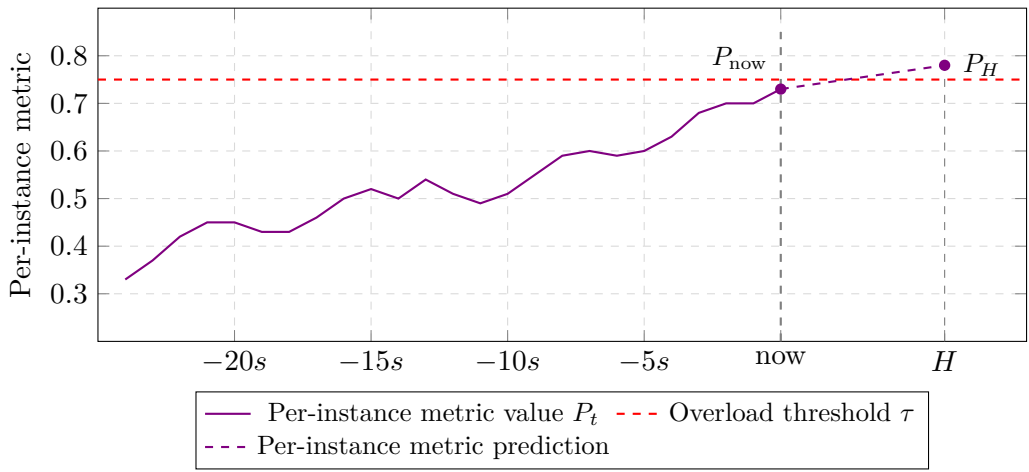

\subsection{Trend Direction}

The algorithm uses different decision logic depending on whether the load is rising, falling,
or nearly flat. To route to the correct path, it classifies the current trend into three
categories. The trend $t_{\text{now}}$ is normalized by the level to obtain the growth rate,
so that the classification is independent of the absolute load:

\[
\gamma = \frac{t_{\text{now}}}{l_{\text{now}}}
\]

\[
D = \begin{cases} \text{UP}         & \text{if } \gamma > \gamma_0 \\ \text{DOWN}       & \text{if } \gamma < -\gamma_0 \\ \text{HORIZONTAL} & \text{otherwise} \end{cases}
\]

The threshold $\gamma_0$ creates a deadband around zero. Small upward or downward movements
are classified as $\text{HORIZONTAL}$ rather than $\text{UP}$ or $\text{DOWN}$, so they are
handled by their own logic instead of being misrouted to a path designed for clear directional trends.

\begin{figure}[H]
\centering
\begin{tikzpicture}
\begin{axis}[
    width=0.92\textwidth,
    height=6cm,
    ylabel={Metric value},
    xmin=-2, xmax=9,
    ymin=0, ymax=1,
    xtick={-1, 0, 6},
    xticklabels={$-1s$, now, $H$},
    ytick={0, 0.2, 0.4, 0.6, 0.8, 1.0},
    grid=major,
    major grid style={dashed, black!15},
    clip=false,
]

\draw[dashed, black!50, thick] (axis cs:0,0) -- (axis cs:0,0.95);

\draw[dashed, black!50] (axis cs:6,0) -- (axis cs:6,0.70);

\addplot[thick, violet, mark=none] coordinates {
    (-2,0.30) (-1,0.35) (0,0.40)
};

\draw[thick, dashed, black!40] (axis cs:0,0.40) -- (axis cs:6,0.40);
\node[anchor=east, font=\small] at (axis cs:-0.15,0.46) {$l_{\text{now}}$};

\addplot[thick, dashed, violet, mark=none] coordinates {
    (0,0.40) (6,0.70)
};

\node[fill=violet, circle, inner sep=1.5pt] at (axis cs:6,0.70) {};
\node[anchor=south west, font=\small] at (axis cs:6.2,0.71) {$A_H$};

\draw[thick, blue!60] (axis cs:2,0.40)
    .. controls (axis cs:2.1,0.42) and (axis cs:2.1,0.48) ..
    (axis cs:2,0.50);
\node[font=\small, blue!60] at (axis cs:2.5,0.45) {$\theta$};

\draw[<->, thick, black!60] (axis cs:6.5,0.40) -- (axis cs:6.5,0.70);
\node[anchor=west, font=\small] at (axis cs:6.7,0.55) {$t_{\text{now}} \cdot \frac{H}{\Delta t}$};

\end{axis}
\end{tikzpicture}
\caption{The trend extrapolation from now to the prediction horizon $H$. The angle $\theta$
between the trend line and the horizontal baseline $l_{\text{now}}$ corresponds to the
growth rate $\gamma = \tan(\theta)$.}
\label{fig:trend-angle}
\end{figure}
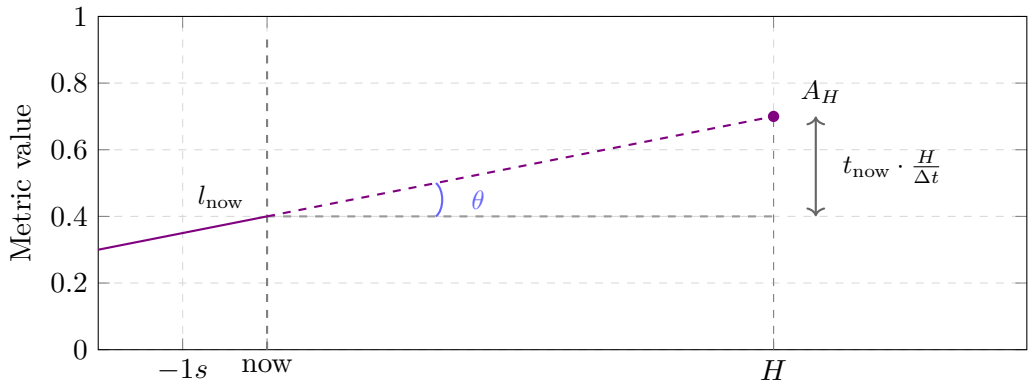

The growth rate $\gamma$ is related to the angle $\theta$ of the trend line by:

\[
\gamma = \tan(\theta)
\]

If the threshold is specified as an angle $\theta_0$ (in degrees), the equivalent growth rate
threshold is:

\[
\gamma_0 = \tan\!\left(\theta_0 \cdot \frac{\pi}{180}\right)
\]

A default of $\theta_0 = 10°$ (equivalently $\gamma_0 \approx 0.176$) classifies trends with
a per-tick growth rate below ${\sim}17.6\%$ of the current level as horizontal.

\subsection{Scaling Direction}

By this point, the algorithm has the following variables describing the current state:
\begin{itemize}[leftmargin=*, itemsep=3pt]
    \item $D$, the trend direction (\textsc{up}, \textsc{down}, \textsc{horizontal})
    \item $P_{\text{now}}$, the current per-instance smoothed level
    \item $P_H$, the predicted per-instance metric value at the horizon
\end{itemize}

Based on these, the algorithm decides which direction to scale, if scaling is needed at all.
Choosing the scaling direction does not mean the algorithm will actually change the instance
count; it still needs to calculate a target and check whether it differs from the current
count. But if the conditions for scaling up or down are not met, the algorithm will maintain
the current target count.

\medskip

\textbf{Scale-up} is considered in one of two cases:
\begin{itemize}[leftmargin=*, itemsep=3pt]
    \item $D = \textsc{up}$, the metric value is rising.
    \item $P_H > \tau$, the predicted per-instance value at the horizon exceeds the
        overload threshold. This catches the case when the application is already overloaded
        and the trend is not falling fast enough to bring the metric below~$\tau$ by the horizon.
\end{itemize}

\textbf{Scale-down} is considered when all of the following hold:
\begin{itemize}[leftmargin=*, itemsep=3pt]
    \item $D \in \{\textsc{horizontal},\, \textsc{down}\}$, the metric value is not rising.
    \item $P_H \leq \tau$, the predicted load at the horizon is within the threshold.
    \item $P_{\text{now}} \leq \tau$, the system is not currently overloaded.
\end{itemize}

\subsection{Scale-Up}

The scale-up logic converts the predicted aggregated load $A_H$ into a target instance count
$N^*$. The idea is to find the minimum number of instances that keeps the per-instance metric
at or below~$\tau$:

\[
N^* = \bigl\lceil \mathcal{N}(A_H,\, \tau) \bigr\rceil
\qquad \left(\,= \left\lceil \frac{A_H}{\tau} \right\rceil \text{ for the default model}\right)
\]

\subsubsection*{Asymmetric Risk}

The predicted aggregated value $A_H = \bar{A}_{\text{now}} + t_{\text{now}} \cdot H$ has two
components with very different certainty. The level $\bar{A}_{\text{now}}$ reflects the confirmed
state of the metric, load that is already happening. The trend contribution
$\Delta A_H = t_{\text{now}} \cdot H$ is a projection, load that \emph{might} happen if the trend
continues.

Every time the algorithm scales up based on the trend, it takes a risk: what if the trend
flattens out right after the decision? The new instances were requested to handle load that
never materialized, and the system ends up over-provisioned. The cost of this mistake depends
on how much of the prediction comes from the uncertain trend versus the confirmed level.

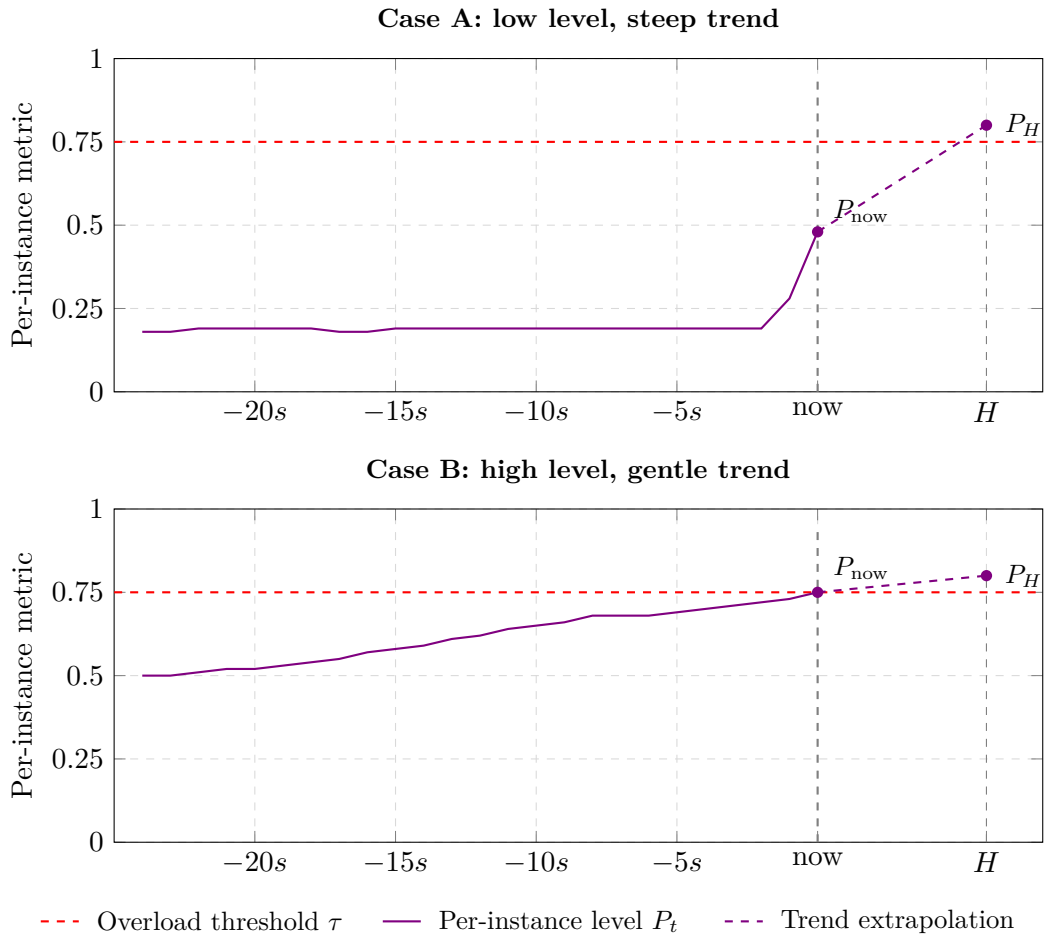
\begin{figure}[H]
\centering

\begin{subfigure}[t]{\textwidth}
\centering
\begin{tikzpicture}
\begin{axis}[
    width=0.92\textwidth,
    height=6cm,
    ylabel={Per-instance metric},
    title={Case~A: low level, steep trend},
    title style={font=\small\bfseries},
    xmin=-25, xmax=8,
    ymin=0, ymax=1.0,
    xtick={-20, -15, -10, -5, 0, 6},
    xticklabels={$-20s$, $-15s$, $-10s$, $-5s$, now, $H$},
    ytick={0, 0.25, 0.5, 0.75, 1.0},
    grid=major,
    major grid style={dashed, black!15},
    clip=false,
]

\addplot[thick, dashed, red, mark=none] coordinates {(-25,0.75) (8,0.75)};

\draw[dashed, black!50, thick] (axis cs:0,0) -- (axis cs:0,0.95);

\draw[dashed, black!50] (axis cs:6,0) -- (axis cs:6,0.80);

\addplot[thick, violet, mark=none] coordinates {
    (-24,0.18) (-23,0.18) (-22,0.19) (-21,0.19) (-20,0.19) (-19,0.19)
    (-18,0.19) (-17,0.18) (-16,0.18) (-15,0.19) (-14,0.19) (-13,0.19)
    (-12,0.19) (-11,0.19) (-10,0.19) (-9,0.19) (-8,0.19) (-7,0.19)
    (-6,0.19) (-5,0.19) (-4,0.19) (-3,0.19) (-2,0.19) (-1,0.28) (0,0.48)
};

\addplot[thick, dashed, violet, mark=none] coordinates {(0,0.48) (6,0.80)};

\node[fill=violet, circle, inner sep=1.5pt] at (axis cs:0,0.48) {};
\node[anchor=south west, font=\small] at (axis cs:0.2,0.48) {$P_{\text{now}}$};

\node[fill=violet, circle, inner sep=1.5pt] at (axis cs:6,0.80) {};
\node[anchor=west, font=\small] at (axis cs:6.3,0.80) {$P_H$};

\end{axis}
\end{tikzpicture}
\label{fig:risk-case-a}
\end{subfigure}

\vspace{0.5em}

\begin{subfigure}[t]{\textwidth}
\centering
\begin{tikzpicture}
\begin{axis}[
    width=0.92\textwidth,
    height=6cm,
    ylabel={Per-instance metric},
    title={Case~B: high level, gentle trend},
    title style={font=\small\bfseries},
    xmin=-25, xmax=8,
    ymin=0, ymax=1.0,
    xtick={-20, -15, -10, -5, 0, 6},
    xticklabels={$-20s$, $-15s$, $-10s$, $-5s$, now, $H$},
    ytick={0, 0.25, 0.5, 0.75, 1.0},
    grid=major,
    major grid style={dashed, black!15},
    clip=false,
]

\addplot[thick, dashed, red, mark=none] coordinates {(-25,0.75) (8,0.75)};

\draw[dashed, black!50, thick] (axis cs:0,0) -- (axis cs:0,0.95);

\draw[dashed, black!50] (axis cs:6,0) -- (axis cs:6,0.80);

\addplot[thick, violet, mark=none] coordinates {
    (-24,0.50) (-23,0.50) (-22,0.51) (-21,0.52) (-20,0.52) (-19,0.53)
    (-18,0.54) (-17,0.55) (-16,0.57) (-15,0.58) (-14,0.59) (-13,0.61)
    (-12,0.62) (-11,0.64) (-10,0.65) (-9,0.66) (-8,0.68) (-7,0.68)
    (-6,0.68) (-5,0.69) (-4,0.70) (-3,0.71) (-2,0.72) (-1,0.73) (0,0.75)
};

\addplot[thick, dashed, violet, mark=none] coordinates {(0,0.75) (6,0.80)};

\node[fill=violet, circle, inner sep=1.5pt] at (axis cs:0,0.75) {};
\node[anchor=south west, font=\small] at (axis cs:0.2,0.76) {$P_{\text{now}}$};

\node[fill=violet, circle, inner sep=1.5pt] at (axis cs:6,0.80) {};
\node[anchor=west, font=\small] at (axis cs:6.3,0.80) {$P_H$};

\end{axis}
\end{tikzpicture}
\label{fig:risk-case-b}
\end{subfigure}

\vspace{0.5em}
\begin{tikzpicture}
\draw[thick, dashed, red] (0,0.15) -- (0.5,0.15);
\node[font=\small, anchor=west] at (0.6,0.15) {Overload threshold $\tau$};
\draw[thick, violet] (4.5,0.15) -- (5.0,0.15);
\node[font=\small, anchor=west] at (5.1,0.15) {Per-instance level $P_t$};
\draw[thick, dashed, violet] (9.0,0.15) -- (9.5,0.15);
\node[font=\small, anchor=west] at (9.6,0.15) {Trend extrapolation};
\end{tikzpicture}

\caption{Per-instance metric view.}
\label{fig:risk-cases}
\end{figure}

Consider two scenarios with $N = 7$ instances and $\tau = 0.75$. Both have an upward trend,
and both reach the same per-instance prediction $P_H = 0.80$ at the horizon, above the
threshold (Figure~\ref{fig:risk-cases}). A naive approach would be to scale up in both cases,
adding an 8th instance.

\begin{figure}[H]
\centering
\hspace*{1.5cm}\begin{tikzpicture}[yscale=0.88]

\begin{scope}[xshift=0cm]
\node[font=\small\bfseries] at (1.5,8.35) {Case A};

\fill[black!5] (0,0) rectangle (3,1.0);
\fill[black!8] (0,1.0) rectangle (3,2.0);
\fill[black!5] (0,2.0) rectangle (3,3.0);
\fill[black!8] (0,3.0) rectangle (3,4.0);
\fill[black!5] (0,4.0) rectangle (3,5.0);
\fill[black!8] (0,5.0) rectangle (3,6.0);
\fill[black!5] (0,6.0) rectangle (3,7.0);
\draw[black!25] (0,1.0) -- (3,1.0);
\draw[black!25] (0,2.0) -- (3,2.0);
\draw[black!25] (0,3.0) -- (3,3.0);
\draw[black!25] (0,4.0) -- (3,4.0);
\draw[black!25] (0,5.0) -- (3,5.0);
\draw[black!25] (0,6.0) -- (3,6.0);

\draw[dashed, black!30, thick] (0,7.0) rectangle (3,8.0);

\node[font=\scriptsize, black!40, anchor=east] at (-0.15,0.5) {inst.\ 1};
\node[font=\scriptsize, black!40, anchor=east] at (-0.15,1.5) {inst.\ 2};
\node[font=\scriptsize, black!40, anchor=east] at (-0.15,2.5) {inst.\ 3};
\node[font=\scriptsize, black!40, anchor=east] at (-0.15,3.5) {inst.\ 4};
\node[font=\scriptsize, black!40, anchor=east] at (-0.15,4.5) {inst.\ 5};
\node[font=\scriptsize, black!40, anchor=east] at (-0.15,5.5) {inst.\ 6};
\node[font=\scriptsize, black!40, anchor=east] at (-0.15,6.5) {inst.\ 7};
\node[font=\scriptsize, black!30, anchor=east] at (-0.15,7.5) {inst.\ 8};

\fill[violet!60] (0.3,0) rectangle (2.7,4.45);
\node[font=\scriptsize, white] at (1.5,2.2) {$\bar{A}_{\text{now}} = 3.34$};

\fill[violet!30] (0.3,4.45) rectangle (2.7,7.47);
\node[font=\scriptsize] at (1.5,5.96) {$\Delta A_H = 2.26$};

\end{scope}

\begin{scope}[xshift=6.0cm]
\node[font=\small\bfseries] at (1.5,8.35) {Case B};

\fill[black!5] (0,0) rectangle (3,1.0);
\fill[black!8] (0,1.0) rectangle (3,2.0);
\fill[black!5] (0,2.0) rectangle (3,3.0);
\fill[black!8] (0,3.0) rectangle (3,4.0);
\fill[black!5] (0,4.0) rectangle (3,5.0);
\fill[black!8] (0,5.0) rectangle (3,6.0);
\fill[black!5] (0,6.0) rectangle (3,7.0);
\draw[black!25] (0,1.0) -- (3,1.0);
\draw[black!25] (0,2.0) -- (3,2.0);
\draw[black!25] (0,3.0) -- (3,3.0);
\draw[black!25] (0,4.0) -- (3,4.0);
\draw[black!25] (0,5.0) -- (3,5.0);
\draw[black!25] (0,6.0) -- (3,6.0);

\draw[dashed, black!30, thick] (0,7.0) rectangle (3,8.0);

\node[font=\scriptsize, black!40, anchor=east] at (-0.15,0.5) {inst.\ 1};
\node[font=\scriptsize, black!40, anchor=east] at (-0.15,1.5) {inst.\ 2};
\node[font=\scriptsize, black!40, anchor=east] at (-0.15,2.5) {inst.\ 3};
\node[font=\scriptsize, black!40, anchor=east] at (-0.15,3.5) {inst.\ 4};
\node[font=\scriptsize, black!40, anchor=east] at (-0.15,4.5) {inst.\ 5};
\node[font=\scriptsize, black!40, anchor=east] at (-0.15,5.5) {inst.\ 6};
\node[font=\scriptsize, black!40, anchor=east] at (-0.15,6.5) {inst.\ 7};

\fill[violet!60] (0.3,0) rectangle (2.7,6.97);
\node[font=\scriptsize, white] at (1.5,3.49) {$\bar{A}_{\text{now}} = 5.23$};

\fill[violet!30] (0.3,6.97) rectangle (2.7,7.47);
\node[font=\scriptsize] at (1.5,7.22) {$\Delta A_H = 0.37$};

\end{scope}

\draw[thick, dashed, violet] (-0.2,7.47) -- (9.2,7.47);
\node[font=\scriptsize, violet, anchor=west] at (9.3,7.47) {Predicted usage ($A_H = 5.60$)};

\draw[thick, dashed, red] (-0.2,7.0) -- (9.2,7.0);
\node[font=\scriptsize, red, anchor=west] at (9.3,7.0) {Capacity threshold ($N \cdot \tau = 5.25$)};

\end{tikzpicture}

\vspace{0.3em}
\begin{tikzpicture}
\fill[violet!60] (0,0) rectangle (0.4,0.3);
\node[font=\small, anchor=west] at (0.5,0.15) {Confirmed usage ($\bar{A}_{\text{now}}$)};
\fill[violet!30] (4.8,0) rectangle (5.2,0.3);
\node[font=\small, anchor=west] at (5.3,0.15) {Predicted usage increase ($\Delta A_H$)};
\end{tikzpicture}

\caption{Capacity comparison.}
\label{fig:risk-capacity}
\end{figure}
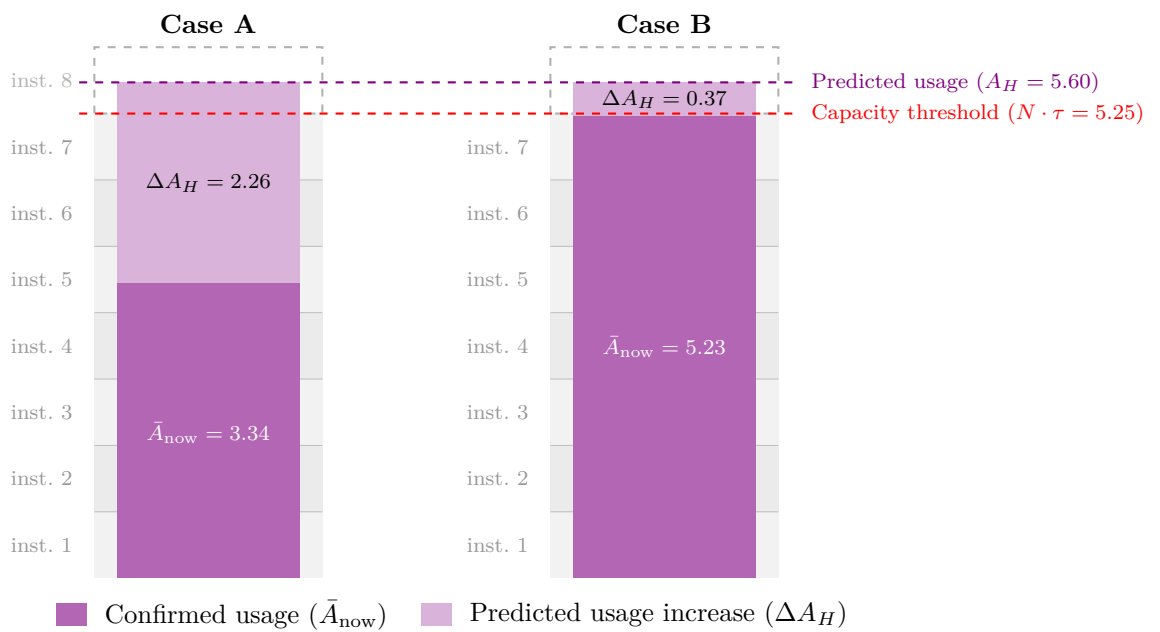

Figure~\ref{fig:risk-capacity} reframes this from the cluster's perspective. Instead of
per-instance metrics, it shows the \emph{aggregated} metric across all $N = 7$ instances stacked
vertically. Each grey segment represents one instance's capacity ($\tau = 0.75$). The red
dashed line marks the current total capacity $N \cdot \tau$.

This view reveals a crucial difference. In Case~A, the confirmed level is $3.34$ (well
below capacity) and the trend contributes $2.26$. In Case~B, the confirmed level is
$5.23$ (nearly filling all 7~instances) and the trend contributes only $0.37$.
Consider what happens if the spike stops right after the scale-up.

In Case~A, if the spike ends the moment we add the 8th instance, the cluster is
over-scaled by 3 instances while in Case~B, it's only over-scaled by 1 instance.
The risk of over-scaling is much higher in Case~A because the trend is a larger portion of the prediction.
The algorithm must account for it.

The algorithm adjusts the trend contribution based on the \emph{growth ratio} ---
how much of the prediction relies on the trend:

\[
\rho = \frac{\Delta A_H}{\bar{A}_{\text{now}}}
\]

In Case~A from Figures~\ref{fig:risk-cases}--\ref{fig:risk-capacity}, $\rho = 0.68$:
the trend adds 68\% on top of the confirmed level, the target count relies heavily
on the trend continuing, and the risk of over-provisioning is substantial if it does
not. In Case~B, $\rho = 0.07$: the trend adds only 7\%, almost the entire
prediction is confirmed load, so the risk is low regardless of what the trend does
next.

When the growth ratio is high, the algorithm reduces the trend contribution to limit
the over-provisioning risk:

\[
\omega = \frac{k}{k + \rho}
\]

\[
\hat{A}_H = \bar{A}_{\text{now}} + \omega \cdot \Delta A_H
\]

When $\rho$ is small, $\omega$ is close to~1 and the trend contribution is preserved
almost entirely, the target was already well-supported by confirmed load. When
$\rho$ is large, $\omega$ decreases and the trend contribution is reduced. The
parameter $k$ (default $k = 2$) controls the balance: under-provisioning is treated
as $k$ times more costly than over-provisioning. Higher $k$ preserves more of the
trend contribution; $k = 1$ would weight both risks equally.

\[
N^* = \bigl\lceil \mathcal{N}(\hat{A}_H,\, \tau) \bigr\rceil
\qquad \left(\,= \left\lceil \frac{\hat{A}_H}{\tau} \right\rceil \text{ for the default model}\right)
\]

\subsubsection*{Trim spillover capacity}

The predicted load $\hat{A}_H$ combines a confirmed component ($\bar{A}_{\text{now}}$)
and an unconfirmed trend extrapolation. Ceiling division on this value guarantees
enough capacity, but it can allocate an extra instance whose need rests entirely on
the trend contribution. If the cluster needs less than 10\% of the extra instance
capacity, the evidence for provisioning it is weak: the trend
may not materialize, and adding the instance now risks overscaling. The algorithm
prefers to wait: if the load genuinely grows, the next evaluation cycle will
confirm the need and provision the instance then.

\begin{figure}[H]
\centering
\begin{tikzpicture}[yscale=1.4]

\begin{scope}[xshift=0cm]

\fill[black!5] (0,0) rectangle (3,1.0);
\fill[black!8] (0,1.0) rectangle (3,2.0);
\draw[black!25] (0,1.0) -- (3,1.0);

\draw[dashed, black!30, thick] (0,2.0) rectangle (3,3.0);

\draw[dashed, black!30, thick] (0,3.0) rectangle (3,4.0);

\node[font=\scriptsize, black!40, anchor=east] at (-0.15,0.5) {inst.\ 1};
\node[font=\scriptsize, black!40, anchor=east] at (-0.15,1.5) {inst.\ 2};
\node[font=\scriptsize, black!40, anchor=east] at (-0.15,2.5) {inst.\ 3};
\node[font=\scriptsize, black!30, anchor=east] at (-0.15,3.5) {inst.\ 4};

\fill[violet!60] (0.3,0) rectangle (2.7,1.5);
\node[font=\scriptsize, white] at (1.5,0.75) {$\bar{A}_{\text{now}}$};

\fill[violet!30] (0.3,1.5) rectangle (2.7,3.3);
\node[font=\scriptsize] at (1.5,2.4) {$\Delta A_H$};

\draw[thick, dashed, black!70] (-0.4,3.0) -- (3.4,3.0);
\draw[thick, dashed, black!70] (-0.4,3.3) -- (3.4,3.3);
\draw[<->, thick, black!70] (3.5,3.0) -- (3.5,3.3);
\node[font=\small, anchor=west] at (3.6,3.15) {$\delta$};

\end{scope}

\node[anchor=west, font=\normalsize] at (5.5,3.15) {%
    $\delta = \mathcal{N}(\hat{A}_H,\, \tau) - (N^* - 1)$};

\end{tikzpicture}
\caption{Trim spillover capacity. $\delta$ is the fractional instance need: the amount by which the
exact (real-valued) required count exceeds $N^* - 1$.
For the default model, $\delta = \hat{A}_H / \tau - (N^* - 1)$.}
\label{fig:instance-margin}
\end{figure}
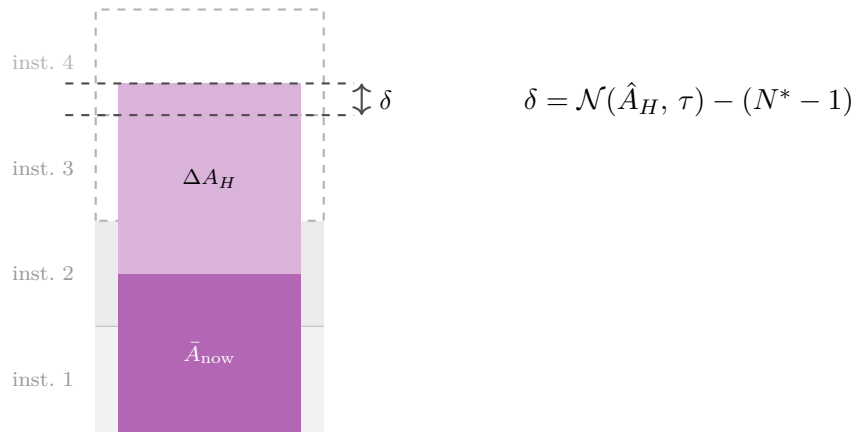

If the system is not currently overloaded ($P_{\text{now}} \leq \tau$) and the fractional
need is small ($\delta \leq 0.1$):

\[
N^* \leftarrow N^* - 1
\]

\subsubsection*{Safety bounds}

The result is clamped to three constraints: scale-up never reduces the count below the
current target (which would be a scale-down through the scale-up path), never exceeds
the configured maximum, and never adds more than $N_{\text{step}}$ instances per decision.

\[
N^* \leftarrow \clamp\!\left(N^*,\;\; N_{\text{target}},\;\; \min(N_{\text{target}} + N_{\text{step}},\;\; N_{\max})\right)
\]

\subsection{Scale-Down}

Scale-down is deliberately cautious. After removing an instance the load is
redistributed across the cluster, raising the per-instance metric value. The algorithm
must ensure that there is enough headroom to handle this increase plus some safety margin
that prevents the immediate scale-up triggered by small fluctuations.

The rule is: after scaling down and redistributing the load, the per-instance
metric should be below the threshold $\tau$ by a $\mu \cdot P_{\text{down}}$ margin,
where $P_{\text{down}} = \mathcal{P}\!\left(\bar{A}_{\text{now}},\; N_{\text{target}} - 1\right)$ is the per-instance load after removal:

\[
N^* = \left\lfloor \mathcal{N}\!\left(\bar{A}_{\text{now}},\; \frac{\tau}{1 + \mu}\right) \right\rfloor + 1
\qquad \left(\,= \left\lfloor \frac{(1 + \mu) \cdot \bar{A}_{\text{now}}}{\tau} \right\rfloor + 1 \text{ for the default model}\right)
\]

The result is clamped so that scale-down never drops below the configured minimum
and never increases the count (which would be a scale-up through the scale-down path).

\[
N^* \leftarrow \clamp\!\left(N^*,\;\; N_{\min},\;\; N_{\text{target}}\right)
\]

\subsection{Output}

The decision stage produces a target instance count $N^*$.

\section{Cooldowns}
\label{sec:cooldowns}

\subsection{Not Required, But Useful}

Cooldowns are \textbf{optional}. The core algorithm already prevents cascading scale-ups by tracking the target
instance count: when the algorithm decides to scale up, it accounts for
instances that have been requested but haven't started yet. If the requested instances are enough to handle the
predicted load, no additional scale-up is triggered.

The algorithm can operate correctly without any cooldowns. So why have them?

\subsection{Trading Efficiency for Stability}

Even if the algorithm makes accurate decisions at every point in time, the resulting scaling behavior can
be too volatile in some cases. A short traffic spike might trigger a scale-up, followed by a drop that triggers a
scale-down a few minutes later. Both decisions are correct (the instances were needed during the
spike and not needed after), but the rapid churn is undesirable. Starting and stopping instances has its own
costs: resource allocation overhead, connection draining, potential brief disruptions.

Cooldowns allow a user to trade some efficiency for stability. By enforcing minimum intervals between scaling actions,
they prevent the system from scaling too often.

\subsection{The Four Cooldown Types}

Each combination of consecutive scaling directions can have its own cooldown:

\begin{itemize}[leftmargin=*, itemsep=4pt]
    \item \textbf{Scale-up after scale-up}: Minimum time since the last scale-up \textbf{decision}. Even
        though pending scale-ups prevent cascading, a user might want additional spacing.

    \item \textbf{Scale-up after scale-down}: Minimum time since the last scale-down \textbf{decision}.
        Prevents rapid back-and-forth oscillation.

    \item \textbf{Scale-down after scale-up}: Minimum time since the last instance \textbf{actually started}
        (not since the scale-up decision). The clock starts when the instance registers, not when the decision
        was made; what matters is how long the instance has been running and absorbing load. The gap between
        the decision and the instance start is already covered: scale-down is blocked while there are pending
        scale-ups that haven't materialized yet.

    \item \textbf{Scale-down after scale-down}: Minimum time since the last scale-down \textbf{decision}.
        Limits how quickly capacity is removed.
\end{itemize}

\subsection{Relationship to Redistribution}

One practical consideration: if the scale-down-after-scale-up cooldown is shorter than $T_R$, a scale-down
could happen while new instances are still ramping up. The redistribution stage smooths the signal during this
period, but the signal hasn't fully settled. It's generally advisable to set the scale-down-after-scale-up
cooldown $\geq T_R$.

\section{Adaptive Init Timeout}
\label{sec:adaptive-init-timeout}

\subsection{Context}

The init timeout $T_I$ (how long a new instance takes to become ready after a scale-up decision) is a
key input to the algorithm. It determines the prediction horizon: how far into the future the algorithm looks
when deciding whether to scale.

A fixed $T_I$ works if startup times are consistent. But in practice, they vary: different machine types,
varying cluster load, image pull times, application warm-up that depends on cache state. The \emph{adaptive
init timeout} tracks these variations automatically.

This is one way to provide the init timeout input. If the startup time is known and stable, a fixed value works
fine. The core algorithm only needs the value; how it's determined is independent.

\subsection{How It's Measured}

Each time the algorithm decides to scale up, it records the decision timestamp $t_j^{\text{dec}}$. When the
new instance actually registers at $t_j^{\text{reg}}$, the difference is the measured startup time:

\[
m_j = t_j^{\text{reg}} - t_j^{\text{dec}}
\]

These measurements are collected in a sliding window $\mathbf{W}$ of configurable size.

\subsection{Requirements}

The calculation must satisfy three properties:

\begin{enumerate}[leftmargin=*, itemsep=4pt]
    \item \textbf{Exclude outliers.} A single abnormally slow or fast startup should not shift the timeout
        significantly. A cold image pull, a scheduling delay, or a network hiccup is not representative of
        typical behavior.

    \item \textbf{Adapt over time.} The real init timeout can change (a new application version may start
        faster or slower, the infrastructure may change). The timeout must converge toward the new reality, not
        stay anchored to historical values.

    \item \textbf{Not decrease too quickly.} If the timeout drops too fast and the real startup time hasn't
        actually improved, the algorithm will look at too short a horizon: it won't scale early enough and
        instances won't be ready in time. Decreasing slowly gives time to confirm the improvement is real.
\end{enumerate}

\subsection{The Calculation}

The timeout converges toward the \textbf{median} of recent measurements, which satisfies requirement
1, since median is inherently resistant to outliers.

Let $\tilde{m} = \median(\mathbf{W})$. The convergence uses clamped step sizes, satisfying requirements
2 and 3:

\begin{align*}
\delta &= \tilde{m} - T_I \\
\delta^+ &= T_I \cdot r \cdot f_{\uparrow} \\
\delta^- &= T_I \cdot r \cdot f_{\downarrow} \\
\hat{\delta} &= \clamp(\delta,\; -\delta^-,\; \delta^+) \\
T_I' &= \lfloor T_I + \hat{\delta} \rceil
\end{align*}

where $\lfloor \cdot \rceil$ denotes rounding to the nearest integer.

The timeout moves toward the median, but no single update can shift it by more than a fraction of its current
value. This keeps the convergence smooth and predictable.

$f_{\uparrow} > f_{\downarrow}$: the timeout increases faster than it decreases. If startup times have
genuinely increased, the algorithm needs to extend its horizon quickly (being too short is dangerous).
If startup times have decreased, the algorithm can shorten its horizon gradually (being too long is merely
cautious).

\section{Metric Model Examples}
\label{sec:metric-model-examples}

The algorithm is parameterized by three functions ($\mathcal{A}$, $\mathcal{P}$,
$\mathcal{N}$) that define how per-instance metric values relate to the
cluster-wide aggregate (Section~\ref{sec:introduction}). The default model (sum /
average) is the simplest choice, but any model satisfying the two required properties
(scaling invariance and per-instance separability) will work.

This section shows the general pattern, gives two non-trivial examples, and explains
what falls outside the framework.

\subsection{The General Pattern}

All supported models share the same structure. Pick any monotonically increasing function
$g$ with an inverse $g^{-1}$. If the aggregate can be written as
$\sum w_i \cdot g(v_i)$ for some per-instance function $g$, it works. If any term
depends on $j \neq i$ or on $N$, it does not (Section~\ref{sec:model-coupling}).
The metric model is:

\begin{align*}
\mathcal{A}\!\left(\{(v_i, w_i)\}\right) &= \sum_i w_i \cdot g(v_i) \\[4pt]
\mathcal{P}(S, N) &= g^{-1}\!\left(\frac{S}{N}\right) \\[4pt]
\mathcal{N}(S, \tau) &= \frac{S}{g(\tau)}
\end{align*}

The default model uses $g(v) = v$ (the identity), giving the familiar sum and average.
The non-linearity, if any, lives entirely in the per-instance transformation $g$.
Because each instance's contribution $w_i \cdot g(v_i)$ depends only on its own value
and weight, the aggregate is separable and invariant under load redistribution.

\subsection{Non-Trivial Models}
\label{sec:model-examples}

\paragraph{Per-instance overhead.}
\label{sec:model-overhead}

Suppose each instance has a fixed baseline cost $b$ that does not redistribute when
instances are added (e.g., a per-instance memory overhead, health-check load, or
background task). Only $v_i - b$ of each instance's metric value represents
redistributable load.

\medskip

\begin{tabular}{@{} l l @{}}
$g(v) = v - b$ & (subtract baseline) \\[4pt]
$\mathcal{A} = \sum w_i \cdot (v_i - b)$ & \\[2pt]
$\mathcal{P}(S, N) = S / N + b$ & \\[2pt]
$\mathcal{N}(S, \tau) = S / (\tau - b)$ & (requires $\tau > b$)
\end{tabular}

\medskip

Using the default model here would underestimate the required instance count: it
assumes the full aggregate can be spread evenly, but the $b$ per instance stays
fixed and does not redistribute.

\bigskip
\hrule
\bigskip

\paragraph{Quadratic per-instance cost.}
\label{sec:model-quadratic}

In some systems, the cost each instance imposes grows non-linearly with its load.
For example, at high CPU utilization request latency grows quadratically due to
queueing effects (an instance at $0.90$ is much more dangerous than one at $0.50$).

\medskip

\begin{tabular}{@{} l l @{}}
$g(v) = v^2$ & (quadratic cost) \\[4pt]
$\mathcal{A} = \sum w_i \cdot v_i^2$ & \\[2pt]
$\mathcal{P}(S, N) = \sqrt{S / N}$ & \\[2pt]
$\mathcal{N}(S, \tau) = S / \tau^2$ &
\end{tabular}

\medskip

Because the aggregate grows quadratically with per-instance load, the default
model (which assumes a linear relationship) would underestimate the required
count (it divides by $\tau$ instead of $\tau^2$).

\subsection{What Does Not Work: Cross-Instance Coupling}
\label{sec:model-coupling}

The algorithm cannot handle models where the aggregate depends on the fleet size $N$
or on interactions between instances. It breaks the core assumptions the algorithm relies on.

\paragraph{Scaling invariance breaks.}
The algorithm assumes $\mathcal{A}$ stays approximately constant when load redistributes
after scaling (no external change). Consider $\mathcal{A} = \sum v_i - c \cdot N^2$:
adding an instance changes the $-c \cdot N^2$ term even though no external load arrived.
The aggregate now moves with scaling decisions, not just with external load.

\paragraph{Prediction becomes circular.}
The prediction stage forecasts future $\mathcal{A}$ using trend smoothing, assuming
$\mathcal{A}$ reflects external load. If $\mathcal{A}$ also depends on $N$, then the
predicted $A_H$ bakes in the current fleet size. Scaling changes $N$, which changes
what $\mathcal{A}$ would be, which invalidates the prediction that motivated the scaling.

\paragraph{$\mathcal{P}$ may not be monotonically decreasing.}
For $\mathcal{P}(S, N) = S/N + c \cdot N$, adding instances eventually makes
per-instance load \emph{worse}. $\mathcal{N}(S, \tau)$ may have no solution or two
solutions, and the scale-up logic has no well-defined target.

\section{Implementation: Intelligent Command Center}
\label{sec:icc-implementation}

We implement and deploy the algorithm in a production Kubernetes environment
using the Platformatic Intelligent Command Center (ICC), a cloud control
plane that manages Node.js applications. ICC monitors resource usage across
the cluster, runs the algorithm pipeline, and adjusts the number of pods to
keep applications healthy under changing load.

\subsection{Deployment Model}

Applications run on \textbf{Watt}, the Platformatic runtime.
A single Watt instance can run multiple applications, each in
its own worker thread within the same process. In a Kubernetes deployment,
each pod runs one Watt instance, and the Deployment is replicated across
multiple pods. An \emph{application} in this context is a logical Node.js
service: it can be a pure Node.js server, a Next.js frontend, or any other
framework supported by Watt.

\subsection{Concept Mapping}

The abstract concepts from the algorithm map to ICC as follows:

\begin{table}[H]
\centering
\begin{tabularx}{\textwidth}{@{} l X @{}}
\toprule
\textbf{Algorithm concept} & \textbf{ICC realization} \\
\midrule
Instance
    & A Kubernetes pod running one Watt instance. \\
Metric
    & Event Loop Utilization (ELU) and heap memory usage, standard
      Node.js runtime metrics measured by Watt per application. Both
      satisfy the metric properties from Section~\ref{sec:introduction}:
      monotonically related to load, distributed across pods, and with
      meaningful overload thresholds. \\
Metric model
    & Default (sum / average). Both ELU and heap redistribute
      approximately linearly across pods when instances are added or
      removed. \\
Threshold $\tau$
    & Configurable per application and per metric. \\
Instance bounds
    & $N_{\min}$ and $N_{\max}$ are configurable per application, and
      can also be set via Kubernetes labels on the Deployment. \\
Scaling action
    & Updating the Kubernetes Deployment replica count. \\
\bottomrule
\end{tabularx}
\caption{Mapping of algorithm concepts to the ICC implementation.}
\label{tab:icc-mapping}
\end{table}

\subsection{Architecture}

\begin{figure}[H]
\centering
\begin{tikzpicture}[
    box/.style={draw, rounded corners=3pt, text centered, font=\small},
    appbox/.style={draw, rounded corners=2pt, minimum width=2.4cm,
                   minimum height=0.5cm, font=\scriptsize, fill=black!5},
    arr/.style={-{Stealth[length=5pt]}, thick},
    lbl/.style={font=\scriptsize, text=black!60}
]

\node[box, minimum width=3.0cm, minimum height=1.2cm] (k8s) at (0,3.5) {Kubernetes API};
\node[box, minimum width=3.0cm, minimum height=1.2cm] (icc) at (7,3.5) {ICC};

\draw[arr] (icc.west) -- (k8s.east)
    node[midway, above, lbl] {target replica count};


\node[box, minimum width=4.2cm, minimum height=2.8cm] (watt) at (0,-0.5) {};
\node[font=\small, anchor=north] at ([yshift=-4pt]watt.north) {Watt};

\node[appbox] (app1) at ([yshift=-2pt]watt.center) {App 1};
\node[appbox, below=0.25cm of app1] (app2) {App 2};

\node[box, minimum width=2.6cm, minimum height=1.2cm] (extra) at (7,-0.5) {Watt-Extra};

\begin{scope}[on background layer]
    \node[draw, dashed, rounded corners=5pt, black!40,
          fit=(watt)(extra),
          inner xsep=12pt, inner ysep=18pt] (pod) {};
\end{scope}
\node[font=\scriptsize, black!50, anchor=south east] at (pod.north east)
    {Pod ($\times N$)};

\draw[arr] (watt.east) -- (extra.west)
    node[midway, above, lbl] {metrics};

\draw[arr] (extra.north) -- (icc.south)
    node[pos=0.25, left=3pt, lbl] {metric batches};

\draw[arr, dashed, black!50]
    (k8s.south) -- (watt.north |- pod.north)
    node[midway, left=3pt, lbl] {pod lifecycle};

\end{tikzpicture}
\caption{Data flow in ICC. Each pod runs a Watt instance hosting one or more
applications. Watt measures per-application metrics; Watt-Extra collects them
into batches and sends them to ICC\@. ICC runs the algorithm pipeline and
updates the Kubernetes Deployment replica count.}
\label{fig:icc-data-flow}
\end{figure}
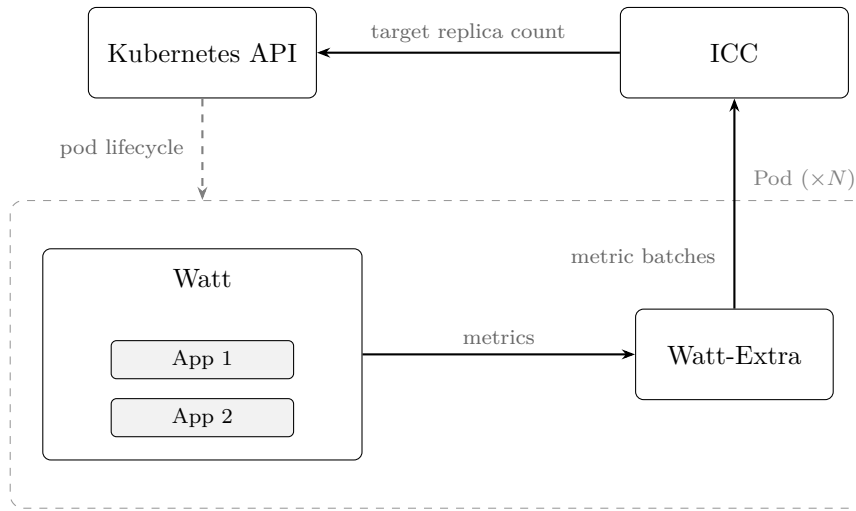

Each Watt deployment is scaled independently: ICC maintains a separate
pipeline instance per deployment, so scaling decisions for one never
interfere with another.

\paragraph{Metric collection.}
Watt measures ELU and heap usage per application at regular intervals.
A companion component, \textbf{Watt-Extra}, subscribes to these measurements,
collects them into batches, and sends them to ICC over HTTP\@. Watt-Extra
implements the dynamic batch timing described in
Section~\ref{sec:architecture}: batches are sent frequently under load
(e.g.\ every 5\,s) and infrequently when idle (e.g.\ every 40\,s).

\paragraph{Instance lifecycle.}
Watt-Extra maintains a persistent connection to ICC\@. When a pod starts,
the connection is established and ICC records the pod's start time ---
required by the redistribution stage (Section~\ref{sec:redistribution}) to
compute stabilization weights. When a pod is terminated (gracefully or
not), the connection drops and ICC detects the loss immediately, allowing
prompt removal of the pod from the active instance set.

\paragraph{Pipeline execution.}
Because metrics arrive in asynchronous batches from different pods, all five
pipeline stages are used: alignment places irregularly-timed samples onto a
uniform grid, imputation estimates values for pods that haven't reported yet,
and redistribution, prediction, and decision operate on the resulting clean
signal. When the pipeline produces a target pod count that differs from the
current count, ICC updates the Kubernetes Deployment replica count.

\paragraph{Adaptive init timeout.}
The init timeout $T_I$ is estimated from observed pod startup times using the
adaptive mechanism from Section~\ref{sec:adaptive-init-timeout}. This allows
the prediction horizon to track the actual Kubernetes scheduling and
initialization latency, which varies across clusters and over time.

\subsection{Multi-Application Deployments}

Because Watt can host multiple applications in the same process, each pod
produces independent metrics for each application it runs. ICC runs the
algorithm pipeline separately for each (application, metric) pair within a
deployment. The final target pod count is the maximum across all pipelines:
the most resource-constrained application drives the scaling decision. This
ensures that no application becomes a bottleneck, even if others in the same
Watt instance are lightly loaded.

\section{Performance Comparison}
\label{sec:performance-comparison}

This section compares the predictive scaling algorithm (running in ICC) against
two widely-used Kubernetes scalers (HPA and KEDA) under the same load, on
the same cluster, with the same application.

\subsection{Test Design}

A Next.js 16 e-commerce application (App Router, Server Components, SSR) runs
on Platformatic Watt with one worker per pod (1~CPU / 2\,GB RAM). An Envoy
proxy with 30\,s linear slow start sits between the load balancer and the pods,
ramping traffic to new pods gradually so that V8 JIT compilation on cold code
paths does not distort the comparison.

All three scalers operate on the same deployment (min~4, max~20 pods):

\begin{itemize}[leftmargin=*, itemsep=3pt]
    \item \textbf{ICC}, the predictive algorithm described in this document,
        scaling on Event Loop Utilization (ELU) with a 0.7 threshold.
    \item \textbf{KEDA}, scaling on the same metric (ELU) via a Prometheus
        query, with the same 0.7 threshold.
    \item \textbf{HPA}, scaling on CPU utilization with a 70\% target.
\end{itemize}

KEDA uses the same metric and threshold as ICC, so the comparison isolates
the scaling algorithm. HPA is included because it is the most widely deployed
Kubernetes scaler, not as a direct comparison: its results reflect the
choice of metric (CPU instead of ELU) in addition to the reactive algorithm.
CPU utilization does not directly measure event loop saturation in Node.js
applications: the event loop can be nearly saturated while CPU reports
moderate usage.

Each scaler is tested under two load profiles:

\begin{itemize}[leftmargin=*, itemsep=4pt]
    \item \textbf{Steady ramp}, traffic grows from 10 to 800\,req/s over
        ${\sim}$2.5 minutes, then holds at 800\,req/s for 90 seconds. This is
        the primary comparison: the most common real-world pattern, where
        traffic grows gradually as users arrive over the course of minutes.
    \item \textbf{Sudden spike}, traffic jumps from 0 to 800\,req/s in 10
        seconds, then holds at 800\,req/s for 120 seconds. This tests behavior
        when there is minimal trend history to extrapolate from.
\end{itemize}

Between tests, the deployment is reset to 4 pods and warmed up to ensure
consistent starting conditions.

\subsection{Scaling Behavior}

Each chart below shows three traces: the average ELU across all pods (purple,
left axis), the pod count (green, right axis), and the target request rate
(grey shaded area). The dashed red line marks the ELU threshold $\tau = 0.7$.

\paragraph{Steady ramp.}
Traffic grows from 10 to 800\,req/s over ${\sim}$2.5 minutes, then holds
at 800\,req/s for 90 seconds.

\begin{figure}[H]
\centering
\begin{tikzpicture}
\begin{axis}[
    width=0.92\textwidth,
    height=7cm,
    xlabel={Time},
    ylabel={ELU (avg across pods)},
    xmin=0, xmax=300,
    ymin=0, ymax=1.05,
    xtick={60, 120, 180, 240},
    xticklabels={1:00, 2:00, 3:00, 4:00},
    ytick={0, 0.2, 0.4, 0.6, 0.8, 1.0},
    axis y line*=left,
    grid=major,
    major grid style={dashed, black!15},
    legend style={at={(0.5,-0.15)}, anchor=north, font=\small, legend columns=4,
        column sep=8pt},
    every axis x label/.style={at={(ticklabel* cs:1.0)}, anchor=west},
    clip=false,
]

\addplot[thick, dashed, red!70, mark=none] coordinates {(0,0.7) (300,0.7)};
\addlegendentry{ELU threshold}

\addplot[cyan!60!black, mark=none, fill=cyan!15, fill opacity=0.3] coordinates {
    (0,0.0125) (30,0.0125) (50,0.25) (70,0.375) (90,0.5) (110,0.625)
    (130,0.75) (150,0.875) (170,1.0) (290,1.0)
} \closedcycle;
\addlegendentry{Target req/s}

\addplot[thick, violet, mark=none] coordinates {
    (0,0.002) (7,0.002) (13,0.002) (20,0.002) (26,0.002) (33,0.127)
    (39,0.317) (46,0.228) (52,0.099) (59,0.066) (65,0.075) (72,0.118)
    (78,0.253) (85,0.407) (91,0.452) (98,0.535) (105,0.607) (111,0.655)
    (117,0.644) (124,0.663) (131,0.650) (137,0.695) (144,0.671)
    (150,0.625) (157,0.636) (163,0.688) (170,0.669) (176,0.754)
    (183,0.802) (189,0.636) (196,0.671) (202,0.740) (209,0.705)
    (215,0.695) (222,0.711) (228,0.640) (235,0.616) (241,0.626)
    (248,0.618) (255,0.629) (261,0.601) (268,0.607) (274,0.611)
    (281,0.623) (287,0.603)
};
\addlegendentry{Avg.\ ELU}

\addlegendimage{semithick, black!50}
\addlegendentry{Pod count}

\end{axis}

\begin{axis}[
    width=0.92\textwidth,
    height=7cm,
    xmin=0, xmax=300,
    ymin=0, ymax=14,
    axis y line*=right,
    axis x line=none,
    ylabel={Pods},
    ytick={4, 6, 8, 10, 12},
]
\addplot[semithick, black!50, mark=none, const plot, forget plot] coordinates {
    (1,4) (103,5) (113,6) (133,7) (174,9) (184,10) (214,11) (300,11)
};
\end{axis}
\end{tikzpicture}
\caption{ICC: ELU and pod count.}
\label{fig:bench-icc}
\end{figure}
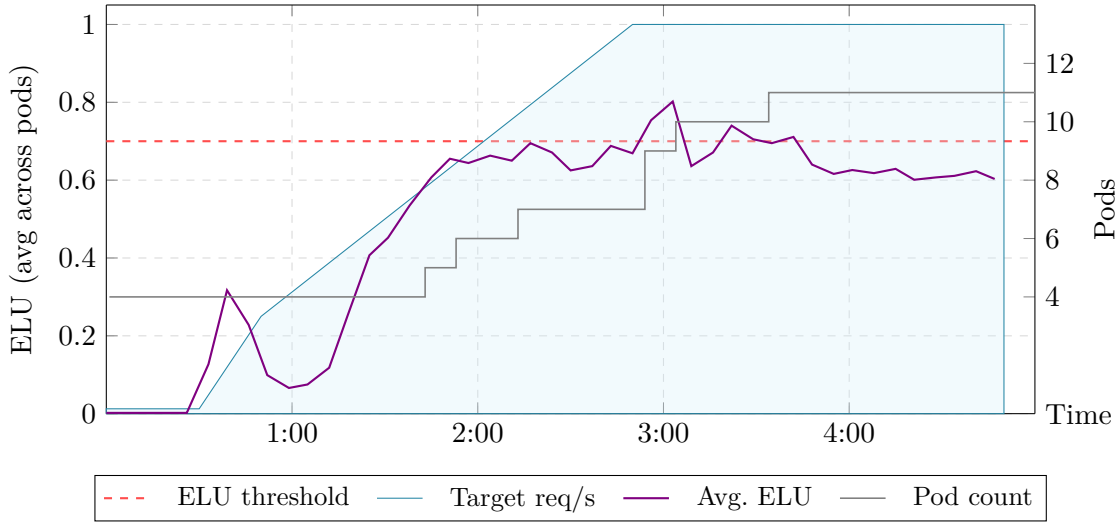

The predictive algorithm in ICC keeps ELU near the 0.7 threshold.
This happens because the algorithm acts on the trend, projects where ELU will be,
and scales up in advance to adjust capacity accordingly. It also does not
over-provision: it uses the minimum number of pods needed to keep ELU near the threshold.

\begin{figure}[H]
\centering
\begin{tikzpicture}
\begin{axis}[
    width=0.92\textwidth,
    height=7cm,
    xlabel={Time},
    ylabel={ELU (avg across pods)},
    xmin=0, xmax=300,
    ymin=0, ymax=1.05,
    xtick={60, 120, 180, 240},
    xticklabels={1:00, 2:00, 3:00, 4:00},
    ytick={0, 0.2, 0.4, 0.6, 0.8, 1.0},
    axis y line*=left,
    grid=major,
    major grid style={dashed, black!15},
    legend style={at={(0.5,-0.15)}, anchor=north, font=\small, legend columns=4,
        column sep=8pt},
    every axis x label/.style={at={(ticklabel* cs:1.0)}, anchor=west},
    clip=false,
]

\addplot[thick, dashed, red!70, mark=none] coordinates {(0,0.7) (300,0.7)};
\addlegendentry{ELU threshold}

\addplot[cyan!60!black, mark=none, fill=cyan!15, fill opacity=0.3] coordinates {
    (0,0.0125) (30,0.0125) (50,0.25) (70,0.375) (90,0.5) (110,0.625)
    (130,0.75) (150,0.875) (170,1.0) (290,1.0)
} \closedcycle;
\addlegendentry{Target req/s}

\addplot[thick, violet, mark=none] coordinates {
    (1,0.020) (7,0.021) (14,0.002) (20,0.002) (27,0.002) (33,0.002)
    (40,0.078) (46,0.136) (53,0.208) (59,0.090) (66,0.049) (72,0.058)
    (79,0.229) (85,0.293) (92,0.429) (98,0.469) (105,0.517) (111,0.555)
    (118,0.622) (124,0.694) (131,0.730) (138,0.812) (144,0.909)
    (151,0.891) (157,0.889) (164,0.870) (170,0.792) (177,0.811)
    (183,0.819) (190,0.916) (196,0.866) (203,0.757) (209,0.806)
    (216,0.850) (222,0.890) (229,0.801) (235,0.816) (242,0.809)
    (248,0.762) (255,0.691) (261,0.636) (268,0.631) (275,0.578)
    (281,0.575) (288,0.578)
};
\addlegendentry{Avg.\ ELU}

\addlegendimage{semithick, black!50}
\addlegendentry{Pod count}

\end{axis}

\begin{axis}[
    width=0.92\textwidth,
    height=7cm,
    xmin=0, xmax=300,
    ymin=0, ymax=14,
    axis y line*=right,
    axis x line=none,
    ylabel={Pods},
    ytick={4, 6, 8, 10, 12},
]
\addplot[semithick, black!50, mark=none, const plot, forget plot] coordinates {
    (1,4) (141,5) (156,6) (186,8) (216,10) (231,12) (300,12)
};
\end{axis}
\end{tikzpicture}
\caption{KEDA: ELU and pod count.}
\label{fig:bench-keda}
\end{figure}
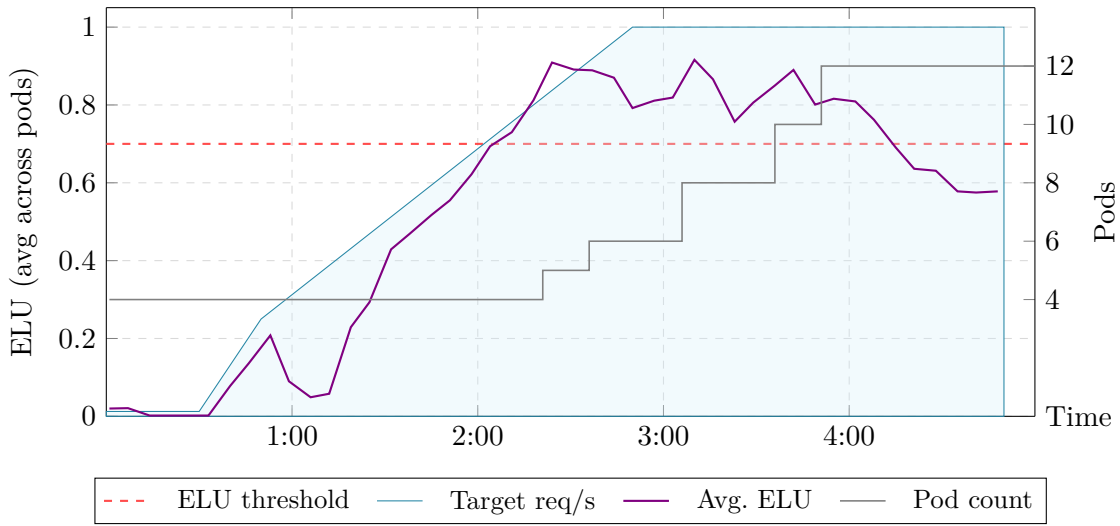

KEDA uses the same metric (ELU) and the same threshold (0.7). Both KEDA and
HPA compute the target replica count from the sum of current metric values
across all instances:

\[
N^* = \left\lceil \frac{S}{\tau} \right\rceil, \qquad S = \sum_{i=1}^{N} v^i
\]

where $v^i$ is the current metric value of instance $i$. This is a reactive approach: it waits for ELU to cross the
threshold before acting.
As a result, it fails to keep ELU under the threshold during the ramp-up period,
and average ELU reaches 0.92 at the peak of the load.

This shows the fundamental limitation of reactive scaling: it can only respond to
changes after they happen, not anticipate them. When load grows, the scaler is
always behind the curve, allowing ELU to climb well above the threshold before
it reacts.

The problem becomes even worse as overloaded instances' performance decreases in a
non-linear way due to queueing effects, which eventually forces KEDA and other
reactive algorithms to over-provision. 

Lowering the threshold does not solve this. A lower threshold does not make the
scaler react \emph{faster}, it makes it react to a \emph{lower value}. The
only difference is that the application now runs at a lower utilization baseline
permanently, using more pods to handle the same load. This trades constant
over-provisioning for slightly more headroom when a spike hits, a cost paid
at all times, not just during spikes.

\begin{figure}[H]
\centering
\begin{tikzpicture}
\begin{axis}[
    width=0.92\textwidth,
    height=7cm,
    xlabel={Time},
    ylabel={ELU (avg across pods)},
    xmin=0, xmax=300,
    ymin=0, ymax=1.05,
    xtick={60, 120, 180, 240},
    xticklabels={1:00, 2:00, 3:00, 4:00},
    ytick={0, 0.2, 0.4, 0.6, 0.8, 1.0},
    axis y line*=left,
    grid=major,
    major grid style={dashed, black!15},
    legend style={at={(0.5,-0.15)}, anchor=north, font=\small, legend columns=4,
        column sep=8pt},
    every axis x label/.style={at={(ticklabel* cs:1.0)}, anchor=west},
    clip=false,
]

\addplot[thick, dashed, red!70, mark=none] coordinates {(0,0.7) (300,0.7)};
\addlegendentry{ELU threshold}

\addplot[cyan!60!black, mark=none, fill=cyan!15, fill opacity=0.3] coordinates {
    (0,0.0125) (30,0.0125) (50,0.25) (70,0.375) (90,0.5) (110,0.625)
    (130,0.75) (150,0.875) (170,1.0) (290,1.0)
} \closedcycle;
\addlegendentry{Target req/s}

\addplot[thick, violet, mark=none] coordinates {
    (0,0.002) (7,0.002) (13,0.002) (20,0.002) (26,0.002) (33,0.112)
    (39,0.163) (46,0.163) (52,0.163) (59,0.080) (65,0.041) (72,0.126)
    (78,0.252) (85,0.310) (91,0.396) (98,0.504) (104,0.533) (111,0.644)
    (117,0.672) (124,0.795) (130,0.877) (137,0.908) (143,0.952)
    (150,0.803) (156,0.838) (163,0.864) (172,0.863) (179,0.973)
    (186,0.898) (192,0.866) (199,0.892) (205,0.943) (212,0.910)
    (218,0.823) (225,0.802) (231,0.848) (238,0.895) (244,0.942)
    (251,0.954) (257,0.763) (264,0.694) (270,0.670) (277,0.700)
    (283,0.648) (290,0.661)
};
\addlegendentry{Avg.\ ELU}

\addlegendimage{semithick, black!50}
\addlegendentry{Pod count}

\end{axis}

\begin{axis}[
    width=0.92\textwidth,
    height=7cm,
    xmin=0, xmax=300,
    ymin=0, ymax=14,
    axis y line*=right,
    axis x line=none,
    ylabel={Pods},
    ytick={4, 6, 8, 10, 12},
]
\addplot[semithick, black!50, mark=none, const plot, forget plot] coordinates {
    (1,4) (141,5) (171,6) (201,8) (246,11) (300,11)
};
\end{axis}
\end{tikzpicture}
\caption{HPA: ELU and pod count.}
\label{fig:bench-hpa}
\end{figure}
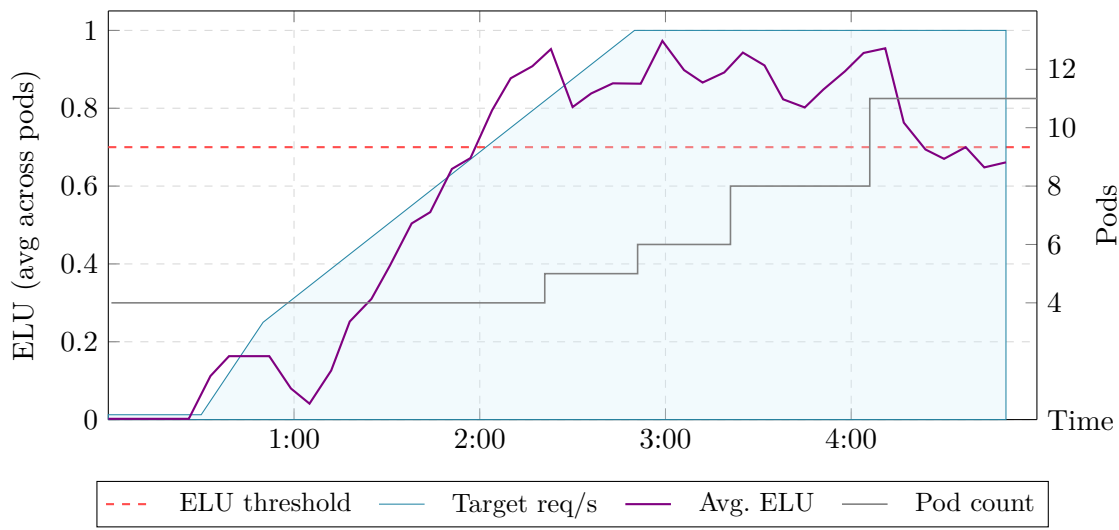

HPA shows the same reactive pattern as KEDA, but scales on CPU utilization
rather than ELU\@. CPU is a coarser indicator of Node.js application health:
the event loop can be nearly saturated while CPU reports a different picture,
or vice versa.

\paragraph{Impact on latency.}
The scaling behavior above directly determines what users experience. When ELU
is below the threshold, the event loop processes requests promptly. When ELU
exceeds the threshold, requests queue and latencies climb into seconds ---
eventually reaching the client timeout.

\begin{table}[H]
\centering
\begin{tabularx}{\textwidth}{@{} l *{3}{>{\centering\arraybackslash}X} @{}}
\toprule
& \textbf{ICC} & \textbf{KEDA} & \textbf{HPA} \\
\midrule
Success rate          & 99.47\%    & 95.11\%    & 90.97\% \\
Avg.\ latency        & 167\,ms    & 1,174\,ms  & 1,499\,ms \\
Median latency        & 26\,ms     & 154\,ms    & 522\,ms \\
p(90) latency         & 317\,ms    & 3,530\,ms  & 4,168\,ms \\
p(99) latency         & 1,970\,ms  & 10,001\,ms & 10,001\,ms \\
Errors                & 718        & 6,591      & 12,039 \\
\bottomrule
\end{tabularx}
\caption{Steady ramp: latency and error rates.}
\label{tab:benchmark-ramp}
\end{table}

ICC kept ELU near the threshold throughout, achieving a 99.47\% success rate
and 317\,ms at p(90). KEDA and HPA spent extended periods well above the
threshold: KEDA lost 5\% of requests, HPA lost 9\%. Their p(99) latencies
hit the 10\,s client timeout because the queue grew faster than the event
loop could drain it.

\paragraph{Sudden spike.}
\label{sec:spike-scenario}

The spike scenario jumps from 0 to 800\,req/s in 10 seconds, then holds.
No scaler can prevent the initial overload: there is no trend history to
predict from and no time for new pods to start. The question is how quickly
each scaler recovers.

\begin{figure}[H]
\centering
\begin{tikzpicture}
\begin{axis}[
    width=0.92\textwidth,
    height=7cm,
    xlabel={Time},
    ylabel={ELU (avg across pods)},
    xmin=0, xmax=170,
    ymin=0, ymax=1.05,
    xtick={30, 60, 90, 120, 150},
    xticklabels={0:30, 1:00, 1:30, 2:00, 2:30},
    ytick={0, 0.2, 0.4, 0.6, 0.8, 1.0},
    axis y line*=left,
    grid=major,
    major grid style={dashed, black!15},
    legend style={at={(0.5,-0.15)}, anchor=north, font=\small, legend columns=4,
        column sep=8pt},
    every axis x label/.style={at={(ticklabel* cs:1.0)}, anchor=west},
    clip=false,
]

\addplot[thick, dashed, red!70, mark=none] coordinates {(0,0.7) (170,0.7)};
\addlegendentry{ELU threshold}

\addplot[cyan!60!black, mark=none, fill=cyan!15, fill opacity=0.3] coordinates {
    (0,0) (10,1.0) (130,1.0)
} \closedcycle;
\addlegendentry{Target req/s}

\addplot[thick, violet, mark=none] coordinates {
    (0,0.010) (7,0.002) (13,0.002) (20,0.002) (26,0.010) (32,0.029)
    (39,0.410) (45,0.838) (52,1.000) (58,0.695) (65,0.605) (72,0.644)
    (78,0.743) (85,0.816) (91,0.880) (98,0.751) (104,0.695) (111,0.668)
    (117,0.634) (124,0.615) (130,0.607) (137,0.567) (143,0.553)
    (150,0.542) (156,0.541) (163,0.543)
};
\addlegendentry{Avg.\ ELU}

\addlegendimage{semithick, black!50}
\addlegendentry{Pod count}

\end{axis}

\begin{axis}[
    width=0.92\textwidth,
    height=7cm,
    xmin=0, xmax=170,
    ymin=0, ymax=16,
    axis y line*=right,
    axis x line=none,
    ylabel={Pods},
    ytick={4, 6, 8, 10, 12, 14},
]
\addplot[semithick, black!50, mark=none, const plot, forget plot] coordinates {
    (1,4) (46,8) (86,12) (96,13) (166,13) (170,13)
};
\end{axis}
\end{tikzpicture}
\caption{ICC: spike scenario.}
\label{fig:bench-spike-icc}
\end{figure}
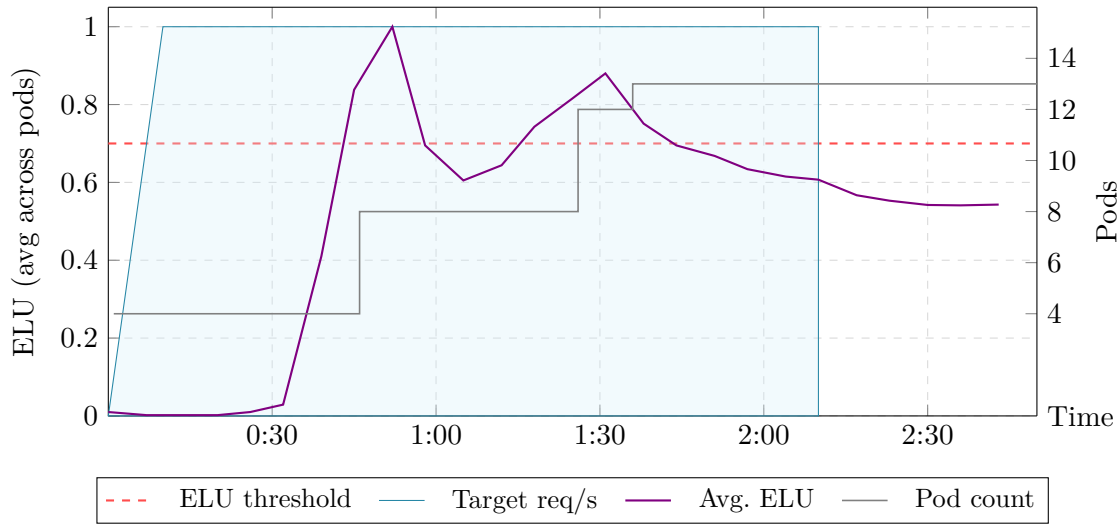

Without trend history, ICC cannot predict the spike. But once the first
samples arrive, the trend estimate builds rapidly. The asymmetric
parameters ($\alpha_{\uparrow} > \alpha_{\downarrow}$) ensure the upward
movement is captured within a few ticks. The saturation mechanism
(Section~\ref{sec:saturation}) preserves the trend even while ELU is
clipped at 1.0, so the algorithm continues scaling despite the flat signal.

\begin{figure}[H]
\centering
\begin{tikzpicture}
\begin{axis}[
    width=0.92\textwidth,
    height=7cm,
    xlabel={Time},
    ylabel={ELU (avg across pods)},
    xmin=0, xmax=170,
    ymin=0, ymax=1.05,
    xtick={30, 60, 90, 120, 150},
    xticklabels={0:30, 1:00, 1:30, 2:00, 2:30},
    ytick={0, 0.2, 0.4, 0.6, 0.8, 1.0},
    axis y line*=left,
    grid=major,
    major grid style={dashed, black!15},
    legend style={at={(0.5,-0.15)}, anchor=north, font=\small, legend columns=4,
        column sep=8pt},
    every axis x label/.style={at={(ticklabel* cs:1.0)}, anchor=west},
    clip=false,
]

\addplot[thick, dashed, red!70, mark=none] coordinates {(0,0.7) (170,0.7)};
\addlegendentry{ELU threshold}

\addplot[cyan!60!black, mark=none, fill=cyan!15, fill opacity=0.3] coordinates {
    (0,0) (10,1.0) (130,1.0)
} \closedcycle;
\addlegendentry{Target req/s}

\addplot[thick, violet, mark=none] coordinates {
    (1,0.002) (7,0.002) (14,0.002) (20,0.002) (27,0.002) (33,0.002)
    (40,0.186) (46,0.606) (53,0.942) (59,0.943) (66,0.667) (72,0.667)
    (79,0.731) (85,0.779) (92,0.827) (98,0.965) (105,0.914) (111,0.761)
    (118,0.745) (124,0.752) (131,0.843) (137,0.862) (144,0.912)
    (151,0.965) (157,0.702) (164,0.646)
};
\addlegendentry{Avg.\ ELU}

\addlegendimage{semithick, black!50}
\addlegendentry{Pod count}

\end{axis}

\begin{axis}[
    width=0.92\textwidth,
    height=7cm,
    xmin=0, xmax=170,
    ymin=0, ymax=16,
    axis y line*=right,
    axis x line=none,
    ylabel={Pods},
    ytick={4, 6, 8, 10, 12, 14},
]
\addplot[semithick, black!50, mark=none, const plot, forget plot] coordinates {
    (2,4) (50,6) (96,9) (141,12) (170,12)
};
\end{axis}
\end{tikzpicture}
\caption{KEDA: spike scenario.}
\label{fig:bench-spike-keda}
\end{figure}
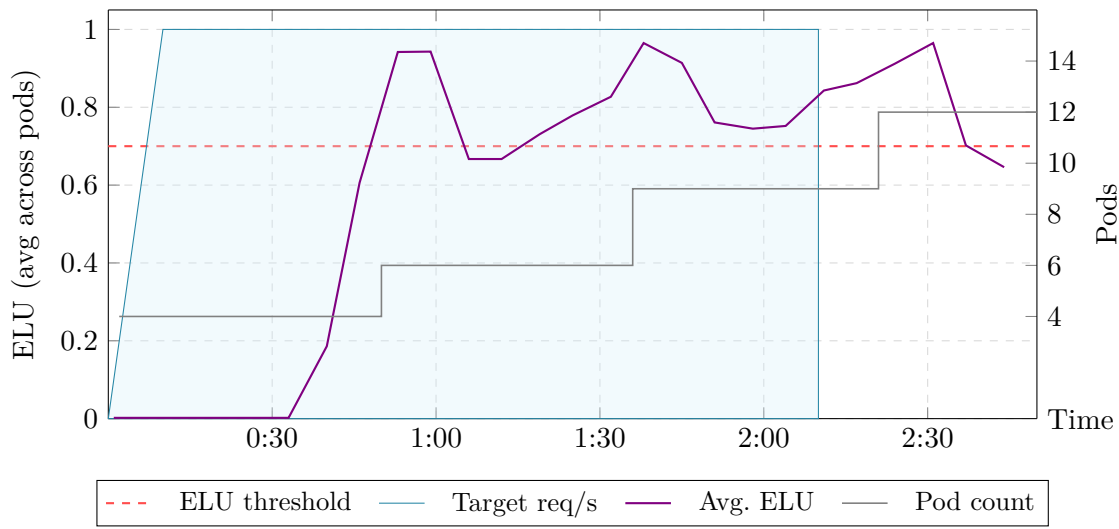

The reactive formula scales in proportion to the current overload ratio, but
each decision is based on a single snapshot. It cannot account for the fact
that load arrived all at once and more capacity is needed than the current
ratio suggests. The result is a staircase of incremental scale-ups, each
insufficient, while ELU remains elevated.

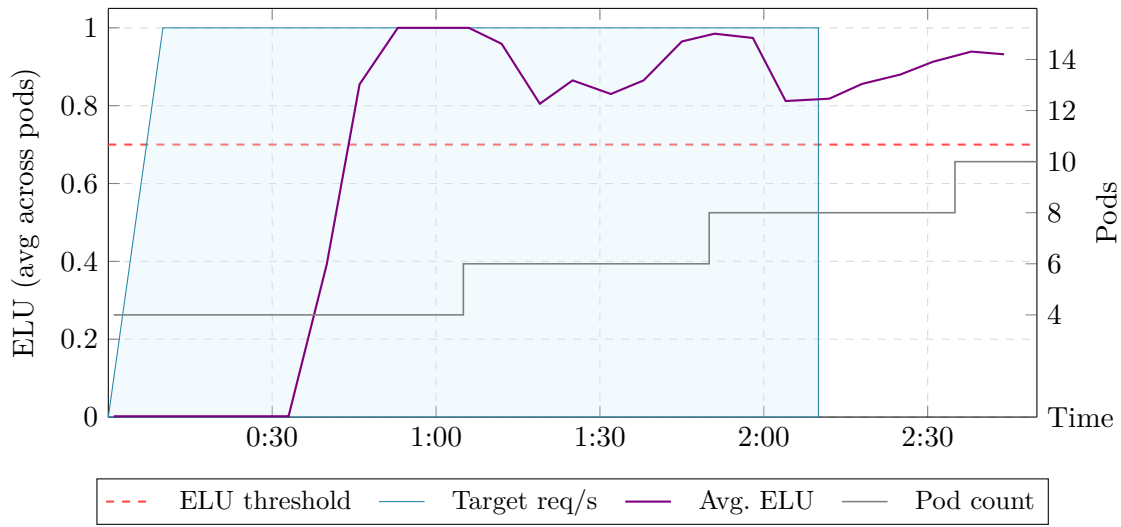
\begin{figure}[H]
\centering
\begin{tikzpicture}
\begin{axis}[
    width=0.92\textwidth,
    height=7cm,
    xlabel={Time},
    ylabel={ELU (avg across pods)},
    xmin=0, xmax=170,
    ymin=0, ymax=1.05,
    xtick={30, 60, 90, 120, 150},
    xticklabels={0:30, 1:00, 1:30, 2:00, 2:30},
    ytick={0, 0.2, 0.4, 0.6, 0.8, 1.0},
    axis y line*=left,
    grid=major,
    major grid style={dashed, black!15},
    legend style={at={(0.5,-0.15)}, anchor=north, font=\small, legend columns=4,
        column sep=8pt},
    every axis x label/.style={at={(ticklabel* cs:1.0)}, anchor=west},
    clip=false,
]

\addplot[thick, dashed, red!70, mark=none] coordinates {(0,0.7) (170,0.7)};
\addlegendentry{ELU threshold}

\addplot[cyan!60!black, mark=none, fill=cyan!15, fill opacity=0.3] coordinates {
    (0,0) (10,1.0) (130,1.0)
} \closedcycle;
\addlegendentry{Target req/s}

\addplot[thick, violet, mark=none] coordinates {
    (1,0.002) (7,0.002) (13,0.002) (20,0.002) (26,0.002) (33,0.002)
    (40,0.393) (46,0.855) (53,1.000) (59,1.000) (66,1.000) (72,0.959)
    (79,0.805) (85,0.865) (92,0.830) (98,0.865) (105,0.965) (111,0.985)
    (118,0.974) (124,0.812) (132,0.818) (138,0.856) (145,0.880)
    (151,0.913) (158,0.939) (164,0.932)
};
\addlegendentry{Avg.\ ELU}

\addlegendimage{semithick, black!50}
\addlegendentry{Pod count}

\end{axis}

\begin{axis}[
    width=0.92\textwidth,
    height=7cm,
    xmin=0, xmax=170,
    ymin=0, ymax=16,
    axis y line*=right,
    axis x line=none,
    ylabel={Pods},
    ytick={4, 6, 8, 10, 12, 14},
]
\addplot[semithick, black!50, mark=none, const plot, forget plot] coordinates {
    (1,4) (65,6) (110,8) (155,10) (170,10)
};
\end{axis}
\end{tikzpicture}
\caption{HPA: spike scenario.}
\label{fig:bench-spike-hpa}
\end{figure}

HPA faces the same reactive limitation as KEDA, compounded by using CPU
utilization, which lags behind event loop saturation in Node.js applications.
The scaler sees less urgency than the actual ELU would suggest, resulting in
even slower scaling.

\begin{table}[H]
\centering
\begin{tabularx}{\textwidth}{@{} l *{3}{>{\centering\arraybackslash}X} @{}}
\toprule
& \textbf{ICC} & \textbf{KEDA} & \textbf{HPA} \\
\midrule
Success rate          & 91.51\%    & 87.47\%    & 77.31\% \\
Avg.\ latency        & 1,126\,ms  & 1,989\,ms  & 2,205\,ms \\
Median latency        & 55\,ms     & 855\,ms    & 1,102\,ms \\
p(90) latency         & 3,385\,ms  & 6,108\,ms  & 7,338\,ms \\
p(99) latency         & 10,001\,ms & 10,001\,ms & 10,001\,ms \\
Errors                & 8,028      & 11,212     & 21,067 \\
\bottomrule
\end{tabularx}
\caption{Sudden spike: latency and error rates.}
\label{tab:benchmark-spike}
\end{table}

All three scalers suffer during the initial burst: the p(99) hits the 10\,s
client timeout across the board. The difference is in recovery. ICC's median
of 55\,ms means most requests after the initial burst were served normally,
while KEDA (855\,ms) and HPA (1,102\,ms) remained degraded throughout the
hold period. HPA lost nearly a quarter of all requests.

\subsection{Test Environment}

The benchmark ran on AWS EKS (us-east-1), Kubernetes v1.35, with 4 worker
nodes (m5.2xlarge: 8~vCPU, 32\,GB RAM each). Load was generated from a
dedicated EC2 instance (c7gn.2xlarge, ARM64) in the same VPC using Grafana k6.
The full benchmark automation, scaler configurations, and raw data are
available in the benchmark repository\footnote{%
\url{https://github.com/platformatic/k8s-watt-performance-demo/tree/scaler}}.

\begin{table}[H]
\centering
\begin{tabularx}{\textwidth}{@{} l l X @{}}
\toprule
\textbf{Parameter} & \textbf{Value} & \textbf{Description} \\
\midrule
$\tau$ (ELU)             & 0.7        & Per-instance overload threshold \\
$\Delta t$               & 1000\,ms   & Sample interval \\
$T_I$                    & 25\,s      & Init timeout \\
$T_R$                    & 30\,s      & Redistribution timeout \\
$\eta$                   & 1.2        & Prediction horizon multiplier \\
$H_{\min}$               & 10\,s      & Minimum prediction horizon \\
$\kappa$                 & 1          & Stabilization weight shape \\
$\alpha_{\uparrow}$      & 0.2        & Smoothing parameter (upward) \\
$\alpha_{\downarrow}$    & 0.1        & Smoothing parameter (downward) \\
$\beta_{\uparrow}$       & 0.2        & Trend parameter (upward) \\
$\beta_{\downarrow}$     & 0.1        & Trend parameter (downward) \\
$\gamma_0$               & $10°$      & Trend direction threshold \\
$k$                      & 2          & Risk dampening parameter \\
$\mu$                    & 0.3        & Scale-down margin \\
\bottomrule
\end{tabularx}
\caption{ICC configuration.}
\label{tab:config-icc}
\end{table}

\begin{table}[H]
\centering
\begin{tabularx}{\textwidth}{@{} l l X @{}}
\toprule
\textbf{Parameter} & \textbf{Value} & \textbf{Description} \\
\midrule
Metric                   & ELU        & Via Prometheus query \\
Threshold                & 0.7        & Same as ICC \\
Polling interval         & 15\,s      & Reduced from default 30\,s \\
\bottomrule
\end{tabularx}
\caption{KEDA configuration.}
\label{tab:config-keda}
\end{table}

\begin{table}[H]
\centering
\begin{tabularx}{\textwidth}{@{} l l X @{}}
\toprule
\textbf{Parameter} & \textbf{Value} & \textbf{Description} \\
\midrule
Metric                   & CPU        & Resource utilization \\
Target                   & 70\%       & Average utilization \\
Polling interval         & 15\,s      & HPA default \\
\bottomrule
\end{tabularx}
\caption{HPA configuration.}
\label{tab:config-hpa}
\end{table}

KEDA's polling interval was reduced from its default of 30\,s to 15\,s to
match HPA and provide a fairer comparison. With the default 30\,s interval,
KEDA's results would be worse.

\newpage
\section{Conclusion}
\label{sec:conclusion}

We presented a predictive scaling algorithm organized as a five-stage
pipeline. It takes a fundamentally different approach to scaling decisions
than existing solutions. Rather than treating the metric as a discrete value
sampled at evaluation time (where each decision is independent, with no
understanding of direction or dynamics), the algorithm treats the metric as
a continuous signal. Every sample contributes to a running estimate of the
level and its rate of change. The algorithm knows not just where the metric
is, but where it is heading. Short-term prediction follows naturally:
because the trend is already maintained, extrapolating to a future horizon
is straightforward: the algorithm forecasts where the load will be by the
time new capacity is ready and scales based on that forecast.

This continuous view works if the underlying signal reflects external
load rather than the side effects of the algorithm's own decisions.
Per-instance metrics fail this: adding an instance redistributes load and
moves every metric, even when traffic is unchanged. The algorithm operates
on a cluster-wide aggregate that is invariant under
scaling: the total load stays roughly the same when instances are added,
even as individual metrics shift. This is what makes the trend estimate
trustworthy.

The five pipeline stages deliver this clean signal. Alignment places
irregularly-timed samples onto a uniform grid. Imputation fills in values
for instances that haven't reported yet. Redistribution smooths the
transient distortion after scale-up through gradual weighted inclusion,
drop absorption, and a redistribution delta that communicates the expected
weight-growth ramp to the prediction stage so it does not contaminate the
trend estimate. By the time data reaches the prediction stage, it is
continuous, complete, and free of scaling artifacts. The decision stage
converts the forecast into a target instance count, with risk-aware
dampening that accounts for the uncertainty in any extrapolation.

Another core concept is the metric model. Different metrics relate to
scaling in different ways. The three functions $\mathcal{A}$,
$\mathcal{P}$, and $\mathcal{N}$ encode this relationship, so that every
calculation in the pipeline, from aggregation to the final instance
count, reflects how the specific metric actually behaves under scaling.

The algorithm is risk-aware. Acting on a forecast carries a risk: if the
predicted load does not materialize, the system ends up over-provisioned.
The algorithm assesses how much each scaling decision depends on projection
versus observed load, and adjusts its response accordingly. When the
observed load already confirms the need, the algorithm acts decisively.
The more the decision relies on the forecast, the more conservatively the
algorithm responds, reducing the risk of committing resources to load
that may never arrive.

In a controlled comparison against HPA and KEDA, the algorithm kept
per-instance load below the target threshold throughout, demonstrating
that short-term prediction on a scaling-invariant signal can outperform
reactive scaling under sustained load growth.


\end{document}